\documentclass{aa}  

\usepackage{graphicx}
\usepackage{txfonts}
\usepackage{upgreek}
\usepackage{xcolor}
\usepackage[colorlinks,citecolor=blue,urlcolor=blue,filecolor=blue,linkcolor=blue]{hyperref}

\newlength{\mywidth}
\DeclareGraphicsExtensions{.png,.pdf}

\begin{document} 
   \title{Analytical simulations of the effect of satellite
     constellations on optical and near-infrared observations}

   \author{C.\ G.\ Bassa\inst{1} \and O.\ R.\ Hainaut\inst{2} \and
     D.\ Galad\'{\i}-Enr\'{\i}quez\inst{3} }

   \institute{ASTRON Netherlands Institute for Radio Astronomy, Oude
     Hoogeveensedijk 4, 7991 PD Dwingeloo, The
     Netherlands\\
     \email{bassa@astron.nl}
     \and
     European Southern Observatory, Karl-Schwarzschild-Strasse 2,
     85748 Garching bei München, Germany\\
     \email{ohainaut@eso.org}
     \and
     Observatorio de Calar Alto, Sierra de los Filabres,
     04550-G\'ergal (Almer\'{\i}a), Spain\\ \email{dgaladi@caha.es}
   }

   \date{Received \today; Accepted \today}

% \abstract{}{}{}{}{} 
% 5 {} token are mandatory
 
  \abstract
  % context heading (optional) leave it empty if necessary  
      { The number of satellites in low-Earth orbit is increasing
        rapidly, and many tens of thousands of them are expected to be
        launched in the coming years. There is a strong concern among
        the astronomical community about the contamination of optical
        and near-infrared observations by satellite trails. Initial
        investigations of the impact of large satellite constellations
        have been presented in \citet{hw20} and \citet{mcd20}, among
        others.  }
      { We expand the impact analysis of such constellations on
        optical and near-infrared astronomical observations in a more
        rigorous and quantitative way, using updated constellation
        information, and considering imagers and spectrographs and
        their very different characteristics.}
      { We introduce an analytical method that allows us to rapidly
        and accurately evaluate the effect of a very large number of
        satellites, accounting for their magnitudes and the effect of
        trailing of the satellite image during the exposure. We use
        this to evaluate the impact on a series of representative
        instruments, including imagers (traditional narrow field
        instruments, wide-field survey cameras, and astro-photographic
        cameras) and spectrographs (long-slit and fibre-fed), taking
        into account their limiting magnitude.}
      { As already known \citep[][]{wal20}, the effect of satellite
        trails is more damaging for high-altitude satellites, on
        wide-field instruments, or essentially during the first and
        last hours of the night. Thanks to their brighter limiting
        magnitudes, low- and mid-resolution spectrographs will be less
        affected, but the contamination will be at about the same
        level as that of the science signal, introducing additional
        challenges. High-resolution spectrographs will essentially be
        immune. We propose a series of mitigating measures, including
        one that uses the described simulation method to optimize the
        scheduling of the observations. We conclude that no single
        mitigation measure will solve the problem of satellite trails
        for all instruments and all science cases.}
      {}


   \maketitle

   \section{Introduction: Satellite constellations}
   \label{sec:introduction}

   In the most general sense, a constellation of artificial satellites
   is a set of spacecraft that share a common design, distributed
   among different orbits to provide a service and/or a land coverage
   that cannot be achieved by means of just one single
   satellite. Satellite constellations have been in use since the
   early times of the space age and their applications range from
   telecommunications (civil or military, e.g.\ the Molniya series),
   meteorology (Meteosat), remote sensing (as the two-satellite
   constellations Sentinel of ESA) or global navigation satellite
   systems (GNSS; GPS, Glonass, Galileo or Beidou systems).
    
   Until recently, the satellite constellation with a largest number
   of elements was the Iridium system, with some 70 satellites in
   low-Earth orbit (LEO, altitude less than 2000\,km), aimed to
   provide cell phone services with worldwide coverage. The so-called
   Iridium flares caused by reflected sunlight from the flat antennae
   of the first generation of Iridium satellites awoke the awareness
   of the astronomical community about possible deleterious effects of
   satellite constellations on astronomical observations, both optical
   and in radio \citep[][]{jam98}. Global navigation satellites
   systems also require worldwide coverage that, in this case, can be
   achieved with smaller constellations, typically of the order of 30
   satellites, placed at higher orbits ($\sim20\,000$\,km).
    
   For the purpose of this paper we consider a {\em
     mega-constellation} to refer to any satellite constellation made
   up from a number of satellites significantly larger than the
   Iridium constellation, say in excess of 100 satellites. Several
   such systems have been proposed during the last decade, in all
   cases with the purpose to provide fast internet access
   worldwide. The technical requirements for that application (large
   two-way bandwidth, short delay time, large number of potential
   users) lead to designs of constellations made up from huge numbers
   of satellites, counted in the thousands, placed at low-Earth orbit.
    
   The contrast with traditional constellations is overwhelming. The
   examples mentioned above, or the plans to launch new constellations
   of nano- and pico-satellites, imply challenges of their own and
   contribute to the issues of orbital crowding, space debris
   management, radioelectric noise, etc., but they imply numbers of
   satellites orders of magnitude smaller and, most often, with
   elements much fainter, than the mega-constellations discussed in
   here.
  
   Several new generation mega-constellations are in the planning
   stages, and some number of satellites of two of these
   constellations, namely SpaceX's Starlink and OneWeb, have already
   been placed in orbit. In Table\,\ref{tab:1} we list some of the
   planned constellation configurations for which information is
   publicly available. Several more satellite operators indicated
   their intentions to build about a dozen of additional similar
   constellations. Many companies are also planning constellations of
   much smaller satellites (cubesats, nanosats...), which are less
   relevant for optical astronomy.  We note that satellite operators,
   even those with satellites already in orbit, frequently modify the
   configuration of their constellation projects. Furthermore, other
   satellite operators have indicated their intentions to launch
   mega-constellations, but without so far submitting any application
   or publishing any details. Therefore, the configurations used in
   this paper should be treated as a representative, and their impact
   on astronomical observations obtained from the following
   simulations and their interpretation are an illustration of the
   situation as it could be in the late 2020's. In first
   approximation, the numbers scale linearly with the total number of
   satellites in the constellation, so the effects be scaled.
    
   Of all these projects, the SpaceX Starlink constellation is clearly
   in the forefront. Since the launch of their first group of
   satellites in May 2019, the {\em trains} of very bright pearls that
   form the satellites illuminated by the Sun have caught the
   attention even of the general public, and have triggered the alarms
   of the astronomical community \citep[see, for
     example][]{Wit19}. These bright trains are formed only at the
   first stages of the deployment of each Starlink launch, with the
   satellites becoming fainter as they later climb to the final
   operational orbits, and attain an operational attitude that reduces
   their apparent brightness.
   
   Our aim in this paper is to assess the impact of satellite
   mega-constellations on optical and near-infrared astronomy from
   computer simulations of two kinds. Such simulations are described
   in \S\ref{sec:simulations}, where we depict the usual discrete
   simulations and, also, a new approach consisting on formulating
   statistical predictions from analytical probability density
   functions that describe the main properties of the
   constellations. The same section includes consideration on how to
   estimate the apparent brightness of satellites illuminated by
   sunlight, with special attention to the concept of {\em effective
     magnitude}.
   
   The effect on observations is addressed in
   \S\ref{sec:results}. There we provide details on the specific
   details of a set of simulations, their implications for several
   observing modes (direct imaging and spectroscopy, including
   multi-fiber instruments) and we discuss some possibilities to
   mitigate the impact in \S\ref{sec:discussion}. The appendix
   includes the details on how the equations of our analytical model
   are derived.

   \begin{table}[t]
     \caption{Orbit configurations for the satellite constellations
       considered in this paper, totaling almost 60 thousand
       satellites, providing the orbital altitude and inclination,
       number of satellites within an orbital plane $n_\mathrm{sat}$
       and the number of orbital planes $n_\mathrm{plane}$ making up
       the constellation (consisting of $n_\mathrm{sat}\times
       n_\mathrm{plane}$ satellites). These are the constellations for
       which supporting information is available. These constellations
       are therefore to be considered as representative of a plausible
       situation by $\sim 2030$, but the exact details of the
       distribution of satellites may be different.
       \label{tab:1}}
     % based on my list at https://europeansouthernobservatory-my.sharepoint.com/:x:/g/personal/ohainaut_eso_org/EZ6wemyLFAdNuepwXXiHBLgBgxmQ1J8mOkW0Rbrtv6kE1Q?e=suI6um
     \centering
     \begin{tabular}{rrrrr}
       \hline
       \multicolumn{1}{l}{Altitude} &
       \multicolumn{1}{l}{Inclination} &
       \multicolumn{1}{l}{$n_\mathrm{sat}$} &
       \multicolumn{1}{l}{$n_\mathrm{plane}$} &
       \multicolumn{1}{l}{$n_\mathrm{sat}\times n_\mathrm{plane}$}\\
       \hline
%grand total: 64526
%\multicolumn{5}{l}{Starlink Generation 0 \hfill 1584 satellites}\\
\multicolumn{5}{l}{Starlink Generation 1 \hfill 11926 satellites}\\
$550$\,km & $53\degr$ & 22 & 72 & 1584 \\[0.5em]
$540$\,km &  $53\fdg2$ &22& 72& 1584\\
$570$\,km &  $70\degr$ &20& 36&  720\\
$560$\,km &  $97\fdg6$ &58&  6&  348\\
$560$\,km &  $97\fdg6$ &43&  4&  172\\
$335.9$\,km & $42\degr$ & 60 & 42 & 2493 \\
$340.8$\,km & $48\degr$ & 60 & 42 & 2478 \\
$345.6$\,km & $53\degr$ & 60 & 42 & 2547 \\[0.5em]
\multicolumn{5}{l}{Starlink Generation 2 \hfill 30000 satellites}\\
$328$\,km & $30\degr$ & 1 & 7178 & 7178 \\
$334$\,km & $40\degr$ & 1 & 7178 & 7178 \\
$345$\,km & $53\degr$ & 1 & 7178 & 7178 \\
$360$\,km & $96\fdg9$ & 50 & 40 & 2000 \\
$373$\,km & $75\degr$ & 1 & 1998 & 1998 \\
$499$\,km & $53\degr$ & 1 & 4000 & 4000 \\
$604$\,km & $148\degr$ & 12 & 12 & 144 \\
$614$\,km & $115\fdg7$ & 18 & 18 & 324 \\
[0.5em]
\multicolumn{5}{l}{Amazon Kuiper \hfill 3236 satellites}\\
$630$\,km & $51\fdg9$ & 34 & 34 & 1156 \\
$610$\,km & $42\degr$ & 36 & 36 & 1296 \\
$590$\,km & $33\degr$ & 28 & 28 & 784 \\[0.5em]
\multicolumn{5}{l}{OneWeb Phase 1 \hfill 1980 satellites}\\
$1200$\,km & $87\fdg9$ & 55 & 36 & 1980 \\[0.5em]
%\multicolumn{5}{l}{OneWeb Phase 2 \hfill 47844 satellites}\\
%$1200$\,km & $87\fdg9$ & 49 & 36 & 1764 \\
%$1200$\,km & $40\degr$ & 720 & 32 & 23040 \\
%$1200$\,km & $55\degr$ & 720 & 32 & 23040 \\[0.5em]
\multicolumn{5}{l}{OneWeb Phase 2 revised \hfill 6372 satellites}\\
$1200$\,km & $87\fdg9$ & 49 & 36 & 1764 \\
$1200$\,km & $40\degr$ & 72 & 32 & 2304 \\
$1200$\,km & $55\degr$ & 72 & 32 & 2304 \\[0.5em]
\multicolumn{5}{l}{GuoWang GW-A59 \hfill 12992 satellites}\\
 $590$\,km & $85\degr$ &  60 &  8 &  480\\
 $600$\,km & $50\degr$ &  50 & 40 & 2000\\
 $508$\,km & $55\degr$ &  60 & 60 & 3600 \\
$1145$\,km & $30\degr$ &  64 & 27 & 1728\\
$1145$\,km & $40\degr$ &  64 & 27 & 1728\\
$1145$\,km & $50\degr$ &  64 & 27 & 1728\\
$1145$\,km & $60\degr$ &  64 & 27 & 1728 \\[0.5em]
       \hline
     \end{tabular}
   \end{table}

   \section{Simulations}
   \label{sec:simulations}
   \subsection{Visibility}
   \label{ssec:visibility}

   \begin{figure*}[t]
     \includegraphics[width=\textwidth]{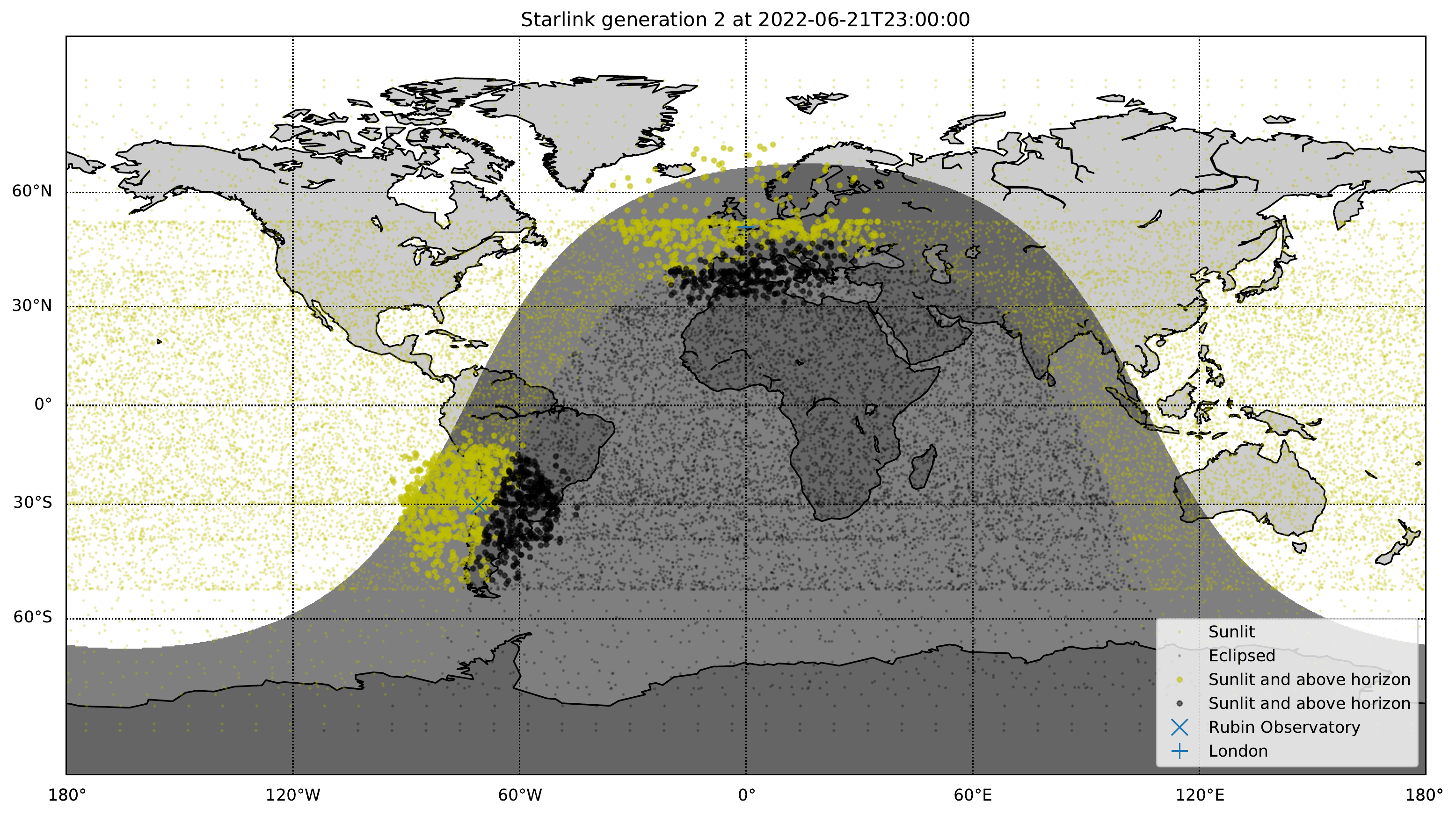}
     \caption{A discrete realization of the Starlink Generation 2
       constellation overlaid on a map of Earth. The day and night
       sides of Earth are shown for 2022 June 21 at 23:00UTC, near the
       June solstice. Satellites are shown as either sunlit (yellow),
       or eclipsed by the Earth (black) and above the horizon for two
       geographical locations, Vera Rubin Observatory in Chile
       (latitude $-30\degr$), and London in the United Kingdom
       (latitude $\sim 50\degr$). Over-densities of satellites in
       geographic latitude $b$ are visible at latitudes
       $|b|=30^\circ$, $40^\circ$ and $53^\circ$, corresponding to the
       inclination of the most populous orbital shells. Due to their
       orbital altitudes, satellites remain visible for locations
       where the Sun is below the horizon, and will remain visible
       throughout all the night for locations at high geographical
       latitudes.}
     \label{fig:earthmap}
   \end{figure*}
   
   \begin{figure*}[t]
     \includegraphics[width=\textwidth]{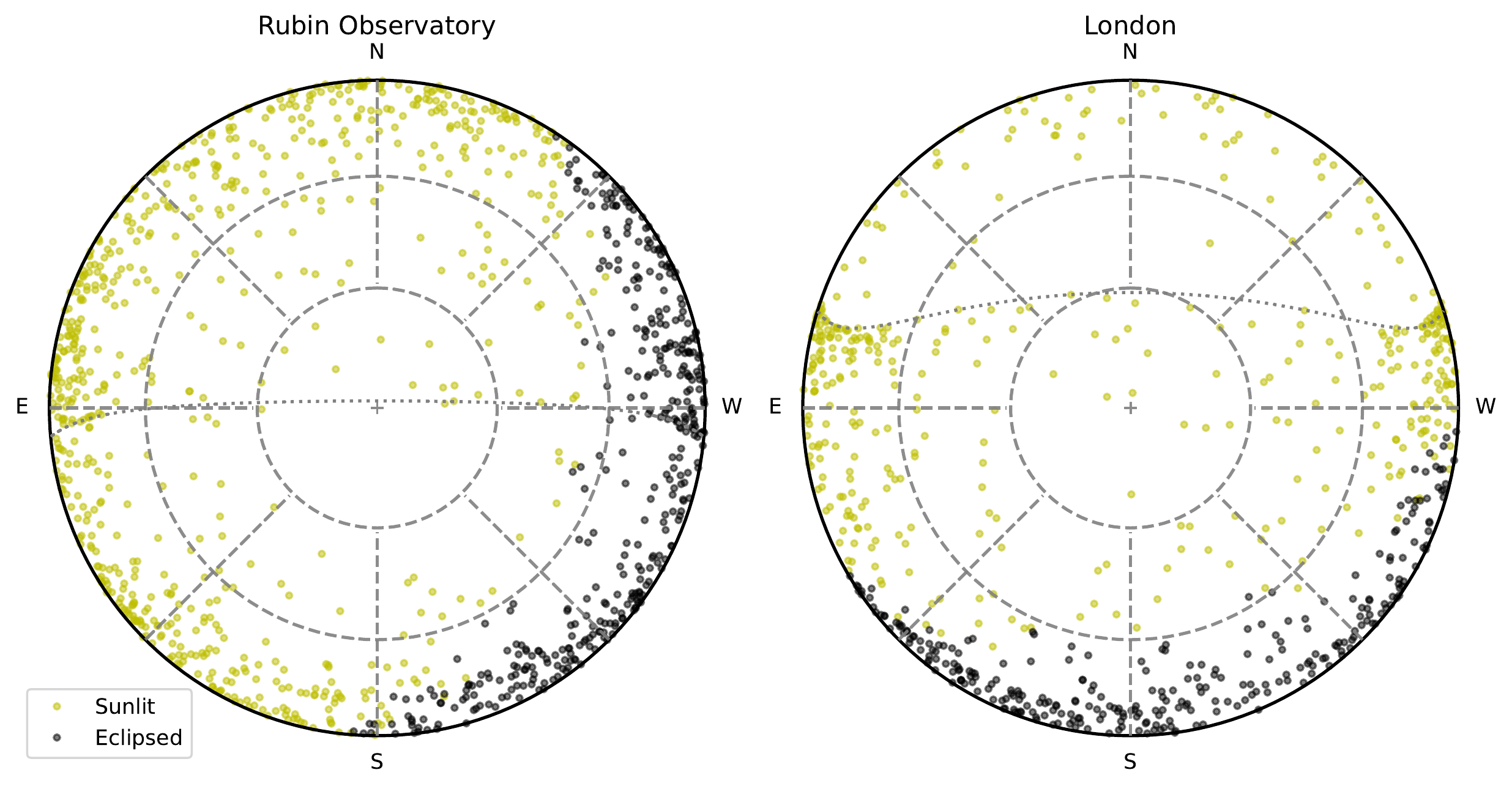}
     \caption{A discrete realization of the Starlink Generation 2
       constellation plotted on all sky maps representing the sky
       above Vera Rubin Observatory (\textit{left}) and London
       (\textit{right}). Satellites are shown as sunlit (yellow) or
       eclipsed by the Earth (black). The number of satellites
       increases towards the horizon because, for a given solid angle,
       looking at lower elevations from the horizon the distance to
       the shell surface increases and, hence, the field of view
       contains larger portions of the orbital shells. Due to the
       orbital inclination and orbital altitude of the orbital shells,
       over-densities of satellites are also present in these
       maps. The map for London shows this very clearly for the two
       orbital shells with $i=53^\circ$ and $h=345$\,km (7178
       satellites) and 499\,km (4000 satellites), where more
       satellites are present south from the dotted line. The dotted
       line in the all sky map for Vera Rubin Observatory is for the
       orbital shell with $i=30^\circ$, $h=328$\,km, with a larger
       number of satellites north from that line.}
     \label{fig:allsky}
   \end{figure*}

   A satellite will be visible at optical and near-infrared
   wavelengths if it is above the horizon for the geographical
   location of the observer and if it is illuminated by the Sun,
   making it to reflect and diffuse sunlight back to the
   observer. Hence, the visibility of a given satellite will depend
   upon its location within its orbit, the geographic location of the
   observer, and the location of the Sun, all as a function of time.
   
   While the motion of a satellite in the Earth's gravitational field
   and through the higher residual layers of the atmosphere is usually
   computed through either numerical integration or using simplified
   analytical models \citep[e.g.][]{mg00, vck06}, these offer an
   accuracy that is not required for the statistical results that we
   want to derive from the simulations presented here. Instead, we
   assume that the satellites move in circular Keplerian orbits, and
   we will neglect the effects of atmospheric drag, the lack of
   spherical symmetry of the Earth's gravitational potential and
   perturbations by the Sun and the Moon. Furthermore, for simplicity,
   satellite positions are defined in an inertial geocentric reference
   system (Celestial Intermediate Reference System), neglecting the
   small effects of precession, nutation and Earth orientation
   parameters. These assumptions allow us to consider only the
   idealised motion of the satellites, the motion of the observer
   (uniform Earth rotation) and the apparent motion of the Sun.

   Satellite constellations for worldwide coverage are generally
   configured as Walker constellations \citep{wal84}. Here, satellites
   are in circular orbits of distinct orbital shells, with each shell
   described by its orbital altitude $h_\mathrm{sat}$ and orbital
   inclination $i$. Within each shell, equally spaced orbital planes
   are populated by satellites, also equally spaced within each
   plane. In Table\,\ref{tab:1} we use $n_\mathrm{sat}$ to denote the
   number of satellites in a single orbital plane, and
   $n_\mathrm{plane}$ to indicate the number of orbital planes. The
   total number of satellites within each shell is $N_\mathrm{sat} =
   n_\mathrm{sat}\times n_\mathrm{plane}$, and the total number of
   satellites in a constellation the sum of the number of satellites
   in each shell.

   In this paper we consider only idealized, complete constellations.
   Individual satellites, as well as the trains of satellites in very
   low orbits right after launch and the satellites in
   near-to-re-entry orbits, are not considered. Even taking into
   account the large number of launches required to replenish the
   constellations, and the hopefully equally large number of
   satellites on end-of-life orbits, the number of such satellites
   will be one to two orders of magnitude smaller than the number of
   satellites in operations. Also, as these satellites are in lower
   orbits their impact is concentrated during the very beginning and
   end of night. Hence, their contribution to the overall situation
   caused by the constellations in operation is therefore small.
   
   We expand on the work by \citet{hw20}, who used a simplistic
   geometric approximation to estimate the density of satellites and
   their effects. That model had the advantage of being extremely fast
   and numerically lean, and to yield acceptable results for
   mid-latitude observatories, but it had the major shortcoming not to
   account for latitude effects. This aspect is rigorously addressed
   in the following sections.
   
   \subsubsection{Discrete simulations}
   \label{ssec:discrete}
   To obtain a realization of a satellite constellation at a given
   time, the positions and velocities of all satellites in the
   constellation are computed using the assumptions listed
   earlier. Figure~\ref{fig:earthmap} shows such a realization for the
   Starlink Generation 2 constellation on a map of the Earth near the
   June solstice when the Sun is at its highest northern
   declination. Hence, for locations at high geographical latitudes in
   the northern hemisphere, satellites will be visible throughout the
   whole night, whereas for lower latitudes and in the southern
   hemisphere satellites will only be visible around twilight. This
   figure depicts over-densities of satellites at geographical
   latitudes near the orbital inclination of different orbital shells.

   These over-densities are also seen in the all-sky maps shown in
   Fig.\,\ref{fig:allsky} for two locations, the Vera~Rubin
   Observatory (VRO) at Cerro Pach\'on in Chile (at latitude
   $-30\degr$), and London in the United Kingdom (at latitude
   $+50\degr$). The increased distance to the orbital shells when
   looking at lower elevations above the horizon leads to larger
   sampling volumes, which explains that these all-sky maps show
   increasing satellite densities towards the horizon, though
   additional over-densities exist near the projected location of the
   northern and southern limits of the most populous orbital shells.
   
   This discrete approach has been followed to assess the impact of
   satellite constellations in terms of the amount of them that are
   visible during the night above a given elevation from the horizon,
   as shown, for instance, in \citet{mcd20}. To this end, a specific
   constellation is selected (in terms of shell number and, for each
   of them, the altitude, orbital inclination and number of
   satellites), an observatory location is specified (in general only
   latitude is relevant) and the illumination conditions are fixed
   through Sun declination and hour angle. The simulation of the
   Keplerian movement of each satellite of the constellation allows
   counting the number of satellites that are above a specific
   elevation and their illumination conditions (sunlit or
   eclipsed). Additional considerations may allow computing some
   photometric estimates (\S\ref{ssec:photometry}). This kind of
   discrete, {\em all-sky} simulations may be iterated, including some
   random initialisation of parameters at each run to average-out
   systematic effects due to the spatial and temporal texture induced
   by the constellation structure.

   \subsubsection{Number of satellite trails in an observation}
   \label{ssec:traildensity}
   Besides the {\em all-sky} simulations, an {\em
     observation-oriented} approach is also needed. This requires
   specifying all the parameters required for the all-sky simulations
   and, also, selecting the observation direction (azimuth,
   elevation), field-of-view (FOV) and exposure time of an
   observation. In these observations, the motion of the satellite
   during the exposure will leave a satellite trail on the images.
   Repeating this simulation leads to the average estimate of the
   number of satellite trails that would affect the observation. The
   geometry of the problem allows also computing the position angle of
   the trails and the apparent angular velocity of each satellite that
   crosses the FOV.

\begin{figure}[t]
     \includegraphics[width=\columnwidth]{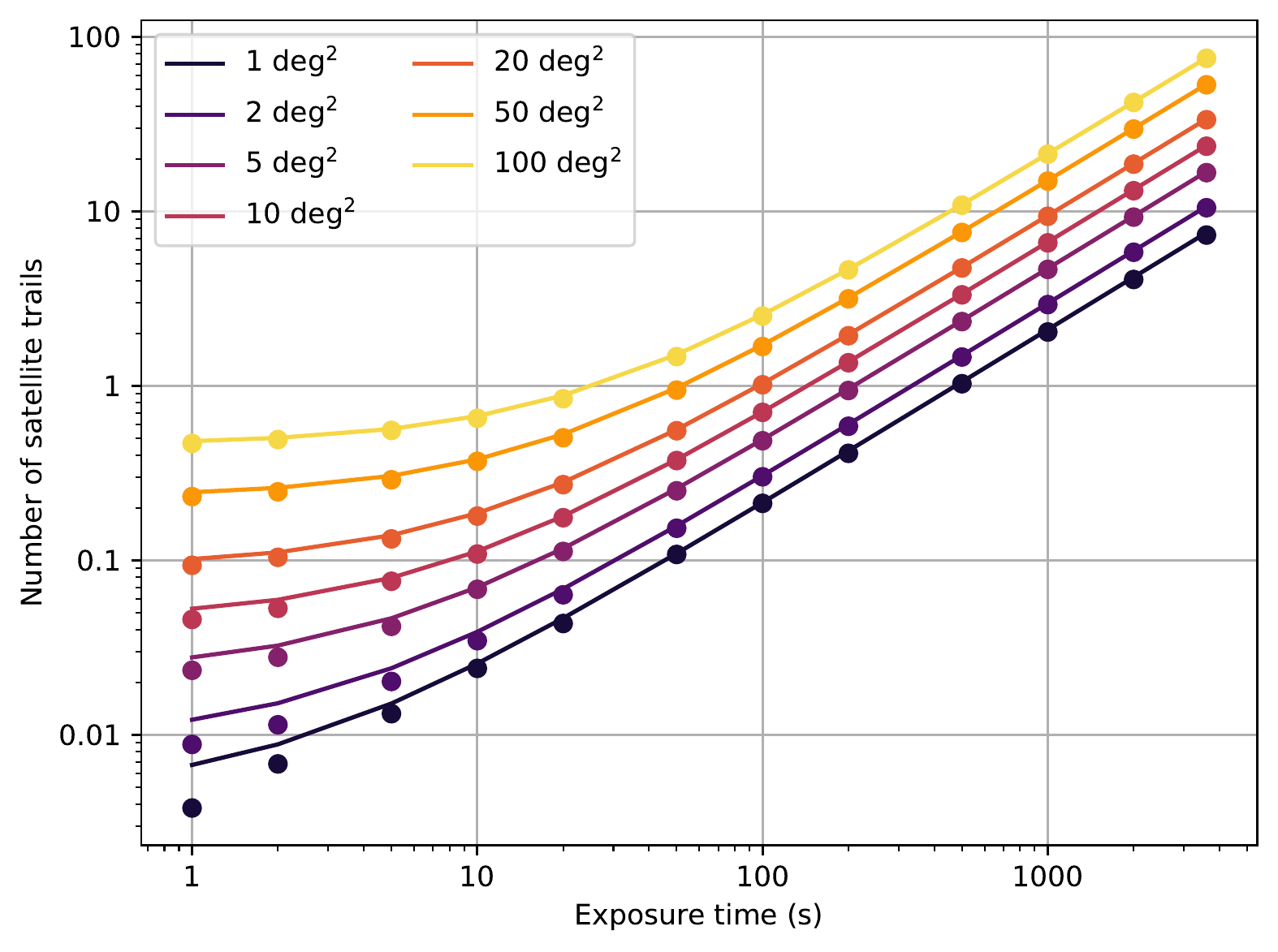}
     \caption{The number of satellites visible in an exposure
       depending on the exposure time $t_\mathrm{exp}$, and the
       instrument field-of-view, specified by its circular radius
       $R_\mathrm{fov}=L/2$. The effect of a discrete constellation of
       $10\,000$ satellites in a single orbital shell of 100 orbital
       planes and 100 satellites per plane at 1000\,km altitude and
       $53\degr$ inclination is simulated for an observer at
       $-30\degr$ latitude, observing towards zenith. The points are
       the results of the discrete simulations for different values of
       exposure time $t_\mathrm{exp}$ and field-of-view ($\pi
       R_\mathrm{fov}^2$). The solid lines are predictions based on
       Eqn.\,\ref{eq:parabolic}. }
     \label{fig:exposure_dependence}
   \end{figure}

   In Fig.\,\ref{fig:exposure_dependence} we show the results of a
   discrete simulation on observations with different exposure times
   and fields-of-view. The discrete simulation used a constellation of
   10\,000 satellites in a single orbital shell of 100 orbital planes
   with 100 satellites within each plane, at 1000\,km altitude and
   $53\degr$ inclination. For an observer at $-30\degr$ latitude, the
   number of satellite trails visible within an observation of an
   exposure time and circular field-of-view towards zenith were
   counted and averaged over 1000 simulations.

   The simulations show a dependence with both field-of-view and
   exposure time (solid dots). This relation can be understood as the
   number of satellites present at the start of the exposure, plus
   those travelling through the field-of-view during the exposure, and
   has the form
   \begin{equation}
     N_\mathrm{trail}=\rho_\mathrm{sat} (A_\mathrm{fov} +
     L_\mathrm{fov} \omega_\mathrm{sat}
     t_\mathrm{exp}).
     \label{eq:parabolic}
   \end{equation}
   Here, $\rho_\mathrm{sat}$ is the instantaneous satellite number
   density, i.e.\ the number of satellites per unit area on the sky,
   and $\omega_\mathrm{sat}$ the angular velocity of the satellites in
   the direction of the exposure. The exposure itself has
   field-of-view area of $A_\mathrm{fov}$ and width $L_\mathrm{fov}$,
   and exposure time $t_\mathrm{eff}$. For comparison with the
   simulations we use a field-of-view with a circular radius
   $R_\mathrm{fov}$, such that $A_\mathrm{fov}=\pi R_\mathrm{fov}^2$
   and $L_\mathrm{fov}=2R_\mathrm{fov}$.

   The simulations provide the instantaneous satellite density
   $\rho_\mathrm{sat}$ and the average angular velocity
   $\omega_\mathrm{sat}$ and using these values,
   Eqn.\,\ref{eq:parabolic} provides the predicted number of trails
   for the different observation parameters, plotted as lines in
   Fig.\,\ref{fig:exposure_dependence}. The predictions match the
   simulations, though for short exposures and/or small
   fields-of-view, the simulations suffer from noise due to the few
   satellites passing through the observation.

   Equation\,\ref{eq:parabolic} is valid for a single shell in a
   constellation, as the instantaneous density $\rho_\mathrm{sat}$ and
   angular velocity $\omega_\mathrm{sat}$ depend on the properties of
   the shell. To obtain the total number of trails in an observation
   for a satellite constellation with multiple shells,
   Eqn.\,\ref{eq:parabolic} can be computed for each shell, and the
   results summed.

   \subsubsection{Analytical simulations}  
   \label{ssec:analytical}
   The averaging over many randomly initialized parameters of a
   satellite constellation for the discrete simulations is
   computationally expensive, and may lead to noise due to
   insufficient satellites passing through observations with small
   fields-of-view and/or short exposures to obtain a valid average
   (see Fig.\,\ref{fig:exposure_dependence}). Given that the averaging
   over randomly initialized parameters has the effect of smoothing
   out the satellite locations within a given orbital shell, we
   instead treat the satellite locations as probability density
   functions, which have the advantage that these are analytical
   expressions.

   Figure\,\ref{fig:earthmap} shows that the satellite locations are
   uniformly spread in geocentric longitude, but strongly peaked
   towards geocentric latitudes $\phi$ close to the values equal to
   the orbital inclination $-i$ and $+i$ of a constellation shell. For
   a satellite at true anomaly (measured from the ascending node)
   $\kappa$, the geocentric latitude is given by $\sin \phi = \sin
   \kappa \sin i$, and its probability distribution follows the
   Arcsine probability
   distribution\footnote{\url{https://en.wikipedia.org/wiki/Arcsine_distribution}}.
   For a single satellite at orbital altitude $h_\mathrm{sat}$, the
   probability density per unit surface area as a function of $\phi$
   is:
   \begin{equation}
     P(\phi, i, h_\mathrm{sat}) = \frac{1}{2\pi^2
       (R_\oplus+h_\mathrm{sat})^2 \sqrt{(\sin i + \sin \phi) (\sin
        i - \sin \phi)}}.
     \label{eq:probdens}
   \end{equation} 
   Here $R_\oplus$ is the radius of the Earth, and this expression is
   valid for $|\phi|<i$, and zero otherwise. A detailed derivation of
   Eqn.\,\ref{eq:probdens} is given in
   Appendix~\ref{sec:ap:probadens}.

   Equation\,\ref{eq:probdens} can be integrated over the surface of
   the orbital shell spanned by the field-of-view of an instrument
   from an observatory located on Earth to obtain the fraction of the
   sample present within the instrument field-of-view. For the
   remainder of the paper, we will make the simplifying assumption
   that $P(\phi, i, h_\mathrm{sat})$ is constant over the instrument
   field-of-view. This assumption will generally be true for the small
   fields-of-view under consideration in the remaining analysis. It
   has the advantage that it removes any dependence on the precise
   shape and orientation of the instrument field-of-view, and instead
   solely depends on the sky area covered by the field-of-view. The
   assumption allows us to evaluate Eqn.\,\ref{eq:probdens} for a
   line-of-sight (specified by azimuth and elevation) of the
   observation from an observatory on Earth intersecting the orbital
   shell at distance $d$ and with an impact angle $\alpha$
   ($\alpha=90\degr$ at zenith and $\alpha<90\degr$ for lower
   elevations). The instantaneous surface density $\rho_\mathrm{sat}$
   can then be obtained by scaling the probability by the surface area
   of the orbital shell covered an angular area $A$ of 1 square
   degree, providing

   \begin{equation}
     \rho_\mathrm{sat}=N_\mathrm{sat} P(\phi, i, h_\mathrm{sat})\frac{{d^2 A}}{\cos \alpha} ,
     \label{eq:dens}
   \end{equation}
   where $N_\mathrm{sat}$ is the number of satellites in the orbital
   shell with inclination $i$ and orbital altitude
   $h_\mathrm{sat}$. Equations to derive the distance $d$ and impact
   angle $\alpha$ given the location of the observatory and azimuth
   and elevation of the observation are given in
   Appendixes~\ref{sec:ap:geopos} and \ref{sec:ap:impact}.
   
   The angular velocity of the satellites in an orbital shell towards
   the line-of-sight can be determined using the equations provided in
   Appendix\,\ref{sec:ap:apparentangvel}. Due to the rotation of the
   Earth within the orbital shell of a satellite constellation, the 
   velocity vectors project differently on the sphere of the sky and 
   hence satellites will have somewhat different angular velocities
   depending on their north- or southbound trajectory. Given that for a
   full constellation, an equal amount of satellites will be on
   northbound as well as southbound trajectories, we can take the average
   of both angular velocities to obtain $\omega_\mathrm{sat}$.

   For a given satellite constellation, observatory latitude and
   observation parameters (field-of-view and exposure time), the
   analytical simulations predict the number of trails as a function
   of azimuth and elevation which is inherently static with time. The
   final step to complete the analytical simulation is taking into
   account the illumination of the different orbital shells by the
   Sun, as this modulates the visibility of a shell as a function of
   time of day and time of year. The impact of solar illumination is
   implemented by computing whether the intersection of a given
   line-of-sight with an orbital shell is in the shadow of the Earth
   or not. If the intersection point is located in the shadow, and the
   satellites are not visible, we set $N_\mathrm{trail}=0$ for that
   shell in the sum of satellite trails over the different orbital
   shells.

\begin{figure}[t]
     \includegraphics[width=\columnwidth]{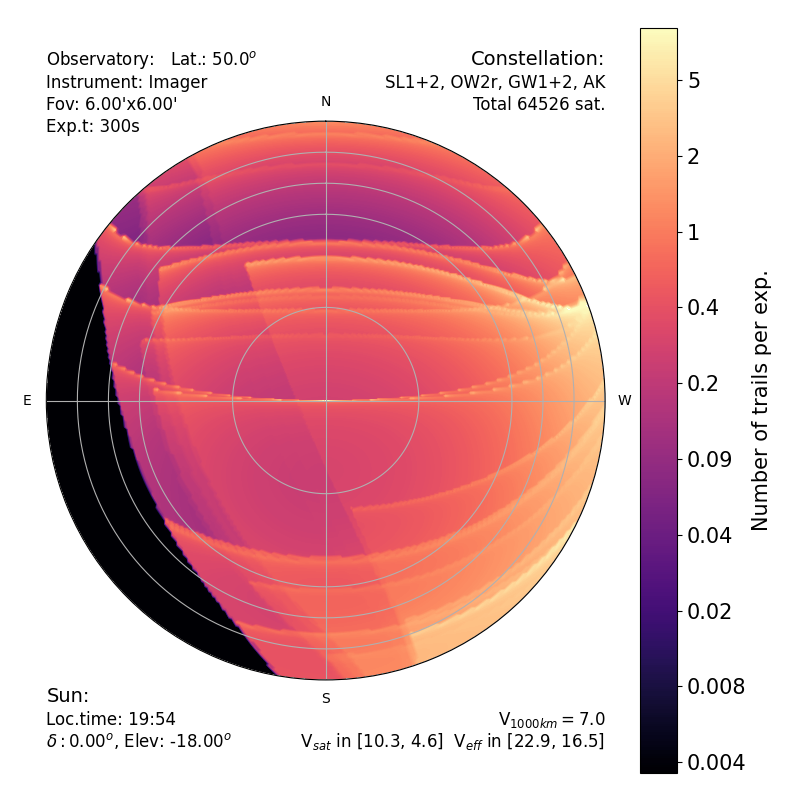}
     \caption{Sky maps with an example of the resulting number of
       trails per exposures. The circles mark $0\degr$, $10\degr$,
       $20\degr$, $30\degr$, and $60\degr$ elevation.  The observatory
       is located at $+50\degr$ latitude; the sun at $-18\degr$
       elevation; the camera has a FOV with diameter
       $L_\mathrm{fov}=6\arcmin$, and the exposure time is 300\,s. The
       satellites are those from Table~\ref{tab:1}. In the black
       region at the South-East, all satellites are already in the
       shadow of the Earth. The edges running from North-East to
       South-West correspond to constellation shells at lower
       altitudes, whose South-East parts are already in the
       shadow. The sharp features running from East to West correspond
       to the edges of constellation shells whose inclinations are
       close to the observatory latitude.}
     \label{fig:densities}
   \end{figure}

   Figure\,\ref{fig:densities} shows all sky maps using the analytical
   simulations, providing the number of satellite trails from
   Eqn.\,\ref{eq:parabolic}. The contribution of the different orbital
   shells of multiple satellite constellations is apparent, as is the
   impact of solar illumination.

   \subsection{Photometry}
   \label{ssec:photometry}

   Providing a reliable estimation of the apparent brightness of the
   satellites is an obvious requirement for any model that intends to
   assess the impact of mega-constellations on astronomical
   observation.  The celestial mechanics part of the problem admits an
   accurate solution that, even in a simplified frame (as described in
   \S\ref{ssec:discrete} and \S\ref{ssec:analytical}), leads to sound
   predictions of the spatial parameters (satellite density and their
   motion). However, the photometric part of the problem faces
   additional difficulties, due to the complex geometrical and
   reflective properties of the satellite that, also, are different
   from one constellation to another.

   In the visible and near-infrared (NIR), the light from the
   satellite is reflected sunlight with a specular and a diffuse
   component. The specular reflection happens on flat panels:
   antennas, satellite bus, possibly also solar panels (although while
   the satellites are in operations, these are perpendicular to the
   Sun). A complete and accurate representation of the reflection by a
   satellite would require detailed knowledge of the shape and
   material of the satellites \citep[see][for a summary of the
     state-of-the-art]{wal20}. A simplified model can be assembled
   from photometric observations covering a range of zenithal distance
   and solar illumination.

   Empirical models \citep[for instance, a flat panel
     model,][]{mal20a} are being developed, and theoretical approaches
   are also used to define photometric parameters of the satellites
   \citep{wal20}. However, as of today, the available observations are
   sufficient only for simplistic photometric models. Hopefully,
   dedicated observation campaigns will refine the characterization of
   the satellites in the coming years.
   
   \begin{table*}[]%------------------------------------
     \caption{Representative magnitudes of the satellites.    
       \label{tab:mag}
     }
     \begin{tabular}{lllllllll}
       \hline
Satellite & Operational& Mag	& Mag	   & Mag & $p r^2$  & $p$& $r$	& Ref.\\  
          & altitude  &at op. &dispersion  & at       &   &  	& 	& 	     \\
		  & [km]      & alt. 	& 	       & 1000km 	 & [m$^2$]  &   &[m]&    \\
\hline
\hline
Starlink original	& 550km	& 4.6	& 0.7	& 5.9	& 	0.085 & 0.25	& 0.58	& 1	\\
	         		& 	    & 4.0	& 0.7	& 5.3	& 	0.152 & 0.25	& 0.78	& 2 \\
			        & 	    & 4.2	&(model)& 5.5	& 	0.125 & 0.25	& 0.71	& 3 \\
Starlink DarkSat	& 550km	& 5.1	&(single)& 6.4	& 	0.056 & 0.08	& 0.71	& 4 \\
Starlink VisorSat	& 550km	& 6.2	& 0.8	& 7.5 	& 	0.023 & 0.25	& 0.30 	& 5 \\
			        & 	    & 5.8	& 0.6	& 7.1	& 	0.028 & 0.25	& 0.33	& 6 \\
OneWeb		        & 1200	& 7.6	& 0.7	& 7.2	& 	0.027 & 0.25	& 0.33	& 7 \\
\hline
     \end{tabular}
    
     Notes: 
     $p$ is the (arbitrary) geometric albedo used for the conversion of the cross-section $pr^2$ into an estimate of the radius $r$.
     References: {\small 
       1: \citet{mal20b};
       2: Krantz in \citet{IAU20};
       3: value used in \citet{hw20};
       4: using the darkening of 0.88 mag on one DarkSat, from \citet{tr20};
       5: median value from \citet{mal21};
       6: average from Krantz in \citet{IAU20};
       7: \citet{mal20c}
     }
   \end{table*}

   \begin{figure}[t]
     \includegraphics[width=\columnwidth]{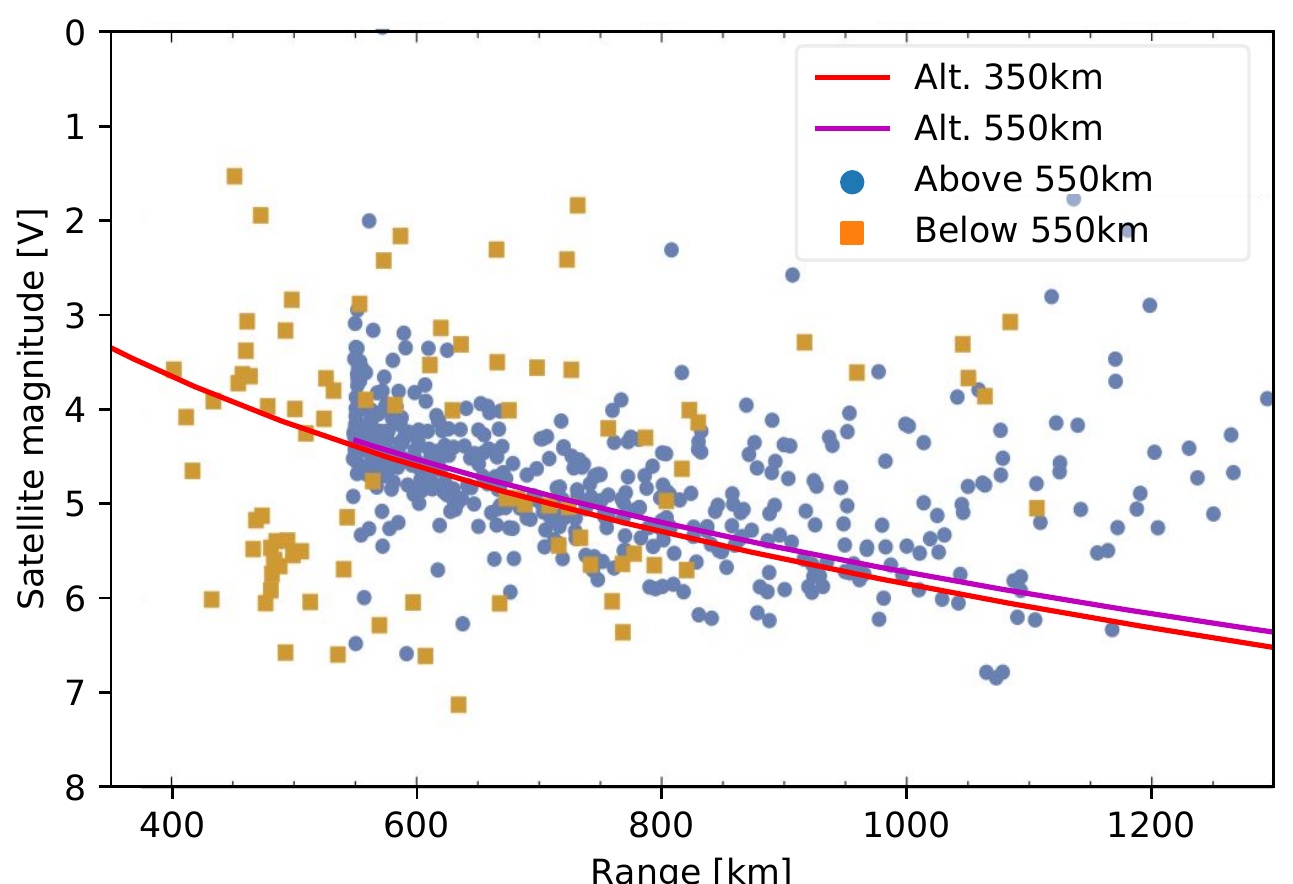}
     \caption{Magnitude of the satellites as a function of the range
       observer-satellite. The dots are measurements of original
       Starlink satellites \citep[Pomenis telescope, from][]{IAU20},
       and the lines are obtained using a simplified Lambertian sphere
       model for two altitudes. The unknown solar phase angle
       contributes to the dispersion of the measurements. }
     \label{fig:mag_observations}
   \end{figure}

   \subsubsection{Apparent magnitude}
   \label{sssec:lambertian}
   The simple photometric model we use is based on the {\em Lambertian
     sphere} model.  The theoretical foundations of the Lambertian
   sphere model can be seen, for instance, in \cite{kar96}. The
   different specific formulations, such as those by \citet{hw20} and
   \citet{mcd20}, can be unified under a single formula:
   \begin{eqnarray}
     \label{eq:unified}
     m_\mathrm{sat} &=& m_{\odot} - 2.5\log_{10}\left( p R_\mathrm{sat}^2 \right)
     + 5 \log_{10} \left ( d_{\mathrm{sat}\odot} d_\mathrm{sat}\right ) \nonumber \\
     &&-2.5\log_{10}\upsilon(\alpha_\odot)
     +k \chi ~.
   \end{eqnarray}
   
   In Eqn.~\ref{eq:unified}, $m_\odot$ is the Sun's apparent magnitude
   as seen from Earth, in the photometric band of interest. Typically,
   this is Johnson's $V$ band, with $m_\odot = -26.75$.  The second
   term considers the object's intrinsic photometric properties: $p
   R_\mathrm{sat}^2$ is the {\em photometric cross-section}, with $p$
   the object's geometric albedo and $R_\mathrm{sat}$ the radius of
   the (spherical) satellite.  The third term includes several
   distances: $d_{\mathrm{sat}\odot}$ is the distance from the
   satellite to the Sun, $d_\mathrm{sat}$ represents the distance from
   the observer to the satellite. For our problem,
   $d_{\mathrm{sat}\odot} = 1$\,AU.  The fourth term is the correction
   for the solar phase $\alpha_\odot$. Finally, $k\chi$ represents the
   extinction term, $k$ being the extinction coefficient (in
   magnitudes per unit airmass), and $\chi$ the airmass (equal to
   $1/\sin {e_\mathrm{sat}}$ in the plane-parallel approximation, with
   $e_\mathrm{sat}$ representing the satellite's elevation above the
   horizon; here, as we know the orbits of the object, we use the
   exact $\chi = d_\mathrm{sat}/h_\mathrm{sat}$). In the $V$ band, $k
   = 0.12$ is a typical value \citep{patat+11}.
   
   For a Lambertian sphere, $\upsilon(\alpha_\odot) = (1 + \cos
   \alpha_\odot)/2 $. However, a large number of photometric
   measurements of Starlink original satellites \citep{mal20b} and
   Starlink VisorSat \citep{mal21} indicate an extremely weak
   dependency of the magnitude (corrected for distance and extinction)
   with the solar phase angle, a circumstance that has to be related
   to the morphology of the satellite, which is very different to a
   sphere. We therefore consider $\upsilon = 1$, leading to
   \begin{equation}
     \label{eq:simplified}
     m_\mathrm{sat} = m_{\odot} + 2.5 \log_{10} d_\mathrm{sat}^{2}-
     2.5\log_{10}\left( p R_\mathrm{sat}^2 \right) +k  d_\mathrm{sat}/h_\mathrm{sat} ~,
   \end{equation}
   where both $d_\mathrm{sat}$ and $R_\mathrm{sat}$ are expressed in
   the same units.
   
   The photometric cross-section is the only parameter that depends on
   the satellite's physical properties in this model. \cite{hw20}
   adopted $pR_\mathrm{sat}^2 = 0.25$ $\mathrm{m}^{2}$ for the
   first-generation Starlink.  In order to facilitate the comparisons
   between satellites, one introduces the {\em absolute magnitude}
   $m_\mathrm{1000km}$, normalized to a standard distance
   $d_\mathrm{sat} = 1000$~km. With that value for $d_\mathrm{sat}$,
   Eqn.~\ref{eq:simplified} becomes
   \begin{equation}
     \label{eq:absmag}
     m_\mathrm{sat} = m_\mathrm{1000km} + 5 \log_{10} \left( d_\mathrm{sat}/1000 \right) 
     + k  d_\mathrm{sat}/h_\mathrm{sat} ~,
   \end{equation}
   with $d_\mathrm{sat}$ expressed in km. Sometimes, $m_\mathrm{550km}
   = m_\mathrm{1000km} - 1.3$ is used instead.
   
   Table~\ref{tab:mag} lists the absolute magnitudes measured for
   different satellite types, and the corresponding photometric
   cross-section and visual magnitude at zenith. The dispersion of the
   measurements of $m_\mathrm{1000km}$ is around 0.7 mag;
   Fig.~\ref{fig:mag_observations} shows an even wider
   dispersion. Consequently, the simplistic model presented above can
   only represent the general trend of the satellite magnitude, as
   illustrated by the comparison with observations
   (Fig.~\ref{fig:mag_observations}). In the simulations described
   below, we will use $m_\mathrm{1000km} = 7$, equivalent to
   $m_\mathrm{550km} = 5.7$.

   \subsubsection{Effective magnitude and limiting magnitude}
   \label{sssec:effective}
   \begin{figure}[t]
     \includegraphics[width=\columnwidth]{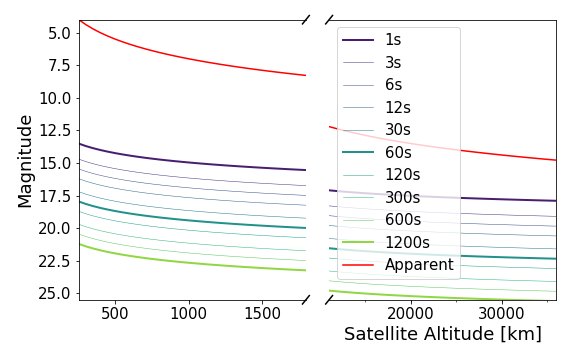}
     \caption{Visual (red) and effective magnitudes of a satellite at
       zenith, as a function of its altitude, for various exposure
       times (see legend). The satellite used is a Starlink VisorSat
       with $m_\mathrm{1000km} = 7$, considered as a trailed point
       source. }
     \label{fig:effectivemag}
   \end{figure}
   
   \begin{figure}[t]
     a\\\includegraphics[width=\columnwidth]{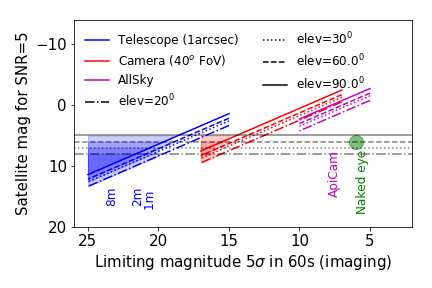}\\
     b\\\includegraphics[width=\columnwidth]{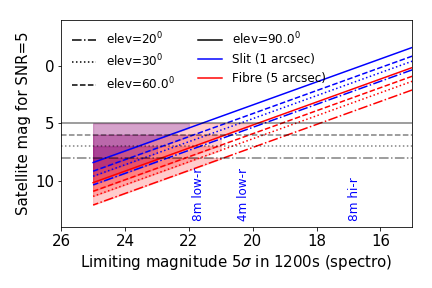}\\
     \caption{Detection limit for the satellite apparent magnitude as
       a function of the limiting magnitude of the instrument. a: the
       for an imager, the limiting magnitude corresponds to 5$\sigma$
       in a 60s exposure time; b: for a spectrograph, the limiting
       magnitude corresponds to $5\sigma$ in a 1200s exposure. In both
       cases, various resolution element sizes are represented in
       different colour. Typical satellite apparent velocities
       corresponding to various elevations are represented with
       different line styles. A satellite will be detected if its
       apparent magnitude is in the shaded area above the coloured
       line corresponding to the considered elevation. The horizontal
       limit correspond to the typical brightest satellite at that
       elevation. Typical limiting magnitudes are indicated.  }
     \label{fig:effimgmag}
   \end{figure}

   During an exposure of duration $t_\mathrm{exp}$, a satellite will
   leave a trail of length $\omega_\mathrm{sat} t_\mathrm{exp}$ (with
   $\omega_\mathrm{sat}$ being the apparent angular speed of the
   satellite), typically much longer than the FOV of the
   instrument. The signal corresponding to the apparent magnitude is
   therefore spread along the length of the trail. The count level on
   the detector amounts to the light accumulated inside an individual
   resolution element (whose size is $r$) during the time
   $t_\mathrm{eff} = r/\omega_\mathrm{sat} $ that the satellite takes
   to cross that element. This leads to the concept of {\em effective
     magnitude}, $m_\mathrm{eff}$, defined as the magnitude of a
   static point-like object that, during the total exposure time
   $t_\mathrm{exp}$, would produce the same accumulated intensity in
   one resolution element than the artificial satellite during a time
   $t_\mathrm{eff}$:
   \begin{equation}\label{eq:effmag}
     m_\mathrm{eff} = m_\mathrm{sat} - 2.5
     \log_{10}\frac{t_\mathrm{eff}}{t_\mathrm{exP}} = m_\mathrm{sat} - 2.5
     \log_{10}\frac{r}{\omega_\mathrm{sat} t_\mathrm{exp}}~.
   \end{equation}
   Figure~\ref{fig:effectivemag} shows the effective magnitude for an
   example. While not directly relevant for low-altitude constellation
   satellites, it is worth noting that the dependency of
   $m_\mathrm{eff}$ with the altitude of the satellites is shallower
   than that of the apparent magnitude.

   In this approximation we are assuming that the satellite's PSF has
   the same shape as the stellar PSF at the telescope focal surface, a
   crude approximation if the distance to the satellite is small
   enough for it to be spatially resolved, or out of focus, or both
   \citep{tyson+20,ragazzoni20}. The apparent angular width of the
   satellite trail is
   \begin{equation}\label{eq:psfsize}
     \theta_\mathrm{sat}^2 = \theta_\mathrm{atm}^2 +
     \frac{D_\mathrm{sat}^2 + D_\mathrm{m}^2}{d_\mathrm{sat}^2}~,
   \end{equation}
   where $\theta_\mathrm{atm}$ is the stellar FWHM (the seeing,
   typically $\sim 0\farcs 8$ from a good site), $D_\mathrm{sat}$ the
   physical diameter of the satellite, $D_\mathrm{m}$ the diameter of
   the telescope mirror, and $d_\mathrm{sat}$ the distance to the
   satellite \citep{tyson+20}. For an 8-m telescope like the ESO Very
   Large Telescope (VLT), or the Simonyi Survey Telescope (formerly
   LSST), a 2\,m satellite at an altitude of 300 to 550km,
   $\theta_\mathrm{sat} \sim 6\arcsec$ to $3\arcsec$ . The spreading
   of the signal from the satellite over this larger area will
   decrease its signal-to-noise ratio (SNR) by up to
   $\theta_\mathrm{sat}/\theta_\mathrm{atm}$, and its peak intensity
   by up to $({\theta_\mathrm{sat}}/{\theta_\mathrm{atm}})^2$, i.e. 2
   to 4 magnitudes fainter than $m_\mathrm{eff}$ from
   Eqn.~\ref{eq:effmag}.

   For imaging, the resolution element is typically the seeing (of the
   order of $1\arcsec$) for telescopic observations, or the pixel (a
   few to a few tens arcsec) for wide-field astrophotography.
   Figure~\ref{fig:effimgmag}.a displays the visual magnitude of the
   faintest satellite that will leave a trail with $\mathrm{SNR} = 5$
   as a function of the limiting magnitude, for $t_\mathrm{exp} =
   60\,$s imaging observations.  This shows that all-sky cameras will
   record only the brightest satellites and flares.  Only the deepest
   wide-field astrophotography (with a limiting magnitude $V \sim 15$
   in $t_\mathrm{exp}=1$~min or fainter) will record the bulk of the
   satellites. Telescopic observations are fully affected by all most
   satellites.
  
   The situation is slightly different for spectroscopy. In the case
   of fibre-fed spectrographs, the resulting data contain no spatial
   information at all; for long-slit spectrographs, the spatial
   information is available in only one direction. Except in the case
   of integral-field spectrographs, the data will therefore not
   include a tell-tale trail indicating the contamination.  For an
   exposure time $t_\mathrm{exp} = 1200$~s, representative of
   individual exposures in the visible, Fig.~\ref{fig:effimgmag}.b
   displays the visual magnitude of the satellite that will reach a
   $\mathrm{SNR}= 5$ as a function of the limiting
   magnitude. Spectrographs having a limiting magnitude brighter than
   $V=20$ in $t_\mathrm{exp}=1200$\,s will essentially be immune: the
   signal from a satellite will be be too faint to be detected. That
   will be the case for low-resolution spectrographs on small to
   medium telescopes, and high-resolution spectrographs on large
   telescopes.
   
   If the SNR of the contamination is much lower than that of the
   science target, the contamination will result in a small increase
   of the background noise, what can probably be neglected for most
   science cases.  Also, cases where the SNR of the contamination is
   much larger than that of the science spectrum are trivial: the
   effect is obvious and the observation is lost.  The situation for
   spectrographs with a limiting magnitude in the $v=20$--23 range in
   $t_\mathrm{exp}=20$~min is more problematic: the satellite trail
   will have a SNR of 2--15, so that contamination caused by the
   satellite will be at a level comparable to that of the science
   signal.  It is therefore plausible that the contamination will {\em
     not} be immediately apparent, and will be discovered only at the
   time of the data analysis, where a solar-type spectrum (reflected
   by the satellite) will be superimposed to that of the science
   target.  These intermediate cases where both SNRs are similar are
   much more problematic and science-case dependent: if the science
   target is a distant galaxy, a solar spectrum will be identified as
   a contamination.  However, if the target was a stellar object, a
   solar contamination might cause spurious conclusions.
   
   \section{Results}
   \label{sec:results}
   \subsection{Time and Solar declination dependence}
   \label{ssec:time_declination_dependence}
\begin{figure}[t]
     \includegraphics[width=\columnwidth]{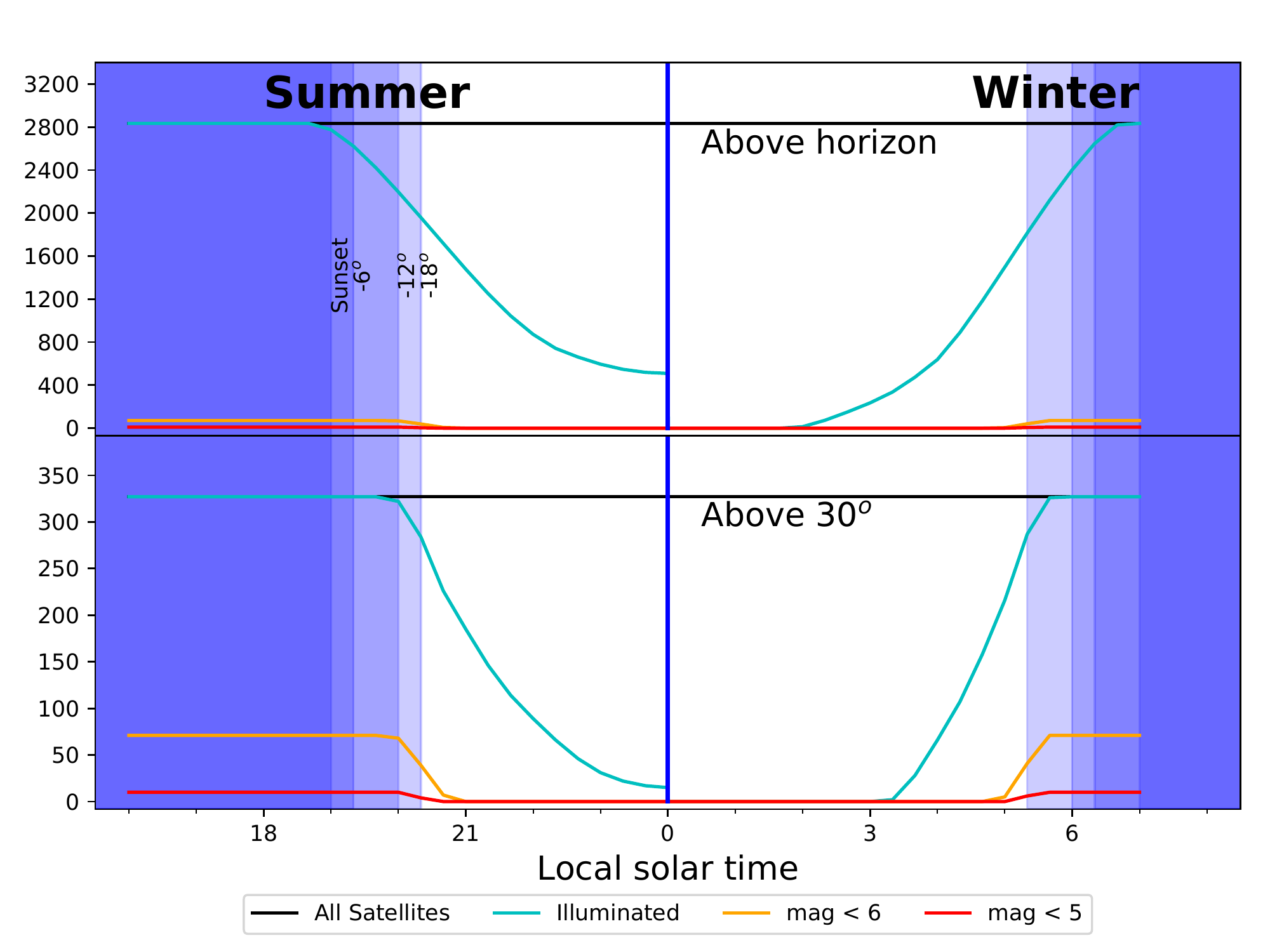}
     \caption{Number of illuminated satellites above the horizon (top
       panel) and above 30$\degr$ elevation, as a function of the
       local solar time, for Paranal (latitude $-24\fdg6$), accounting
       for all the satellites from Table~\ref{tab:1}. Left is for the
       summer solstice ($\delta_{\mathrm{Sun}}=+23\fdg4$), and right
       for the winter solstice ($\delta_{\mathrm{Sun}}=-23\fdg4$). The
       twilights are indicated with blue shading. The black line marks
       the total number of satellites above the elevation considered,
       the blue line those that are illuminated, and the orange and
       red lines those brighter than magnitude 6 and 5, respectively,
       using the photometric model described in
       \S\ref{ssec:photometry}.}
     \label{fig:night}
   \end{figure}

\begin{figure}[t]
     \includegraphics[width=\columnwidth]{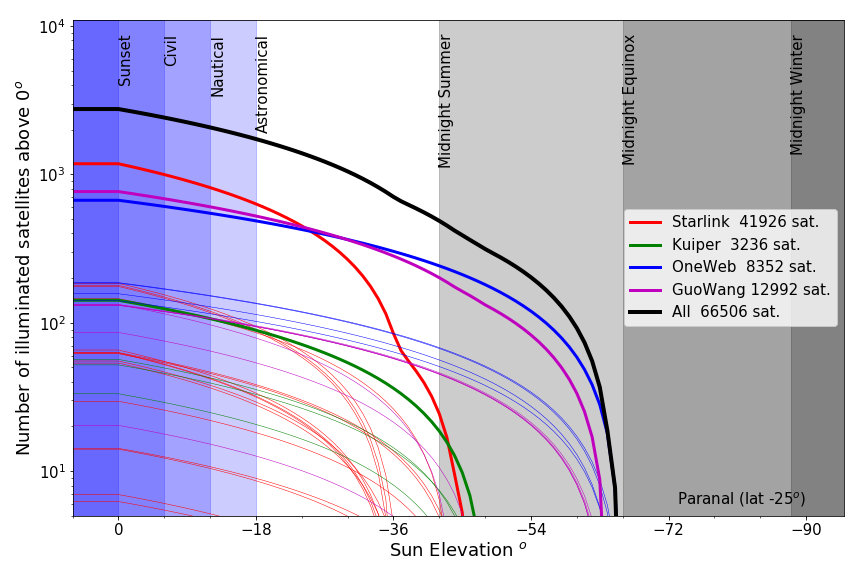}
     \caption{Number of illuminated satellites above the horizon as a
       function of the Sun elevation, for the constellations listed in
       Table~\ref{tab:1} seen from Paranal (latitude $-24\fdg6$; the
       dependency with latitude is not strong). The twilights are
       indicated with blue shadings, and the elevation of the sun at
       midnight by grey shadings for the solstices and equinoxes. The
       thin lines represent the individual shells, and the thick lines
       the totals for each constellation. The upper thick black line
       is the grand total.}
     \label{fig:elevation}
   \end{figure}

   The analytical models for visibility and photometry allow us to
   compute their dependence on time of day as well as time of
   year. Similar results were already presented using a simplified
   geometric model by \citet{hw20}, or discrete simulations by
   \citet{mcd20}.  Figure\,\ref{fig:night} shows the number of
   satellites illuminated by the Sun for the local summer and winter
   seasons for an observatory at latitude $-24\fdg6$. During local
   summer, satellites remain visible above $30\degr$ elevation
   throughout the night. Figure\,\ref{fig:elevation} shows the
   visibility dependence as a function of solar elevation, for each
   shell (exposing the importance of the shell altitude) and for the
   total populations from Table~\ref{tab:1}.

   \subsection{Spatial fine structure}   
   \label{ssec:finestructure}
   An outstanding effect of the orbital shells of satellite
   constellations is the amount of spatial fine structure that arises
   in the quantity of satellites visible on the local celestial
   sphere, as illustrated in Fig.\,\ref{fig:allsky} and
   Fig.\,\ref{fig:densities}. Of course, first of all we find the
   effect of the Earth's shadow, whose behaviour is, as expected,
   dominated by the diurnal rotation of the planet and by the
   interplay between observatory latitude and Sun declination. But the
   structure of satellite shells, combined with orbital mechanics,
   adds a far from negligible spatial fine structure in the quantity
   of satellites visible on the local celestial sphere. In particular,
   as Eqn.~\ref{eq:probdens} shows, each individual shell induces an
   unavoidable over-density high in the sky over observatories placed
   at geographic latitudes $\phi$ whose absolute value is close to the
   orbital inclination $i$. The Northern or Southern boundary of each
   shell lies along a line on the local celestial sphere that crosses
   zenith if $\phi = i$, what incidentally happens for Vera Rubin
   Observatory and some shells currently considered in several
   constellation designs. These over-densities may coincide with the
   culmination elevation of some key objects (let us say, for
   instance, LMC or SMC in the South, or M31 and M33 in the
   North). Given the inclination $i$ of one orbital shell and the
   latitude $\phi$ of the observatory, the shell boundary cuts the
   local meridian at a declination $\delta$ that can be deduced from
   the following equation (justified in
   Appendix~\ref{sec:ap:constedge}):
   \begin{equation}
     \sin{(\delta - \phi)} = \frac{R_{\oplus}+h_\mathrm{sat}}{R_{\oplus}} \sin{(\delta - i)}
     \label{eq:boundary}
   \end{equation}
   
   \subsection{Contribution to the sky brightness}
   \label{ssec:skymag}
   
   \begin{figure}[t]
     a\\\includegraphics[width=\columnwidth]{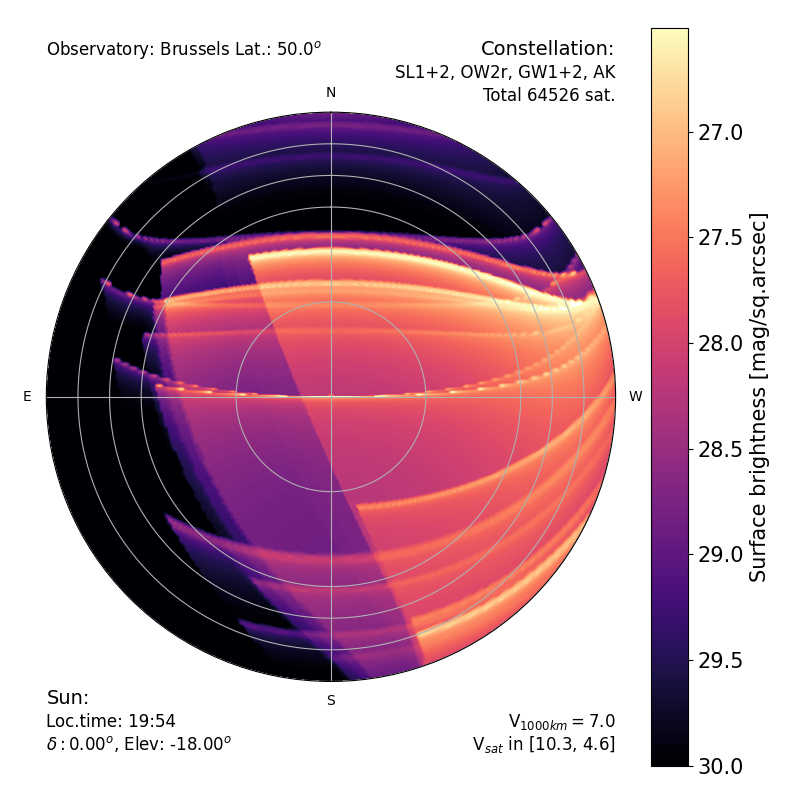}\\
     b\\\includegraphics[width=\columnwidth]{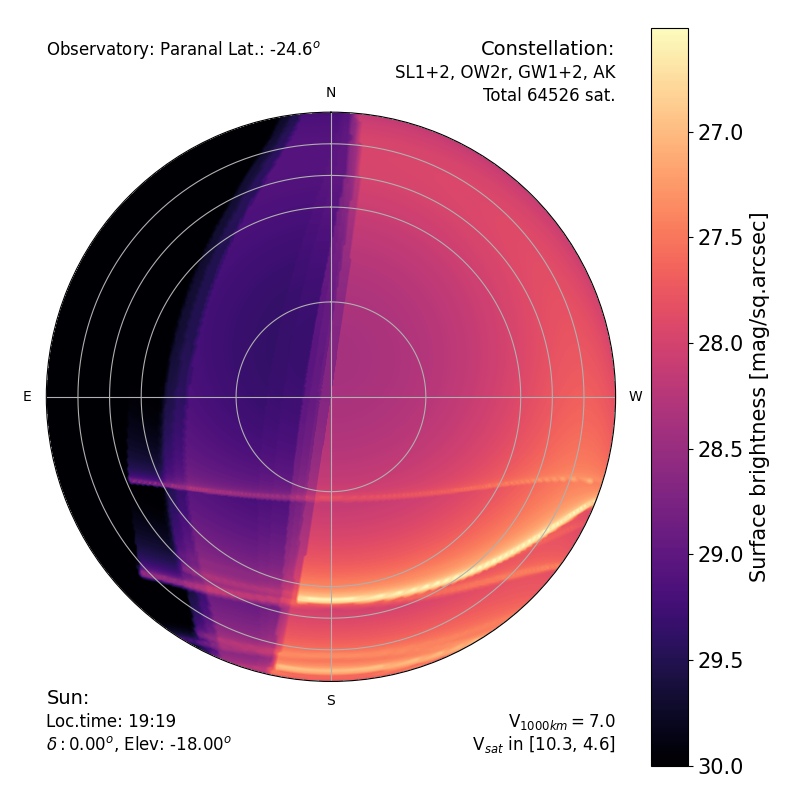}\\
     \caption{Sky brightness contribution from the satellites (using
       the constellations from Table~\ref{tab:1}), at astronomical
       twilight at latitudes $+50\degr$ (a) and $-25\degr$
       (b). Typical sky surface brightness in the visible is 21.7
       mag/sq.arcsec or brighter. The satellites used all have
       $v_\mathrm{1000km}=7$, resulting in visual magnitudes in the
       range indicated at the bottom right corner.}
     \label{fig:skymag}
   \end{figure}
   The satellites, including those that are not directly detected,
   contribute to the sky background. To evaluate this effect, surface
   brightness maps were computed. The magnitude of a satellite from a
   constellation was evaluated using Eqn.~\ref{eq:absmag}, then
   converted into flux, and finally used to weight the satellite
   density map described in \S\ref{ssec:analytical}. A total flux
   density map was obtained summing the contributions of all satellite
   shells, and transformed into surface brightness in
   mag/sq.arcsec. An example is presented in Fig.~\ref{fig:skymag}. In
   the illuminated part of the shells, the satellites contribute to a
   surface brightness in the 28--29~mag/sq.arcsec range
   (0.3--0.7\,$\upmu$cd~m$^{-2}$), with peaks around 26.5
   mag/sq.arcsec (2.7\,$\upmu$cd~m$^{-2}$) at the cusps of the
   constellations. The surface brightness of the dark night sky is
   around $V=21.7$ \citep{patat08} (225\,$\upmu$cd~m$^{-2}$), which
   means that the satellites from Table~\ref{tab:1} will contribute at
   most an additional $\sim1\%$ to the sky brightness in the worst
   areas of the sky. The contribution to the sky brightness is
   therefore small, and the simulations and mitigation focus on the
   discrete contamination by individual satellites.
   %conversions: cd/m2  =  10.8e4 * 10^(-0.4 magpersqarcsec)
   %http://unihedron.com/projects/darksky/magconv.php
   %28mag/sq.arcsec = 0.68 mucd/m2
   %29: 0.27
   %26.5:  2.71 mucd/m2
   %21.7: 225mucd/m2
   
   \citet{Kocifaj+21} have evaluated the increase in diffuse sky
   brightness caused by all current space objects with sized between
   $5\times 10^{-7}$~m to 5~m at altitudes above 200~km. They estimate
   that this excess is 16.2\,$\upmu$cd~m$^{-2}$ (24.6~mag/sq.arcsec)
   and can reach 21.1\,$\upmu$cd~m$^{-2}$ (24.3\,mag/sq.arcsec)at
   astronomical twilight, corresponding to about 10\% of the natural
   sky brightness. That excess is dominated by the small objects (mm
   and below), i.e.\ space debris. The macroscopic satellites
   composing the constellations discussed in this paper will not
   contribute much to the diffuse sky brightness provided they are not
   ground into microscopic debris.

   \subsection{Effect on observations}
   \label{ssec:effectonobservations}
   
   \begin{table*}
     \caption{Characteristics of instruments and exposures used for the simulations.\label{tab:instruments}}
     \begin{tabular}{p{3cm}clclrllp{5cm}}
       \hline
       Inst. Tel. Obs.   & $l$    & $D$  &   &   Field                 & $t_\mathrm{exp}$ [s]   & $r$ [arcsec] & Mag. \\
       \hline
       \multicolumn{6}{l}{\bf Visible and near-IR Imagers }\\
       \hline
       EFOSC   NTT  ESO (La Silla)  & $-29\fdg25$ & 3.6  & Vis & $4\arcmin$                  & 300   &  1   &    24.2 &
       Focal reducer \citep{EFOSC84}\\
       FORS    VLT  ESO (Paranal)  & $-24\fdg6$  & 8.2    & Vis & $6\arcmin$                  & 300   &0.8   &  25.2  &
       Focal reducer \citep{FORS94}\\
       HAWKI   VLT  ESO (Paranal)  & $-24\fdg6$  & 8.2    & NIR & $7\farcm5$                &  60   &0.6   &  21.4   &
       Near-IR imager \citep{HAWKI08}\\
       MICADO  ELT ESO (Armazones)   & $-24\fdg6$  & 39.   & NIR & $50\arcsec$                &  60   &0.015 &  24.9   &
       Visible and near-IR imager with adaptive optics on the 
       ELT$^a$\\
       \\
       %WFI MPE 2.2m   & -29.25 & 2.2   & Vis & $0.5\degr$            & 300   &  1   &    23.3 &
       %Survey wide-field imager\\
       OmegaCam VST  ESO (Paranal) & $-24\fdg6$  & 2.4   & Vis & $1.0\degr$            & 300   & 0.8  &   23.9 &
       Survey wide-field imager \citep{OMEGACAM02}\\
       1.5m Catalina U.AZ (Mt. Lemmon) & $32\fdg4$  & 1.52  & Vis & $2.2\degr$            &  30   &1.5   & 21.4    &
       Survey wide-field 
       imager$^b$\\
       LSST Cam. SST VRO (Pachon)& $-30\fdg2$  & 8     & Vis & $3.0\degr$            &  15   &0.8   &  24.6   &
       Survey wide-field imager$^c$\\
       0.7m Catalina U.AZ (Mt. Bigelow)   & $30\fdg4$  & 0.7   & Vis & $4.4\degr$            &  30   & 3    &   19.8  &
       Survey wide-field imager$^b$\\
       Photo          & $-30\degr$   & 0.07 & Vis & $75\degr\times 55\degr$    &  60   & 60   &   10    &
       Photographic camera with a wide-angle lens from a good site.\\
       \hline
       \multicolumn{6}{l}{\bf Visible and near-IR Spectrographs }\\
       \hline
       FORS 	VLT	 ESO (Paranal)& $-24\fdg6$	& 8.2	& Vis & $6\arcmin \times 1\arcsec$  &	1200 &   0.8 & 	22.0 & Long-slit low-resolution spectrograph \citep{FORS94}	\\	
       UVES	VLT	 ESO (Paranal)& $-24\fdg6$	& 8.2	& Vis & $10\arcsec \times 1\arcsec$ &	1200 &   0.8 &  17.0 & High-resolution echelle spectrograph \citep{UVES00}\\
       4MOST-L 	VISTA  ESO (Paranal)& $-24\fdg6$& 4	& Vis & $4\fdg 1$ &		    1200 &   0.8 &  20.5 & Multi fibre$^1$ spectrograph (low res.) \citep{4MOST10}\\
       4MOST-H 	VISTA  ESO (Paranal)& $-24\fdg6$	& 4	& Vis & $4\fdg 1$ &	    1200 &   0.8 &  18.6 & Multi-fibre$^1$ spectrograph (med res.) \citep{4MOST10}\\
       ESPRESSO VLT  ESO (Paranal)	& $-24\fdg6$	& 8.2	& Vis & $0\farcs5$ &	1200 &   0.5 &  15.8 & High-resolution echelle \citep{ESPRESSO21}\\
       \hline
       \multicolumn{6}{l}{\bf Thermal IR }\\
       \hline
       VISIR VLT & $-24\fdg6$	& 8.2	& ThIR  &  $60\arcsec$&  10 & $0\farcs2$ & -- & Imager \citep{VISIR04} \\
       METIS ELT & $-24\fdg6$  & 39    & ThIR  &  $10\arcsec$&  10 & $0\farcs03$& -- & 
       Adaptive Optics Imager on ELT$^{d, 2}$ \\
       \hline
     \end{tabular}

     $l$: latitude; $D$: diametre of the telescope; Field: field of
     view of the instrument; $t_\mathrm{exp}$: exposure time [s]; $r$:
     resolution element [arcsec]; Mag: $5\sigma$ limiting magnitude
     for a point source for an exposure of duration $t_\mathrm{exp}$.

     References: {\small 
       a, \url{https://elt.eso.org/instrument/MICADO/} - 
       b, \url{https://catalina.lpl.arizona.edu/about/facilities/telescopes} -
       c, \url{https://www.lsst.org/about} -
       d, \url{https://elt.eso.org/instrument/METIS/} }
     
     Notes: 1: 4MOST is a multi-object spectrograph equipped with 2436
     fibres. Monte-Carlo simulations showed that, on average, a
     satellite crossing the field of view will affect 1.3 fibres. 2:
     METIS also has a high-resolution spectrograph.
   \end{table*}

   To evaluate in more detail the effects of satellite constellations
   on observations, we studied a series of representative instruments
   and telescopes. For each of them, we consider the field of view
   (size or diameter in case of 2D field, length and width for a slit,
   diameter of the aperture in case of a fiber), a typical exposure
   time, and the limiting magnitude (obtained from the Exposure Time
   Calculators\footnote{\url{https://etc.eso.org}} for ESO
   instruments, documentation or private communications for
   others). We also estimate the magnitude causing heavy saturation
   either as 5~mag brighter (i.e. 100 times) than the saturation level
   or from publications. Table~\ref{tab:instruments} lists the
   parameters of the exposures and instruments.

   \begin{figure*}
     \centering
     \setlength{\mywidth}{3.4cm}
     \begin{tabular}{p{\mywidth}p{\mywidth}p{\mywidth}p{\mywidth}}
       \multicolumn{4}{l}{\bf a: Trails per exposure}\\
       \includegraphics[width=\mywidth]{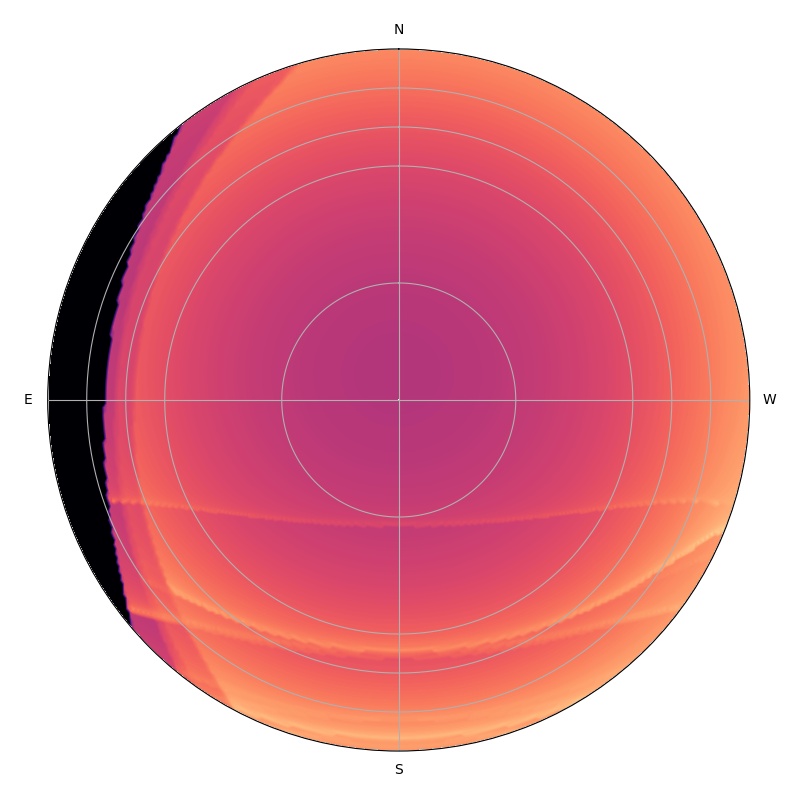}
       Sun Elevation:~$-12\degr$,    Average: 0.20 trail&
       \includegraphics[width=\mywidth]{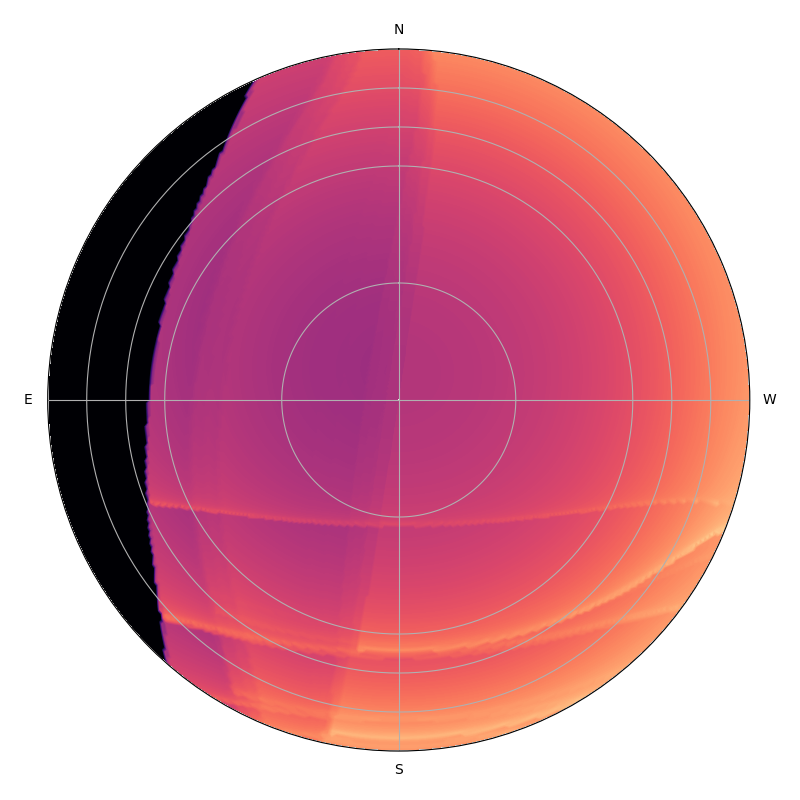} $-18\degr$, 0.16&
       \includegraphics[width=\mywidth]{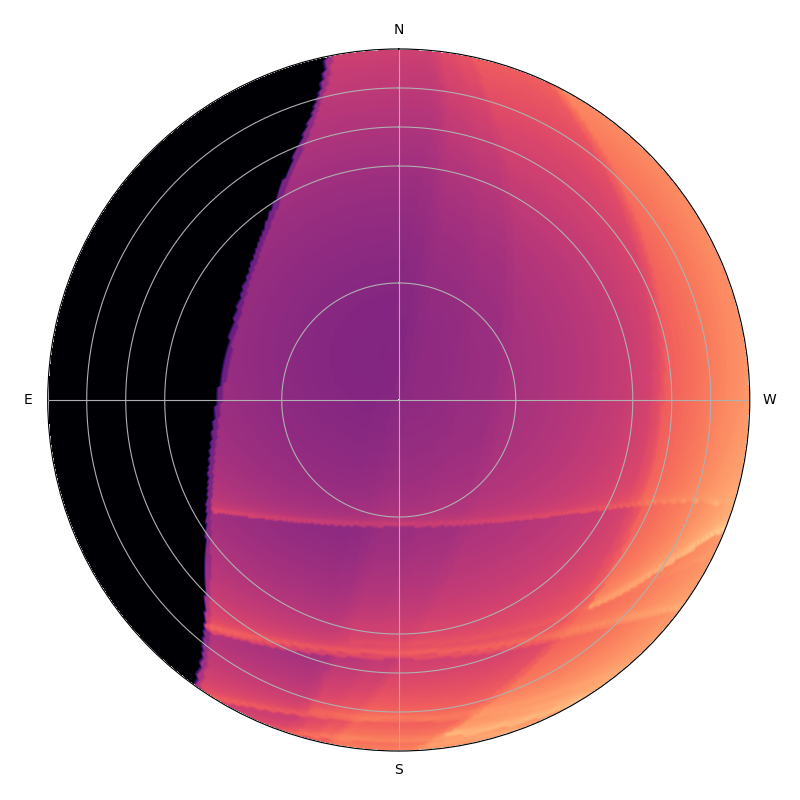} $-24\degr$, 0.081&
       \includegraphics[width=\mywidth]{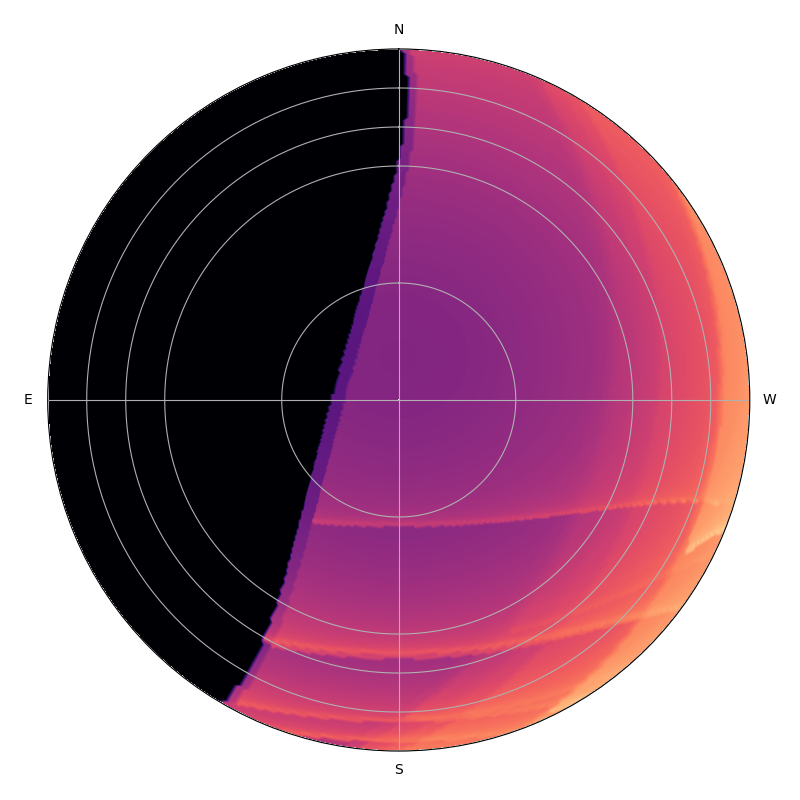} $-30\degr$, 0.045\\
       \includegraphics[width=\mywidth]{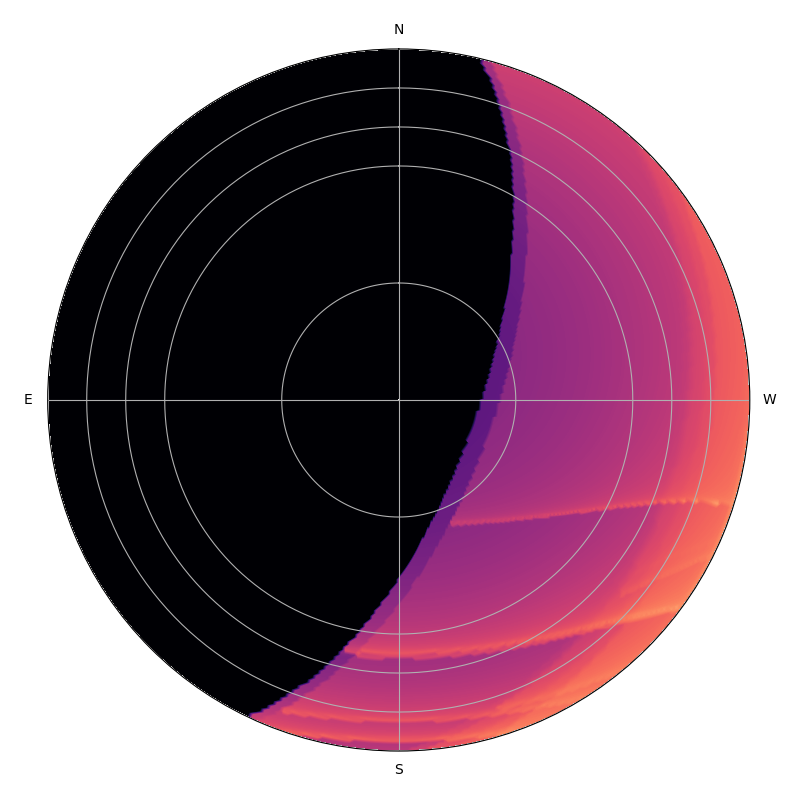} $-36\degr$, 0.023&
       \includegraphics[width=\mywidth]{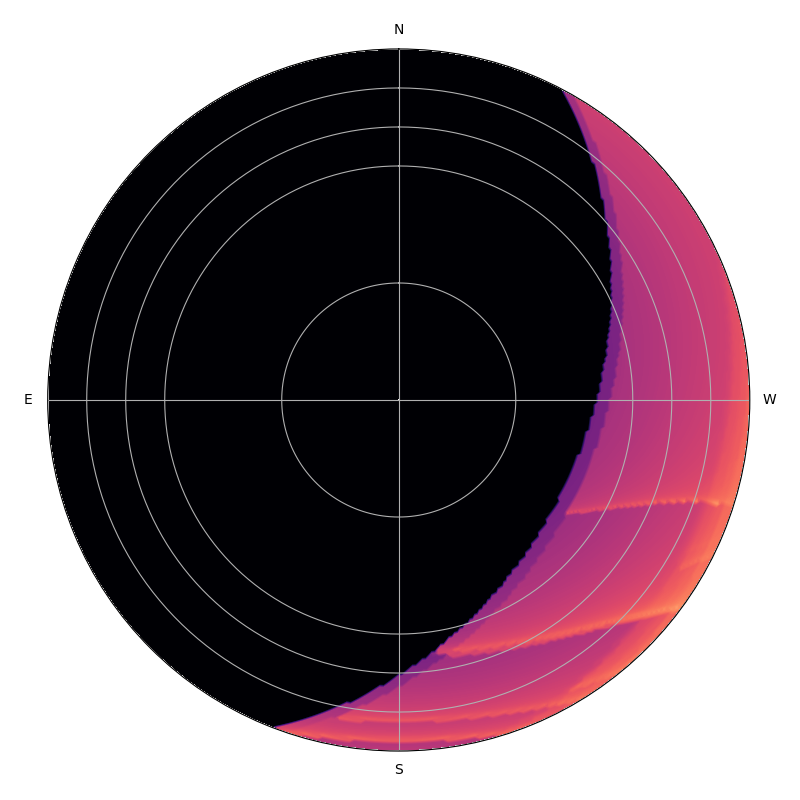} $-42\degr$, 0.005&
       \includegraphics[width=\mywidth]{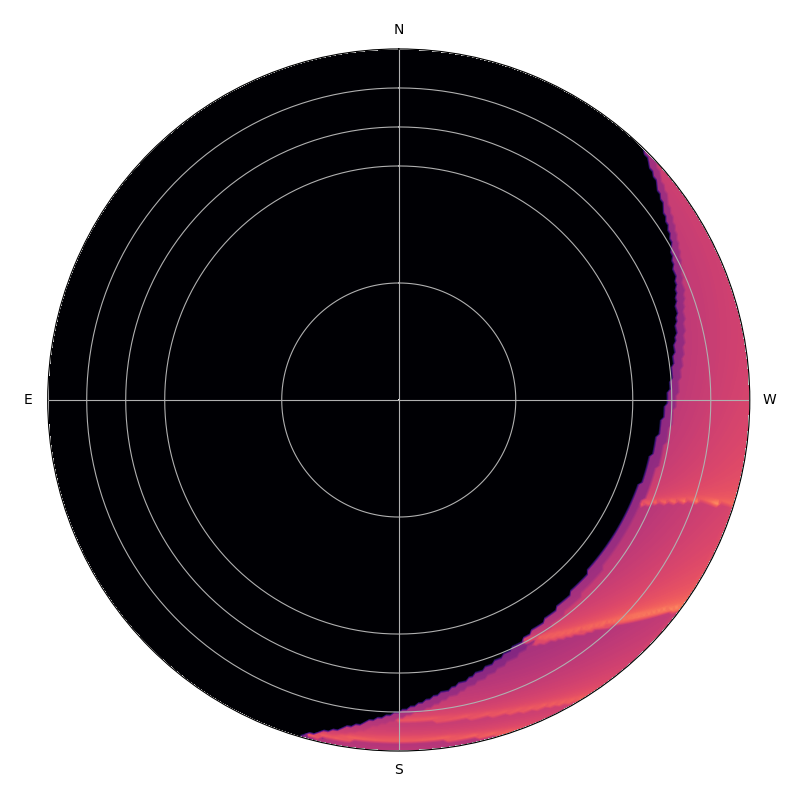} $-48\degr$, 0&
       \includegraphics[width=\mywidth]{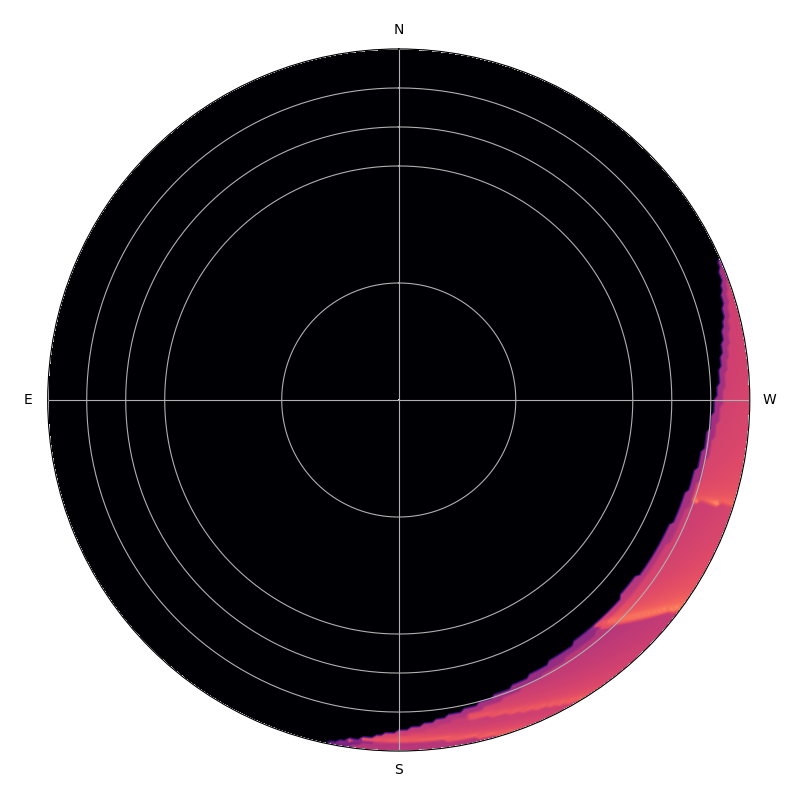} $-54\degr$, 0\\
       \multicolumn{4}{c}{\includegraphics[width=12cm]{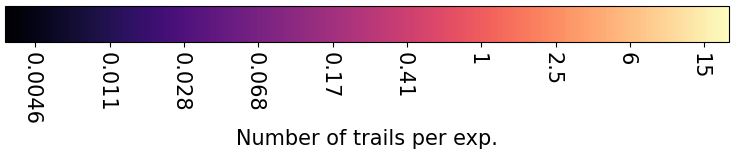}}\\
        ~\\
       \multicolumn{4}{l}{\bf b: Fraction of observations lost to satellite trails}\\
       \includegraphics[width=\mywidth]{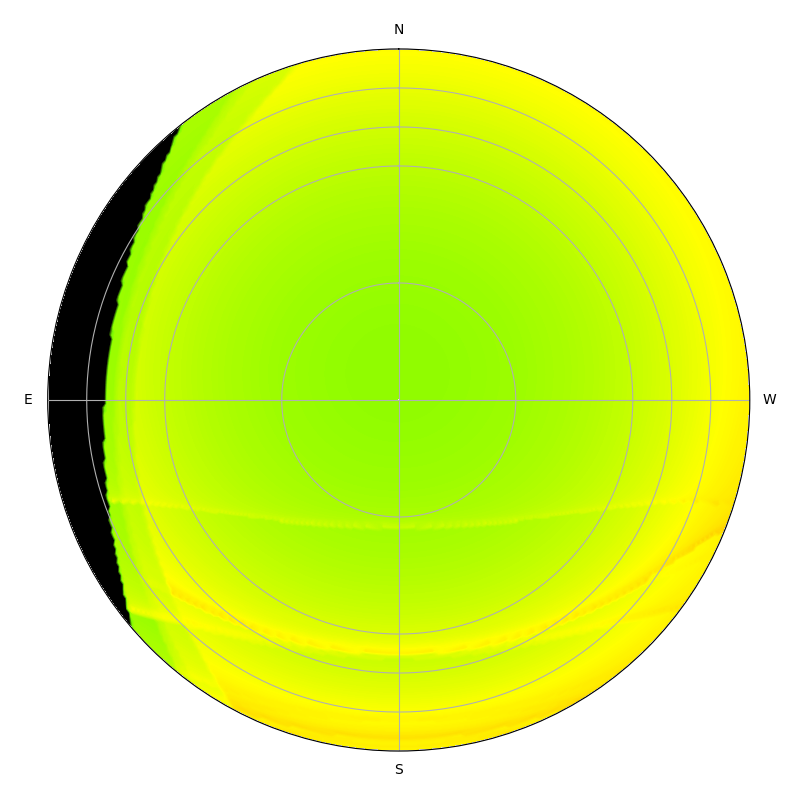}
       Sun Elevation:~$-12\degr$, Average: 0.56\% &
       \includegraphics[width=\mywidth]{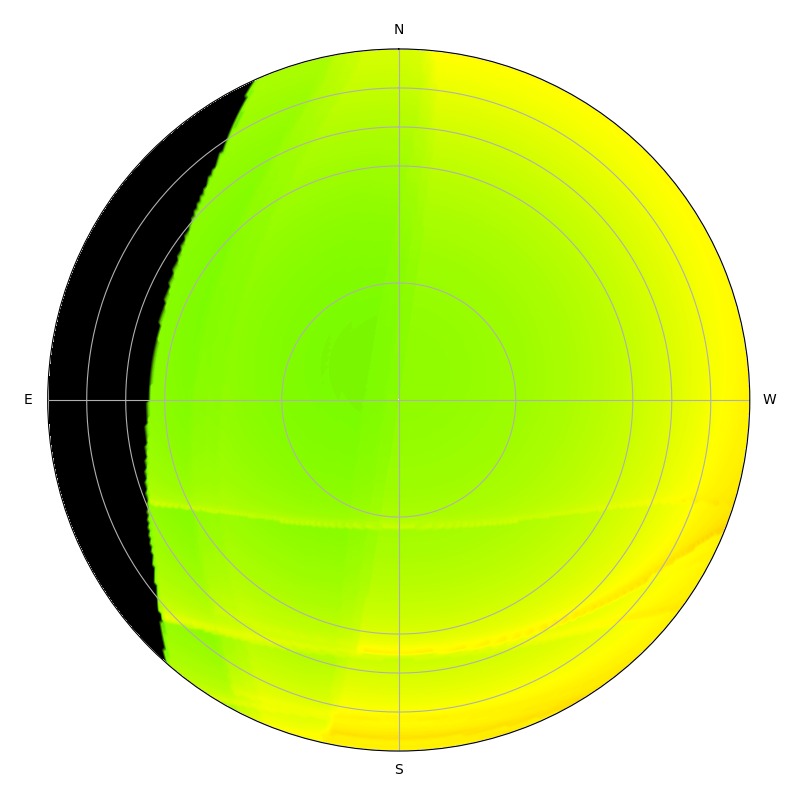} $-18\degr$, 0.44\% &
       \includegraphics[width=\mywidth]{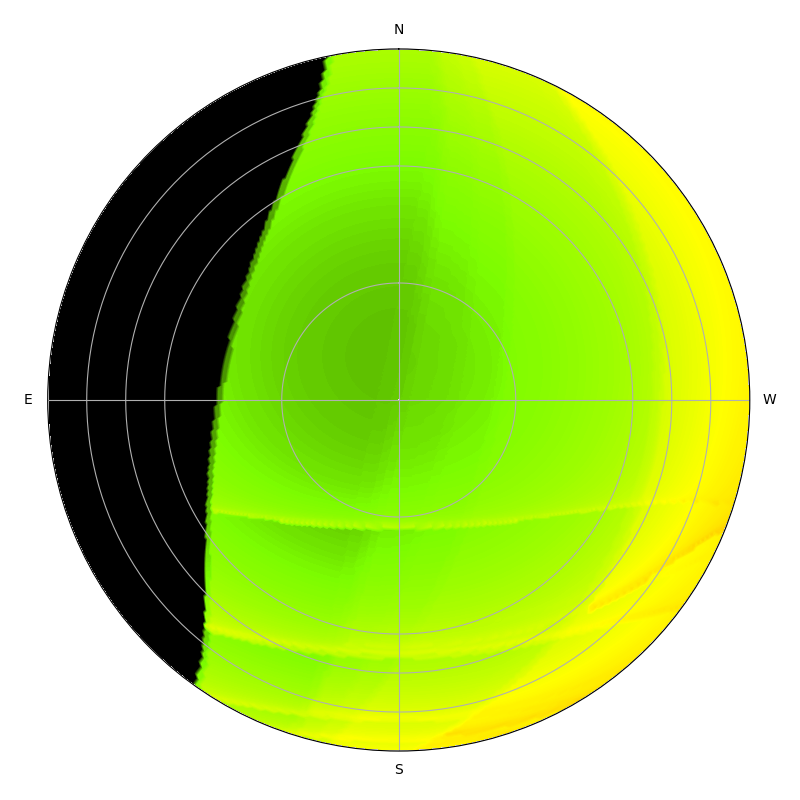} $-24\degr$, 0.23\%&
       \includegraphics[width=\mywidth]{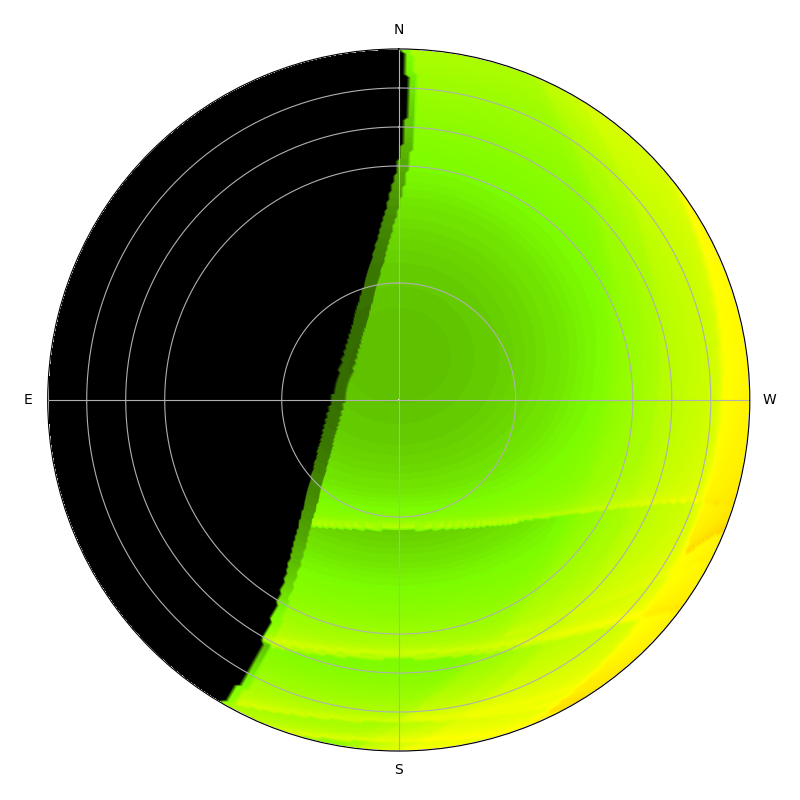} $-30\degr$, 0.12\%\\
       \includegraphics[width=\mywidth]{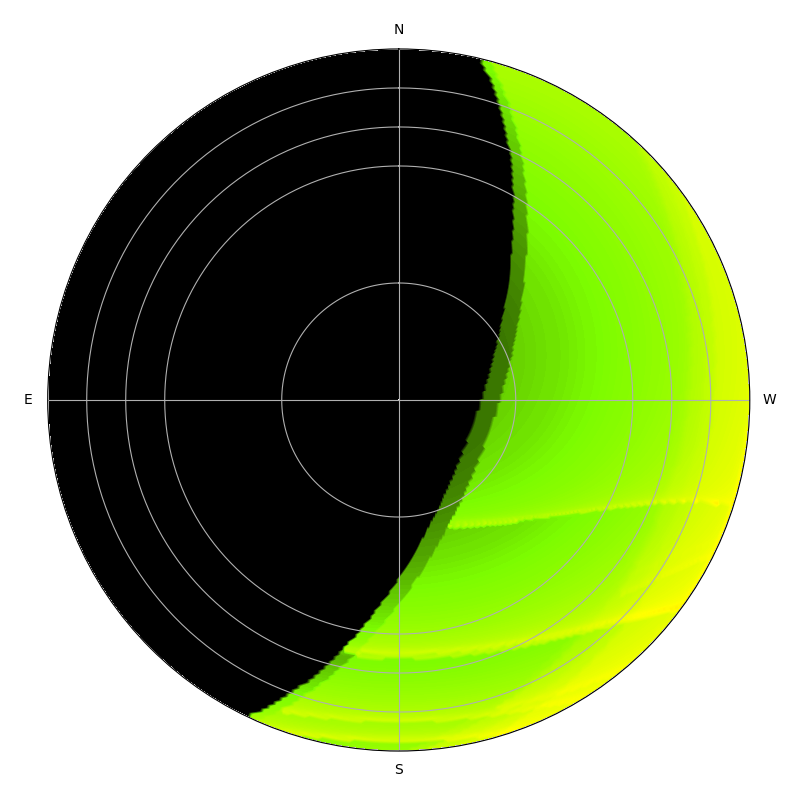} $-36\degr$, 0.06\%&
       \includegraphics[width=\mywidth]{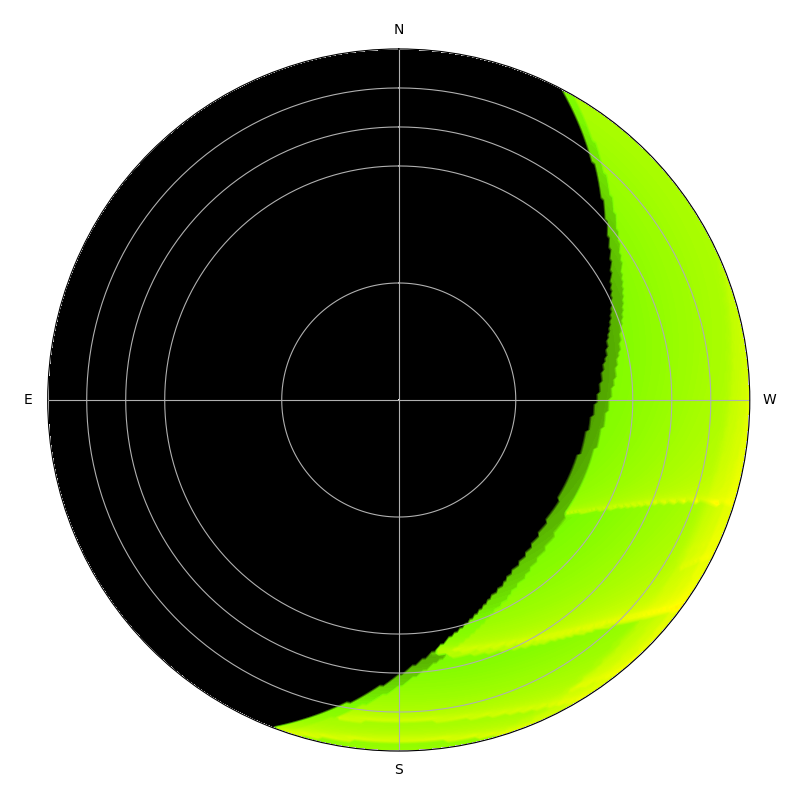} $-42\degr$, 0.01\%&
       \includegraphics[width=\mywidth]{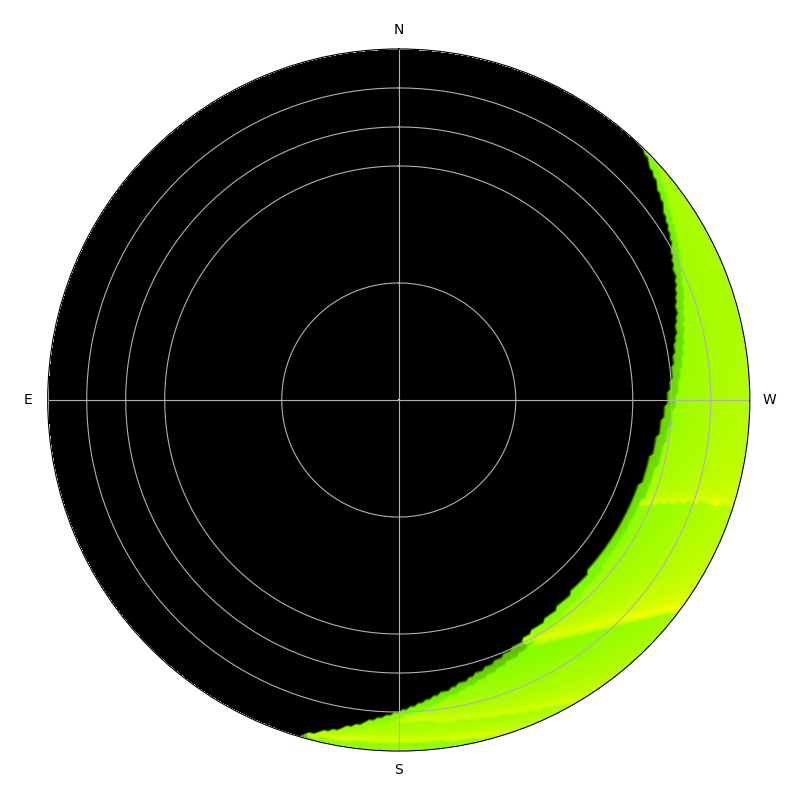} $-48\degr$, 0\%&
       \includegraphics[width=\mywidth]{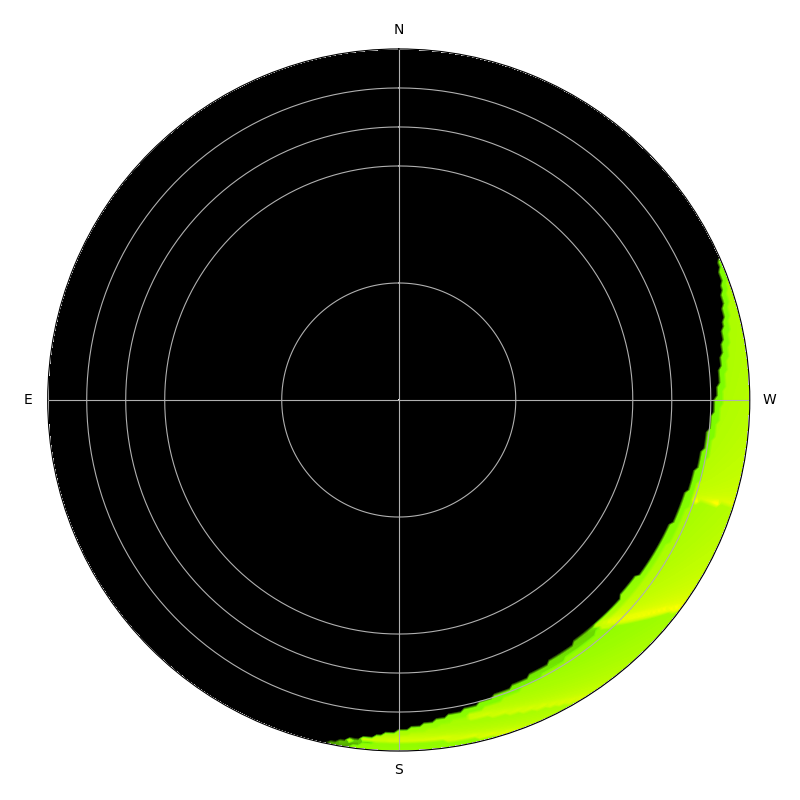} $-54\degr$, 0\%\\
       \multicolumn{4}{c}{\includegraphics[width=12cm]{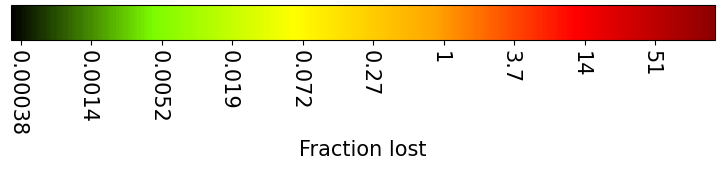}}\\
     \end{tabular}
     \caption{Sky maps of the number of detectable satellite trails
       (a) and effect on the observations (b), for all the satellites
       from Table~\ref{tab:1} on a FORS2 image ($6\arcmin$ field of
       view, 5 min exposure time) on Paranal ($-24\fdg6$ latitude) at
       equinox. The circles indicate elevations $0\degr$, $10\degr$,
       $30\degr$, and $60\degr$. The legend of each plot gives the Sun
       elevation and the average number of trails (a) and the losses
       they cause (b) for observations above $30\degr$ elevation All
       satellites are brighter than the detection limit, and none is
       bright enough to cause heavy saturation. }
     \label{fig:example}
   \end{figure*}
   
   For each constellation shell, the instanteneous satellite density,
   angular velocity and apparent and effective magnitudes were
   estimated. The number of trails affecting an exposure was obtained
   using Eqn.~\ref{eq:parabolic}, using $A_\mathrm{fov}=L_1 \times
   L_2$ for the field of view (with $L_1$ the length and $L_2$ the
   width of the FOV, $L_1>L_2$) and $L_\mathrm{fov}=L_1$ in the second
   term of Eqn.\,\ref{eq:parabolic} -- this maximises the
   cross-section for trails. $N_\mathrm{sat}$ was computed for each
   shell accounting for the effective magnitude of the satellites in
   that shell.

   \begin{figure}[!ht]
     \setlength{\mywidth}{4.0cm}
     \begin{tabular}{p{\mywidth}p{\mywidth}}
       \multicolumn{2}{l}{\bf a: Imagers }\\
       \includegraphics[width=\mywidth]{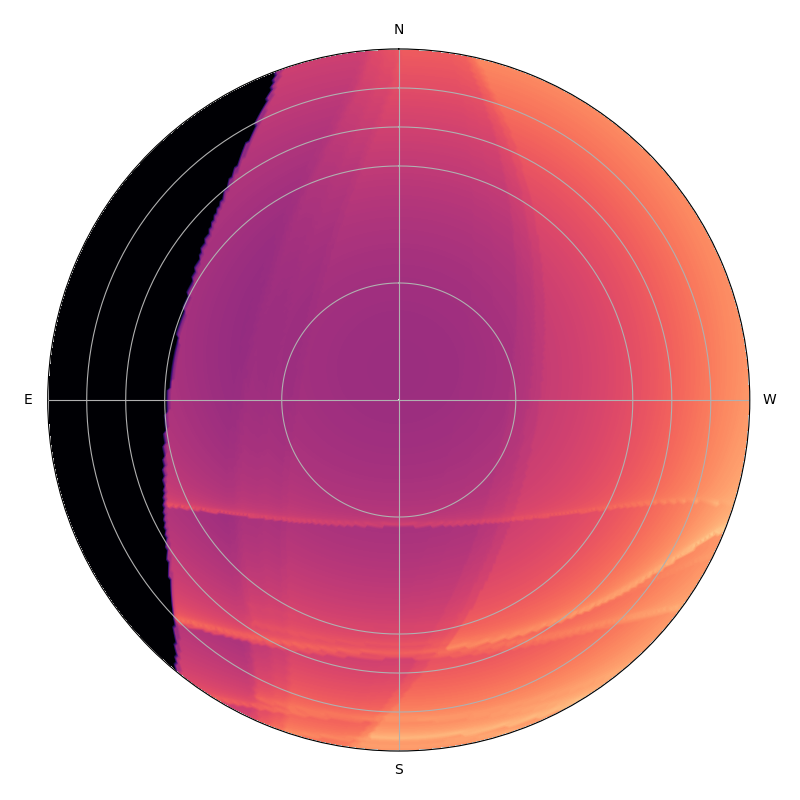} &
       \includegraphics[width=\mywidth]{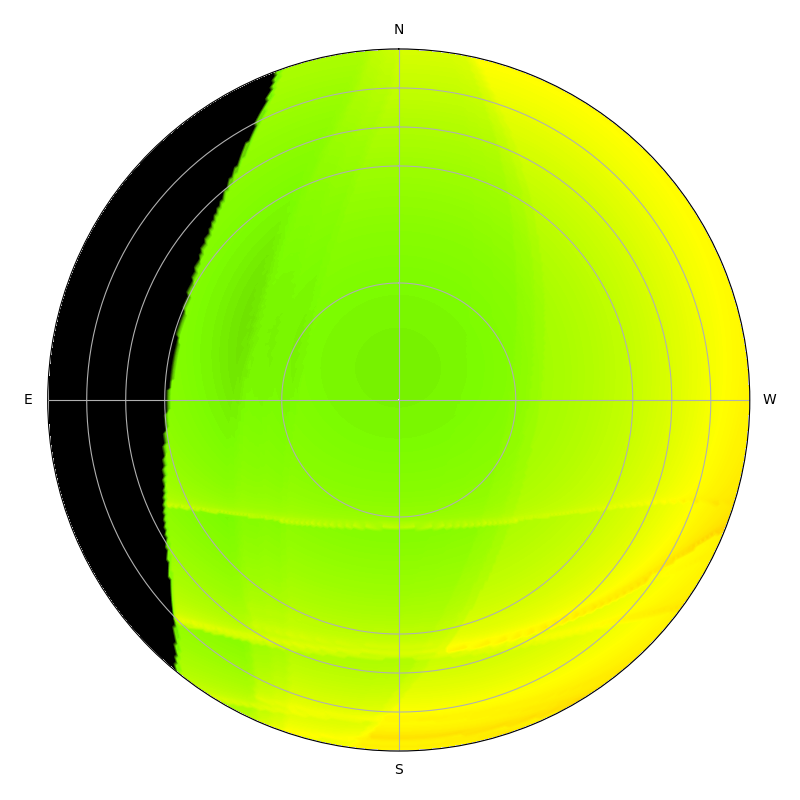} \\
       \multicolumn{2}{c}{
         \shortstack{FORS~(VLT, ESO, Paranal)~25.2\\ Average:~0.16~trail,~0.44\%~loss }}\\
       \includegraphics[width=\mywidth]{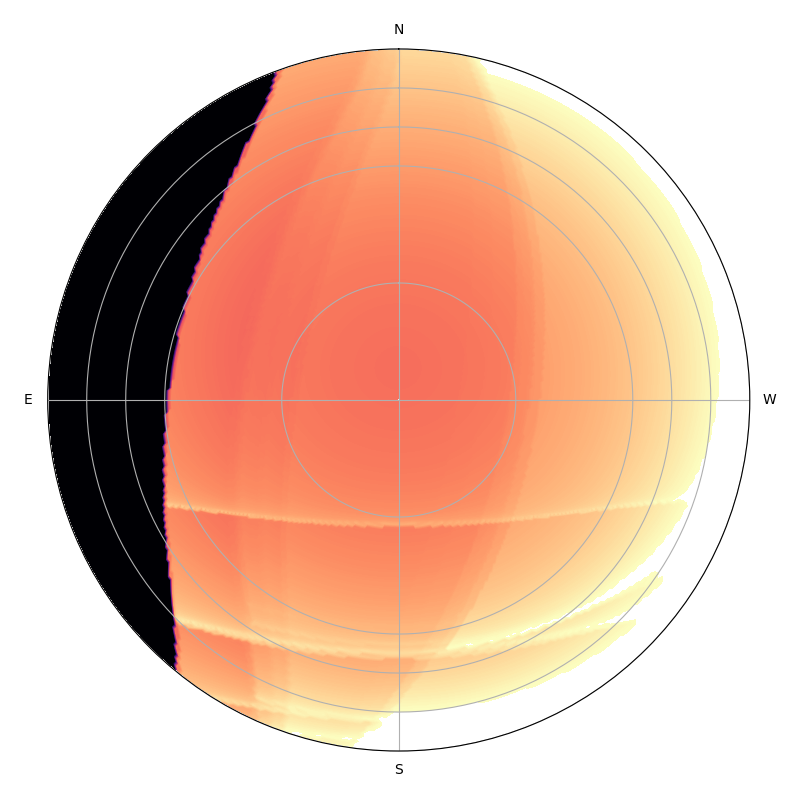} &
       \includegraphics[width=\mywidth]{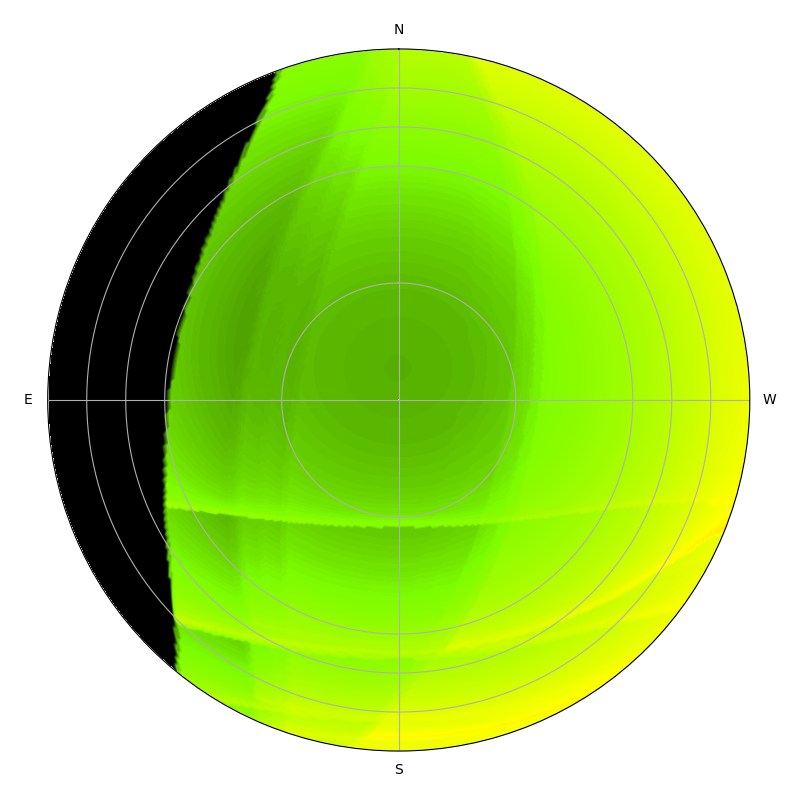} \\
       \multicolumn{2}{c}{
         \shortstack{OmegaCam (VST, ESO, Paranal) 23.9,\\ 1.60 trail, 0.22\% loss }}\\
       \includegraphics[width=\mywidth]{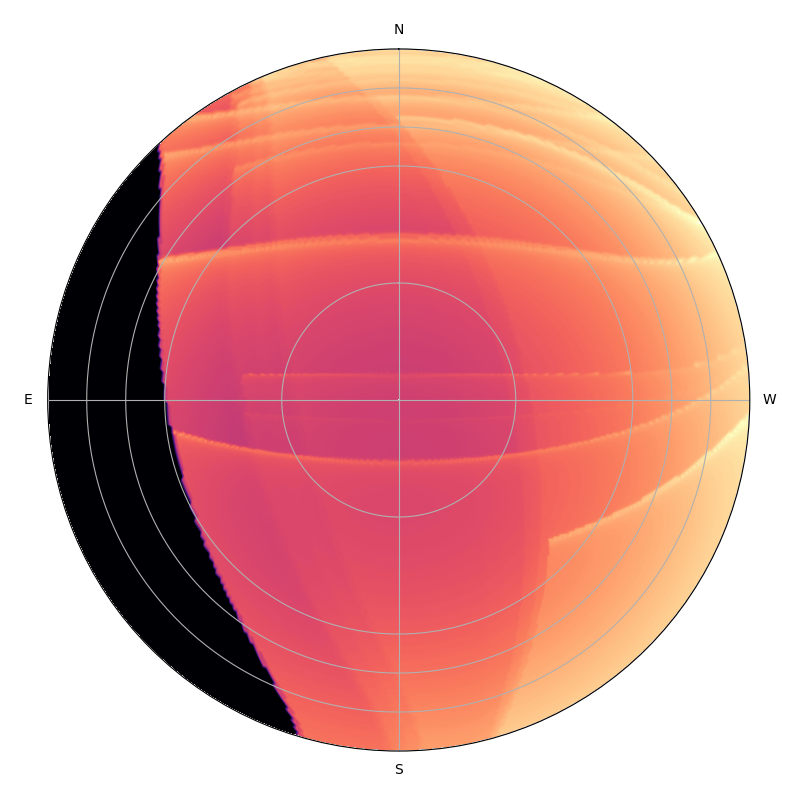}  &
       \includegraphics[width=\mywidth]{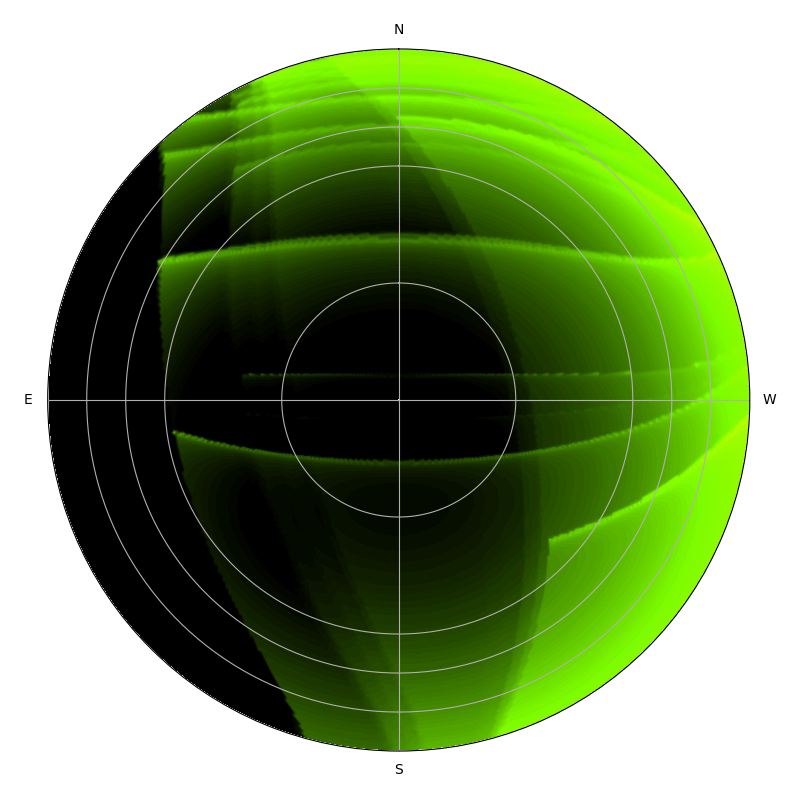}  \\
       \multicolumn{2}{c}{
         \shortstack{1.5m G96 (Catalina, U.AZ, Mt. Lemmon) 21.4,\\ 0.47 trail, 0.030\% loss}}\\
       \includegraphics[width=\mywidth]{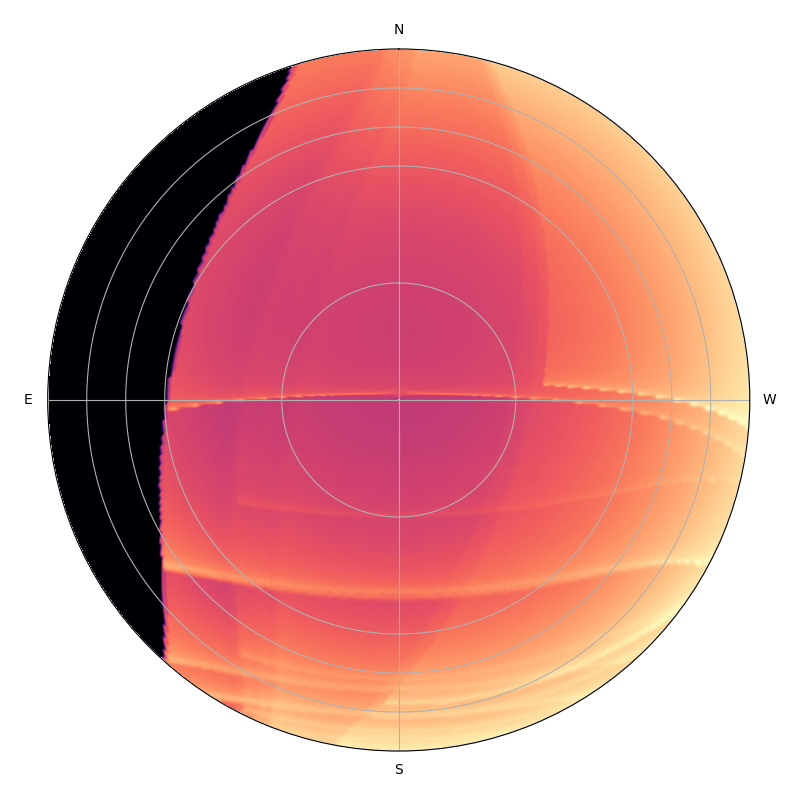}  &
       \includegraphics[width=\mywidth]{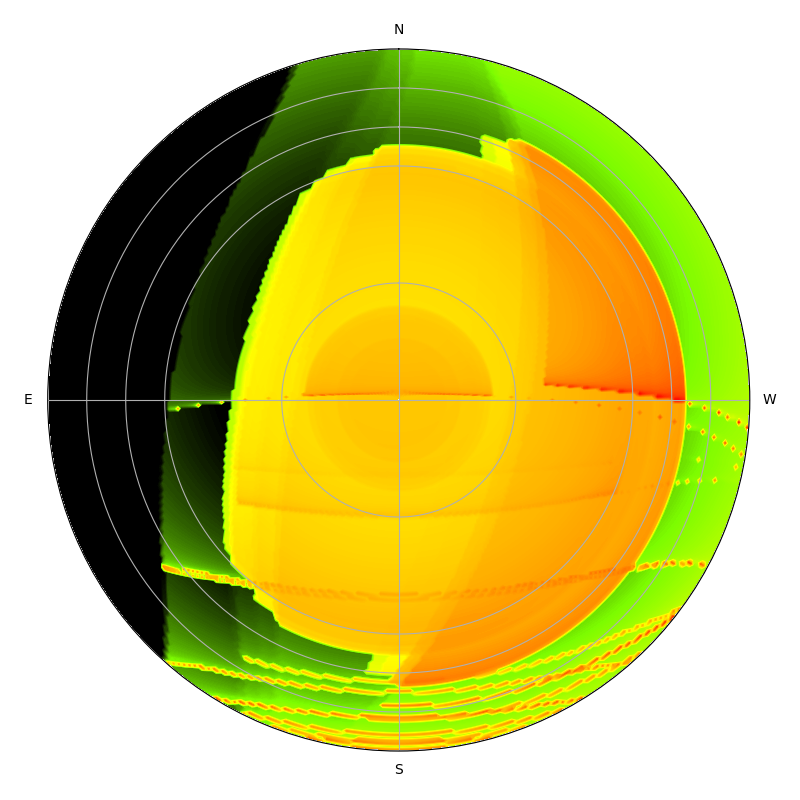}  \\
       \multicolumn{2}{c}{
         \shortstack{SST Cam. (SST, VRO, Pachon) 21.4,\\ 0.41 trail, 22.0\% loss}}\\
       \includegraphics[width=\mywidth]{figures_example/trailperexp.png} &
       \includegraphics[width=\mywidth]{figures_example/fractionlost.png} \\
     \end{tabular}
     \caption{Sky maps of the number of detectable satellite trails in
       an exposure (left) and effect on the observations (right) for a
       series of imagers (see Table~\ref{tab:instruments} for their
       characteristics). The legend of each plot also lists the
       average number of trails above $30\degr$ elevation. The Sun
       declination is $0\degr$, and its elevation $-20\degr$. }
     \label{fig:instruments}
   \end{figure}

   \begin{figure}[!ht]
     \setlength{\mywidth}{4.0cm}
     \begin{tabular}{p{\mywidth}p{\mywidth}}
       \multicolumn{2}{l}{\bf b: Spectrographs }\\
       \includegraphics[width=\mywidth]{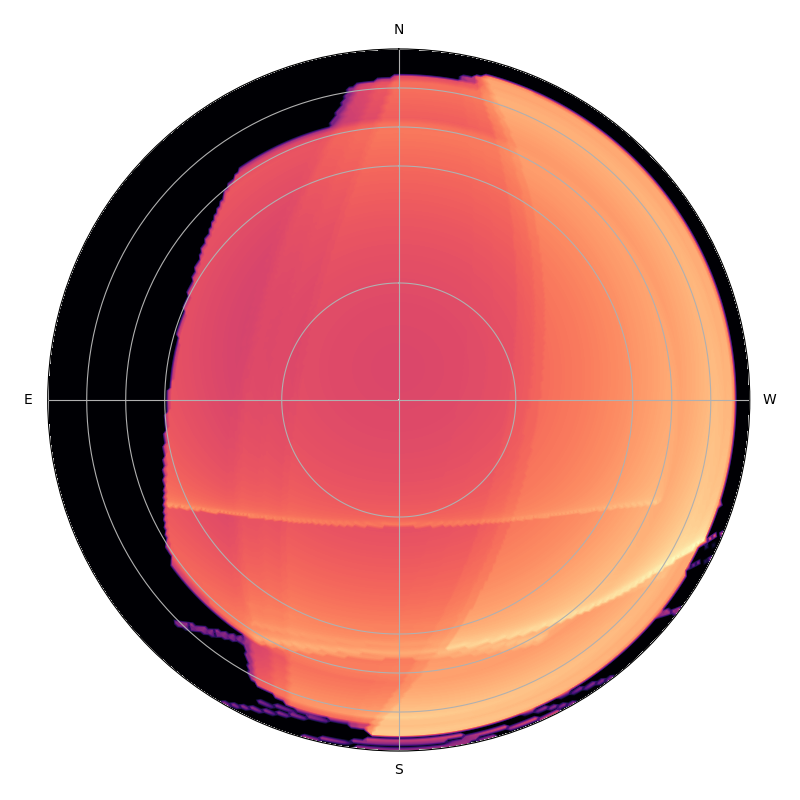} &
       \includegraphics[width=\mywidth]{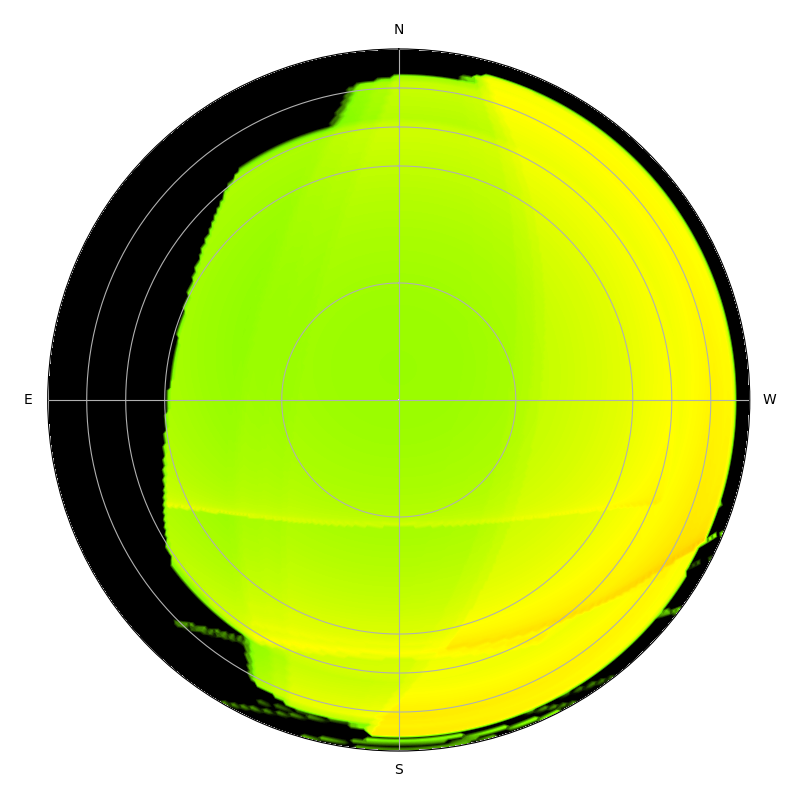} \\
       \multicolumn{2}{c}{
         \shortstack{FORS (VLT, ESO, Paranal)~25.2 Average:~0.64~trail,~8.8\%~loss}}\\ ~\\
       \includegraphics[width=\mywidth]{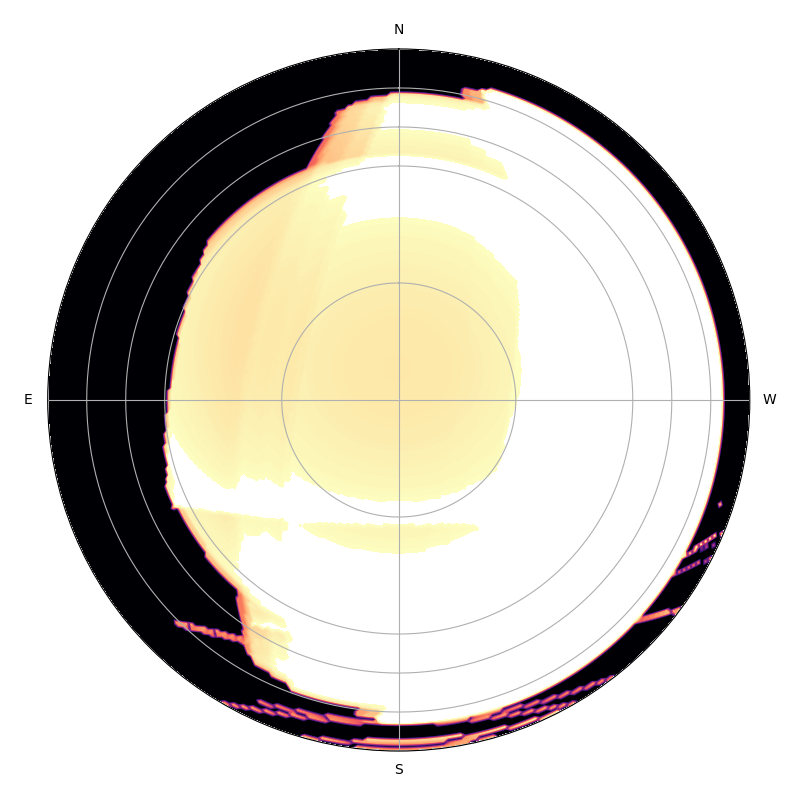} &
       \includegraphics[width=\mywidth]{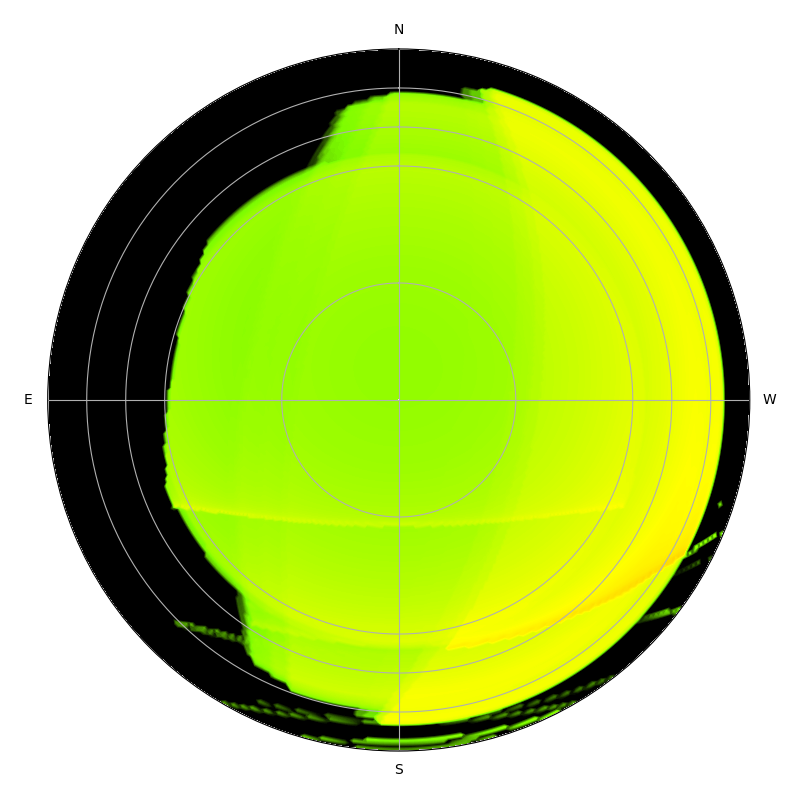} \\
       \multicolumn{2}{c}{\shortstack{
           4MOST-LowRes (VISTA, ESO, Paranal), 14.7 trails, 0.78\% loss }}\\~\\
       \includegraphics[width=\mywidth]{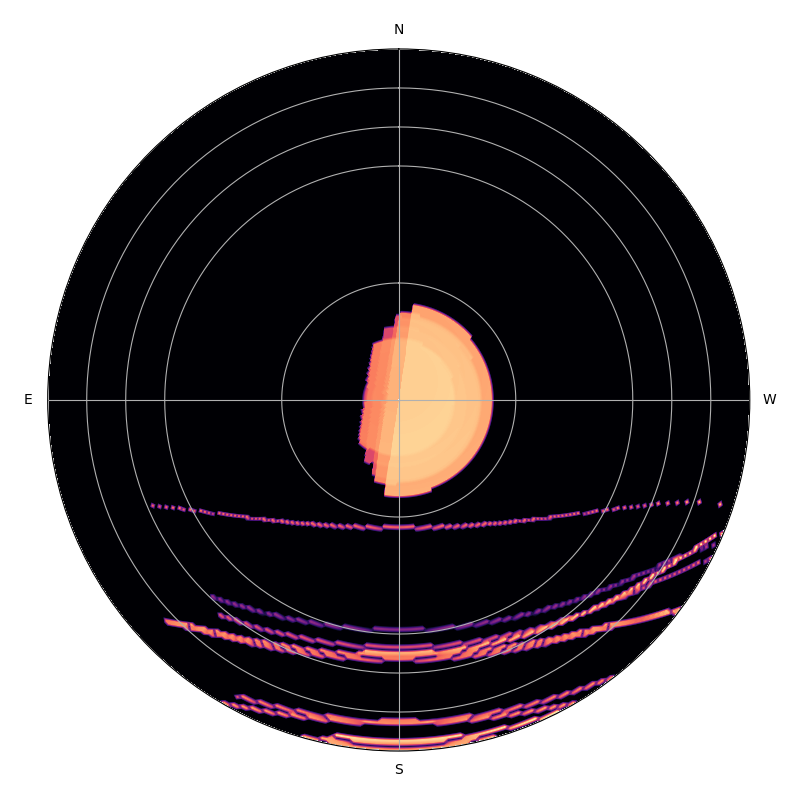}  &
       \includegraphics[width=\mywidth]{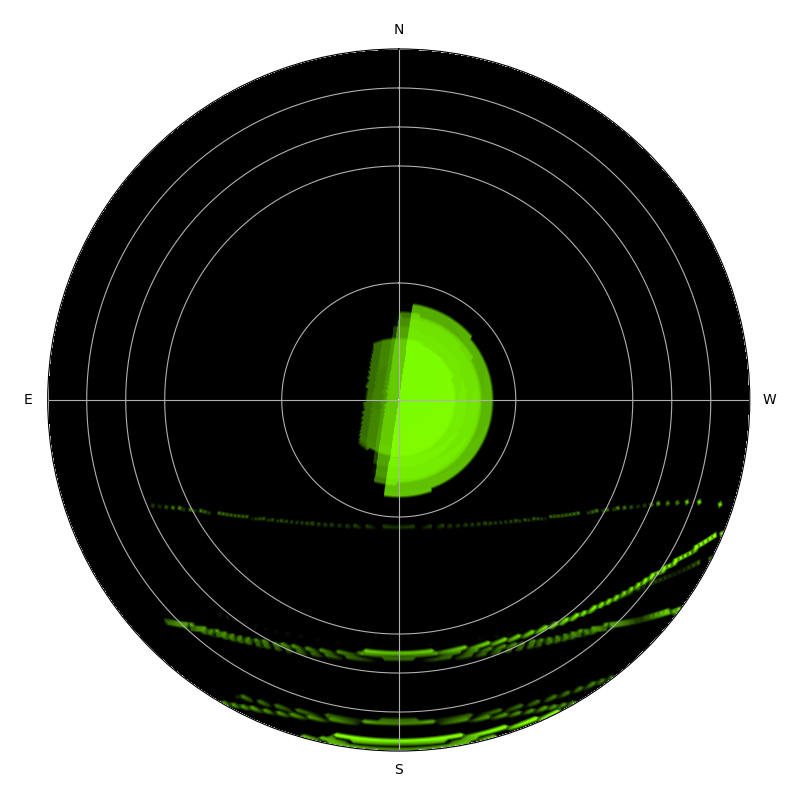}  \\
       \multicolumn{2}{c}{
         \shortstack{4MOST-HiRes (VISTA, ESO, Paranal), 0.33 trail, 0.018\% loss}}\\~\\
       \includegraphics[width=\mywidth]{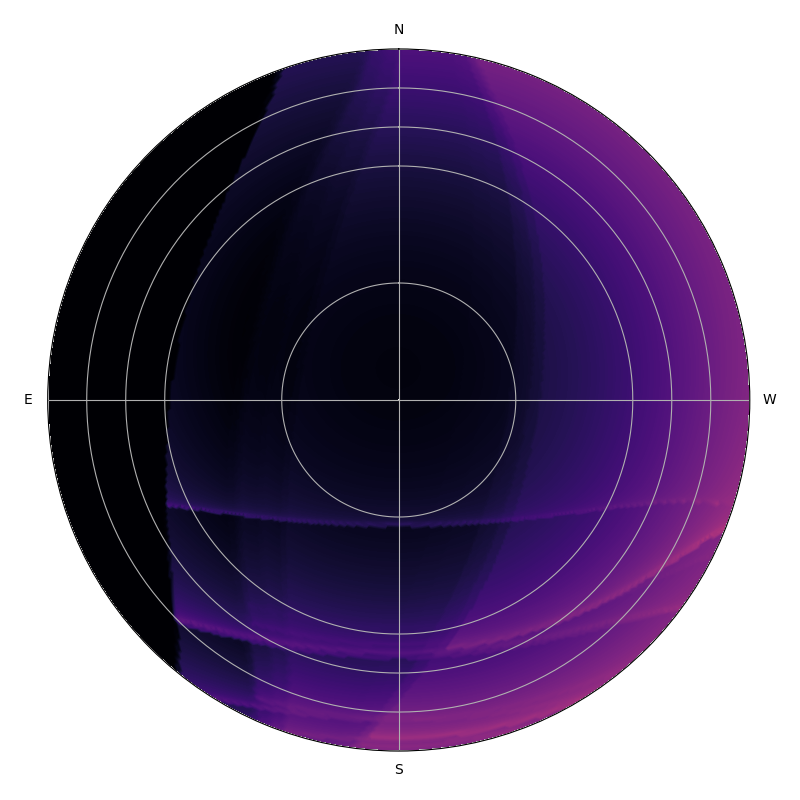}  &
       \includegraphics[width=\mywidth]{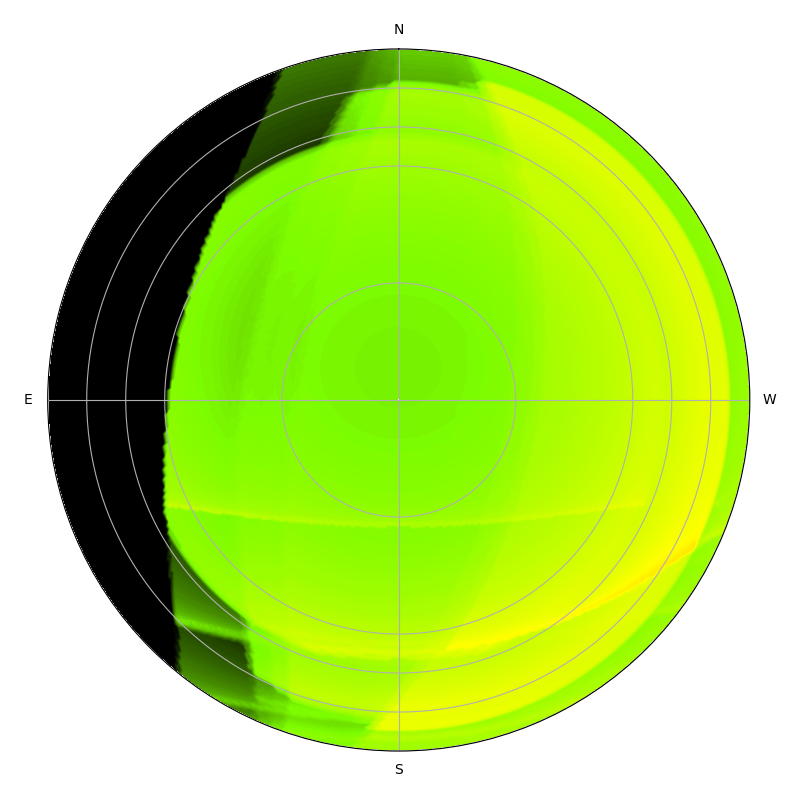}  \\
       \multicolumn{2}{c}{
         \shortstack{HARMONI (ELT, ESO, Armazones), 0.007 trail, 0.70\% loss}}\\~\\
       \includegraphics[width=\mywidth]{figures_example/trailperexp.png} &
       \includegraphics[width=\mywidth]{figures_example/fractionlost.png} \\
     \end{tabular}
     \caption{Sky maps of the number of detectable satellite trails in
       an exposure (left) and effect on the observations (right) for a
       series of spectrographs (see Table~\ref{tab:instruments} for
       their characteristics). The legend of each plot also lists the
       average number of trails above $30\degr$ elevation. The Sun
       declination is $0\degr$, and its elevation $-20\degr$
       ($-18\degr$ for 4MOST-HiRes; the values for $-20\degr$ are 0
       trail and 0\%). }
     \label{fig:instruments2}
   \end{figure}

\begin{figure}[t]
     {\bf a}\includegraphics[width=\columnwidth]{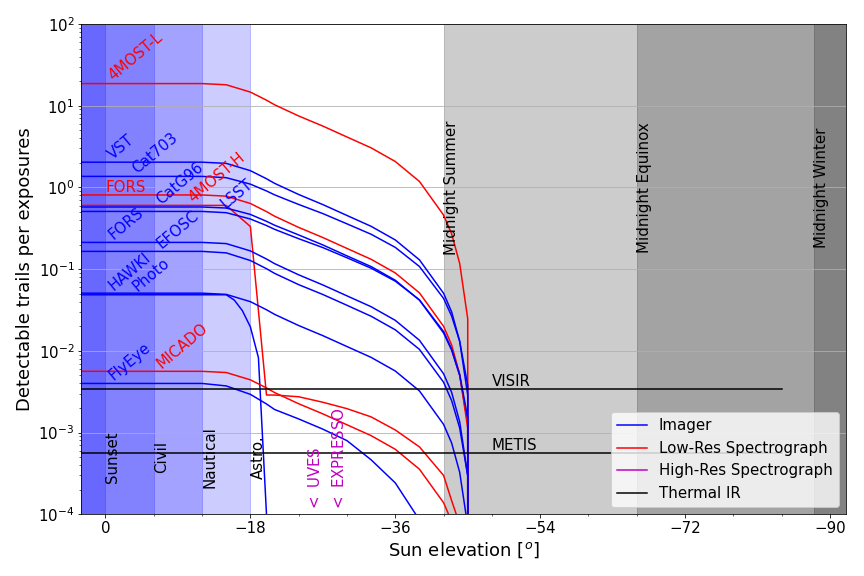}\\
     {\bf b}\includegraphics[width=\columnwidth]{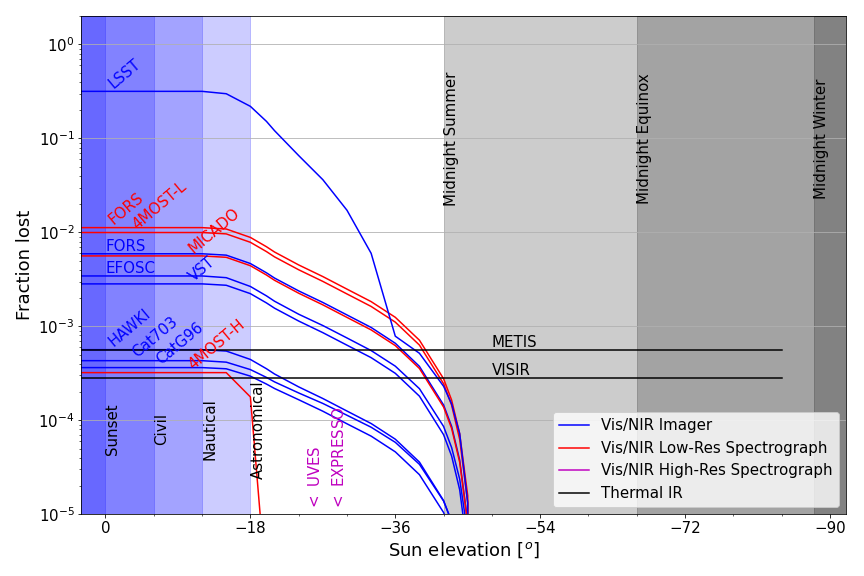}
     \caption{Average number of trails per exposure (a) and average
       fraction of the exposure lost (b) as a function of the sun
       elevation, for representative exposures at elevation $>30\degr$
       on the instruments listed in
       Table~\ref{tab:instruments}. Twilights are shaded in blue;
       inaccessible solar elevations are shaded in grey for the
       equinoxes and solstices for Paranal latitude ($-24\fdg6$).}
     \label{fig:trailsandlosses}
   \end{figure}

   The effect on the observations is computed as follows: Those
   satellites with an effective magnitude fainter than the $1 \sigma$
   detection limit were ignored, considering that their trail would be
   lost in the background noise. Those between the detection limit and
   heavy saturation limit were counted, and each one was considered to
   ruin a $5\arcsec$-wide trail across the whole detector. In case of
   a long slit, they ruin $5\arcsec$ of the slit. In real
   observations, it is plausible that all or part of the data below a
   non-saturated trail could be recovered, so this is a pessimistic
   limit. In the case of a fiber contaminated by a satellite, we
   consider that the whole spectrum is lost.  For trails brighter than
   the heavy saturation limit, the whole exposure is considered
   damaged by the charge bleeding and/or electronic and/or optical
   ghosts.  This was repeated for each shell in the constellation, and
   the effects were summed, resulting in maps of lost fractions.  A
   value of, say, 50\% indicates that either 50\% of the individual
   exposures are entirely lost, or that 50\% of the pixels in each
   exposure are lost or, more likely, a combination in between.  This
   was then repeated for several solar elevations ranging from
   twilight to midnight. Figure~\ref{fig:example} displays the
   resulting sky maps of trail count and fraction lost for an example,
   with instrument specific all sky plots provided for imagers in
   Fig.\,\ref{fig:instruments} and spectrographs in
   Fig.\,\ref{fig:instruments2}.  The average number of trails per
   exposure and the average fraction of the exposure lost were
   computed for the region of the sky above $30\degr$ elevation by
   integrating the results over that region of the sky. These averages
   are shown in Fig.~\ref{fig:trailsandlosses}.
   
   Our software to predict the effect of satellites on observations is
   available at \url{https://github.com/cbassa/satconsim} and can be
   queried online at
   \url{https://www.eso.org/~ohainaut/satellites/simulators.html}.
  
   \section{Discussion}
   \label{sec:discussion}
   As apparent from Fig.~\ref{fig:trailsandlosses} and expected from
   Eqn.~\ref{eq:parabolic}, the number of trails in an exposure
   increases with the size $L_\mathrm{fov}$ of the field-of-view and
   with exposure time $t_\mathrm{exp}$. The effect of the effective
   magnitude (Eqn.~\ref{eq:effmag}) is less intuitive: for the
   altitudes of the constellations considered in this study,
   $m_{\mathrm 1000km}$ and exposure times, the effective magnitudes
   fall in the range 13--23, and become {\em fainter} as the exposure
   time increases, with a linear dependency in $t_\mathrm{exp}$. As
   the limiting magnitude for an exposure goes fainter with a
   dependency in $\sqrt{t_\mathrm{texp}}$ (considering the simple
   sky-noise dominated case) there is, for each instrument, an
   exposure time beyond which a satellite trail will no longer be
   detectable. In other words, the contribution from a satellite to
   the intensity in a resolution element is independent of the
   exposure time, while the noise increases with
   $\sqrt{t_\mathrm{texp}}$. Overall, the SNR of the satellite trail
   decreases with $\sqrt{t_\mathrm{texp}}$; if an exposure
   $t_\mathrm{exp}$ is immune to satellite trails, longer exposures
   will also be immune.

   Imagers on all but the smallest telescopes typically have a
   limiting magnitude fainter than the faintest satellite effective
   magnitude: they are, therefore, affected to some extent by all
   satellite constellations. For many science cases, the presence of a
   trail will only result in a loss of useful imaged area (of the
   order of 0.1 to 1\% for a $5\arcsec$ wide trail crossing a $1\degr$
   or $8\arcmin$ field of view). There will be, however, some science
   cases in which even a faint trail will ruin the whole exposure
   (e.g.\ photometry of a faint trans-Neptunian object overrun by a
   satellite), leaving no other choice than repeating the exposure, if
   this were possible at all (sometimes the repetition is not possible
   as, for instance, for the photometry of a transient gamma-ray
   burst). Furthermore, for the most sensitive cameras, some
   satellites have an effective magnitude brighter than the heavy
   saturation limit, wreaking havoc in the affected exposures, as on
   the LSST camera at the Vera Rubin Observatory (VRO), and resulting
   in much heavier losses \citep{tyson+20}.

   For astrophotography wide-field cameras, the limiting magnitude for
   satellite trails scales inversely with the focal length of the
   lens, with every other parameter remaining constant. A wide-angle
   camera with 30~mm focal length will therefore be 5~mag less
   sensitive than a 3~m focal length telescope with the same focal
   ratio. As a consequence, astrophotography will be immune to most
   satellites in their operational orbits. They can, however, be
   affected by brighter satellites, e.g.\ larger satellites, or
   telecommunication satellites in low altitude transfer orbits (such
   as the bright strings-of-pearls of 60 very bright satellites, as
   observed after the early Starlink launches), or specular
   reflections. Fortunately, these are not numerous: it is foreseen
   that there will be of the order of 10 trains of satellites around
   the Earth at any time to replenish the constellations. While
   potentially spectacularly damaging, these are statistically
   unlikely and visible only during the brightest parts of twilights.

   For spectrographs, the limiting magnitude for a single exposure
   often falls in the range of the satellites effective magnitudes. As
   a consequence, those fainter than the limit are not detected and
   only slightly contribute to the background noise. This is the case
   for all satellites for high-resolution spectrographs or \'echelle
   spectrographs, even on very large telescopes (see the examples of
   UVES and ESPRESSO on the VLT). However, low- to medium-resolution
   spectrographs on medium to large telescopes will detect all or many
   satellites. Furthermore, contrary to imagers, where a satellite
   leaves a tell-tale trail in the data, slit and fibre spectrographs
   do not record spatial information. While high-SNR contamination
   would be easy to notice (e.g.\ the exposure level is much higher
   than expected, and the spectral shape does not match that expected
   for the target), many satellites will leave a signal with a low to
   moderate SNR. In many cases, the contamination will be at a level
   comparable to or below that of the science target and therefore
   unlikely to be detected in real-time. Unless the contamination is
   flagged using other means (see below, \S\ref{ssec:mitigation}),
   there will be cases for which it will become apparent only at the
   time of the scientific analysis of the spectra. As the
   contamination will have a solar spectrum, some science cases will
   be better protected (e.g.\ study of distant quasars) than others
   (e.g.\ study of double stars, where a solar spectrum may not be
   surprising).

   In the thermal IR domain, the overall signal is dominated by the
   very strong thermal emission from the sky and the telescope. The
   individual exposure time is therefore kept extremely short (few
   tens of milliseconds), and the background is registered by chopping
   (i.e.\ performing a small position offset by tilting the secondary
   mirror of the telescope) at about 1~Hz, and nodding (another small
   offset by moving the whole telescope) every few seconds. In
   \citet{hw20}, we conservatively estimated the flux from a satellite
   at 2000\,km at zenith up to 100\,Jy in $N$-band (8--13\,$\upmu$m)
   and up to 50\,Jy in the $M$ and $Q$-bands (5 and 18-20\,$\upmu$m,
   respectively). The variations between an illuminated and a shadowed
   satellite are negligible. These fluxes are well above the detection
   threshold of the thermal IR instruments in
   Table~\ref{tab:instruments}, even accounting for trailing. Because
   of the extremely short exposure time and small field of view, on
   average only $6\times10^{-6}$ trails would be found in a single
   exposure. However, for most observations, the images are not
   individually recorded but averaged over a nodding cycle. While the
   SNR of the trail will be washed away by this average, the values
   plotted in Fig.~\ref{fig:trailsandlosses} correspond to the
   duration of these averages (10\,s). In the case of VISIR on the
   VLT, it is considered that a satellite ruins a $5\arcsec$ trail
   across the detector, and for METIS on the larger ELT, the full
   (smaller) image is ruined by the (broader) trail. Even with these
   extremely pessimistic assumptions, a negligible fraction of the
   thermal IR data is affected. For spectroscopy, even at low
   resolution, the satellite effective fluxes will be below the
   detection limit.

   The case of stellar occultations was discussed in \citet{hw20}. The
   effect was found to be small: a 0.02 to 10 millimag for 10\,s and
   0.1\,s exposure time, and extremely improbable: $10^{-4}$ to
   $10^{-6}$ exposures would be affected. The simulations presented
   here do not change these estimates. As the eclipse by a satellite
   would affect only one measurement in a series, even if it could be
   measured, it would {\em not} be similar to the occultation by an
   exoplanet or a trans-Neptunian object.

   The case of visual observations -- either naked eye, or through
   binoculars or telescope -- will be considered in a separate
   paper. In summary, 15 to 50 satellites would be visible in the sky
   with the naked eye when the sun elevation is between $-12\degr$ and
   $-24\degr$. When the sun is higher than $-12\degr$, the sky is too
   bright and no satellite is visible. When the sun is lower than
   $-24\degr$, no satellite is bright enough to be visible.

   \subsection{Mitigation}
   \label{ssec:mitigation}

   \begin{figure*}[!ht]
     \includegraphics[width=\textwidth]{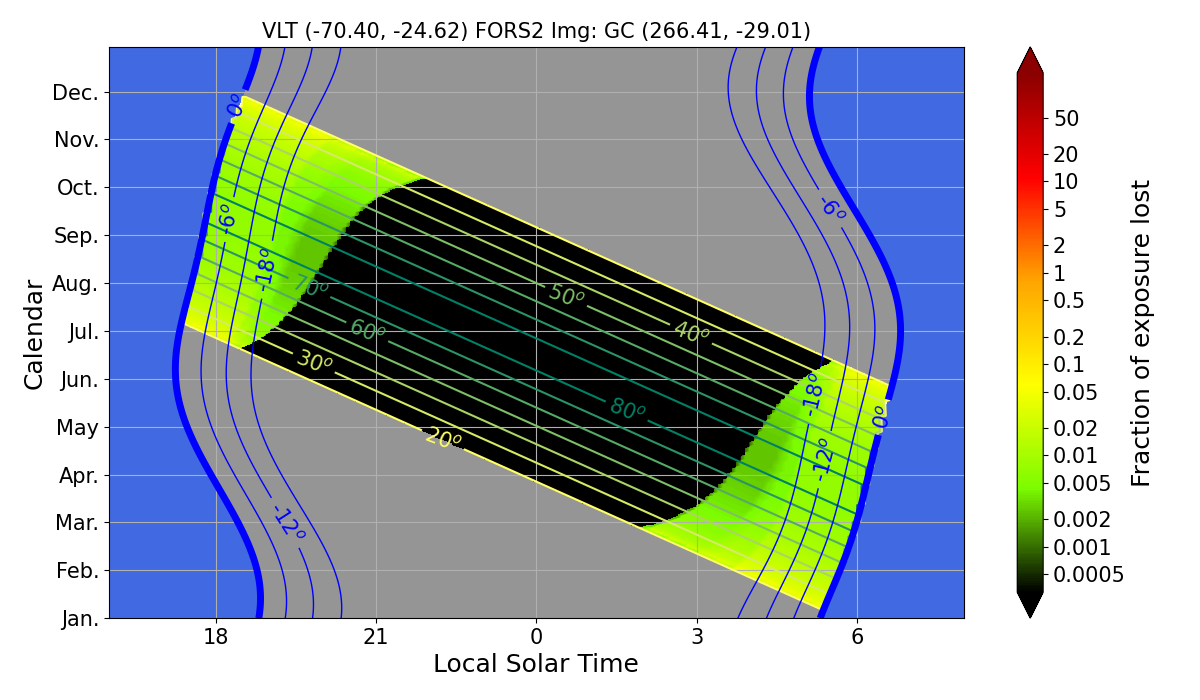}\\
     \includegraphics[width=\textwidth]{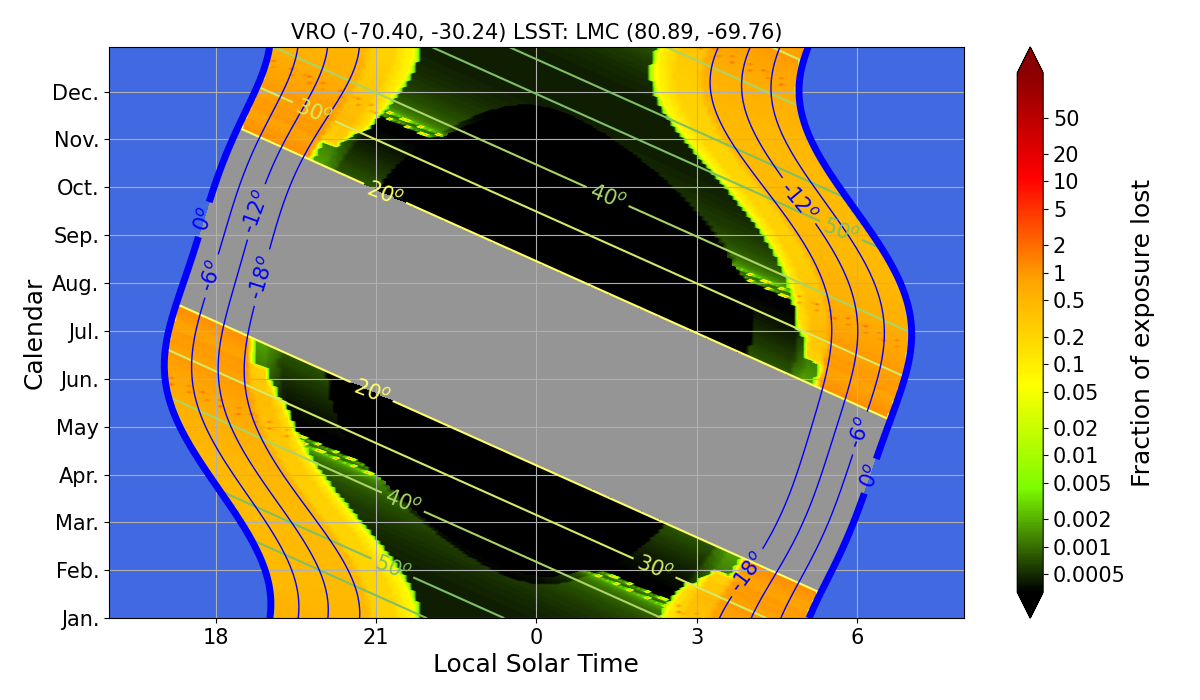}
     \caption{Two examples of calendars showing the visibility of an
       object from an observatory (Galactic Centre from the VLT, left,
       and Large Magellanic Cloud from VRO, right). The fraction of
       observing time lost due to satellites is indicated by the
       colour scale (1 indicating that all exposures are damaged), for
       300\,s exposures with FORS2 and 15\,s on the LSST camera,
       respectively. The satellites are the 60+ thousand from
       Table~\ref{tab:1}.  The blue shading marks daytime, and the
       blue contours indicate the twilights. The elevation of the
       object is indicated by the greenish contour lines, and the grey
       shading indicates the times when the object is below 20$\degr$
       elevation. }
     \label{fig:calendar}
   \end{figure*} 
 
   First order of mitigation refers to the satellites themselves. 

   {\em Number and altitude:} As seen in Fig.~\ref{fig:elevation}, the
   number of illuminated satellites in sight is of course a function
   of the number of satellites in the constellations, but also of the
   altitude of the constellation. Furthermore, high-altitude
   constellations remain illuminated by the Sun much longer than
   low-altitude ones. The apparent magnitude of a high-altitude
   satellite will be fainter than that of the same satellite on a
   lower altitude (Eqn.~\ref{eq:absmag}), which is advantageous for
   small telescopes. However, for large telescopes, the satellite
   appears extended (Eqn.~\ref{eq:psfsize}), so that its surface
   brightness will not decrease much. Overall, a constellation at
   1000~km will be more damaging than a three times larger
   constellation at 500~km altitude.
    
   {\em Brightness of the satellites:} Obviously, keeping the
   brightness of all satellites below the detection limit of all
   telescopes would be ideal. With the increasing size of the
   telescopes and sensitivity of the instruments, this is not
   realistic. An achievable goal would be to reduce the effective
   cross-section of the satellite so that they remain always below the
   saturation threshold of the most sensitive instrument. Today and in
   the foreseeable future, this threshold is set by the SST camera at
   VRO, at $V_{\mathrm 550km}>7$ \citep{tyson+20}, or $V_{\mathrm
     {1000\,km}} > 8.3$. The changes introduced by
   SpaceX\footnote{\url{https://www.spacex.com/updates/starlink-update-04-28-2020/}}
   with VisorSat and modified attitude of the solar array are very
   promising steps in the right direction: most of the satellites are
   now below the heavy saturation threshold, while still causing
   electronic cross talk \citep{tyson+20}. More systematic
   measurements of the satellites are needed \citep[as suggested by
     \textsc{SatCon1} Recommendation 8, see][]{wal20}, and awareness
   of the satellite operators is a must. \textsc{SatCon1}
   Recommendation 5 and \textsc{Dark \& Quiet Skies} Recommendation
   15\footnote{See p. 153 of the report:
   \url{https://www.iau.org/static/publications/dqskies-book-29-12-20.pdf}}
   formalize the brightness limit at $V_{\mathrm {550\,km}} > 7$.

   Once the satellites are in orbit, the next level of mitigation is
   at the time of preparation and scheduling of the
   observations. Because of the progression of the Earth shadow
   through the constellation shells, and of the fine structure in the
   apparent density of satellites (\S\ref{ssec:finestructure}), the
   fraction of losses can change dramatically by pointing the
   telescope in a slightly different direction. For a given time at a
   given observatory, sky maps such as those in Fig.~\ref{fig:example}
   would allow an observer to pick objects in the region of the sky
   that are least affected by satellites. As these maps can be
   generated on-the-fly, they could be integrated, for instance, in a
   queue observation optimization algorithm. Another way to consider
   the scheduling is to pick the best time slots to observe a given
   object with a specific instrument. Using the same methodology as
   for the sky maps, a calendar can be populated with the expected
   density of satellites for an object seen from an observatory. Such
   calendars, some examples of which are displayed in
   Fig.~\ref{fig:calendar}, could be used when allocating specific
   telescope time to an observation program in traditional visitor
   mode. Obviously, both the sky maps and calendars will introduce
   additional complexity in the scheduling process, and will not
   resolve all issues. For instance, some observations must be
   performed at a given time (e.g.\ an exoplanetary transit
   observation).

   A more aggressive mitigation would be to close the shutter of the
   instrument just before a damaging satellite enters the field of
   view, and re-open it just after it left. As the satellites move at
   apparent speeds of $\sim 0.1$ to $\sim 1\degr$\,s$^{-1}$, the
   interruption would be extremely short, virtually nullifying the
   losses of exposure time. The challenge is to send the signal to the
   shutter at the right time. Two methods can be envisioned:

   The first would rely on a complete, accurate and up-to-date
   database of orbital elements so that the position of all satellites
   can be computed at any time, and offending ones identified in
   times.  However, the accuracy of ephemerides, both in position and
   in timing, must be of the order of the field-of-view. An accuracy
   of a fraction of a degree (and 1\,s) may be achievable, making this
   viable for imagers. An accuracy of $\sim 1\arcsec$, required for
   spectrographs, would imply predicting the position of the
   satellites at 2--5~m accuracy, with a timing precision of
   $1/100$~s. This method presents various challenges: the database
   must be complete; it must be up-to-date (as non-keplerian effects,
   including active orbital corrections, modify the orbit with a
   time-scale of up to a few days); it must be precise (the current
   standard Two-Line Elements do not provide the required precision);
   the computations must be done with the appropriate precision, and
   scan the whole database for each observation. Furthermore, this
   method would not work for all instruments: large survey cameras
   tend to be fairly heavy and slow, precluding rapid and repetitive
   shutting and opening.

   The second method would rely on an auxiliary camera mounted in
   parallel with the main telescope, with a field of view of 5 to
   $10\degr$ (a few degrees larger than the field of the main
   instrument). This camera would take an image of the field about
   every second, and an analysis system would detect any transient
   object. An object moving towards the science field of view would
   trigger the closing of the shutter. The challenge here is to build
   a system fast enough to process the data in real-time, and a camera
   sensitive enough to detect the satellites. For instruments
   sensitive to all the satellites down to the faintest (e.g.\ imagers
   on large telescopes), it may not be realistically
   feasible. However, for spectrographs, which have a brighter
   limiting magnitude, a 30\,cm auxillary telescope with a fast
   read-out detector is promising. Spectrographs would strongly
   benefit from this mitigation, as their exposures tend to be much
   longer than those of imagers (and therefore the loss of an exposure
   more costly), and because they lack spatial information, what makes
   it possible that the contamination will remain unnoticed until the
   data are analysed.

   The final stage of mitigation is an a posteriori subtraction of the
   satellite trail from the data. This also comes with some
   limitations. Saturated trails (e.g.\ on a survey camera on a large
   telescope like the SST Cam at VRO) are un-recoverable. Fainter
   trails may be identified on images, modelled and
   subtracted. However, atmospheric scintillation will make these
   trails irregular, and difficult to model. Even assuming that they
   can be cleanly subtracted (and that this subtraction is trusted by
   the scientist), they will result in an increase of the photon
   noise. It may be safer to mask them and filter them out by
   combining several exposures of the same field. This is already
   systematically done for various types of blemishes affecting the
   images, such as cosmic ray hits and gaps between chips in the
   detector mosaic, and would result in a well-quantified loss of
   total exposure time in the affected strips, as already taken into
   account for detector gaps.

   The case of slit or fibre spectroscopy is again more difficult: no
   tell-tale trail reveals the passage of a satellite, which can only
   be detected as an additional solar-type spectrum added to the
   data. A solar spectrum can be iteratively subtracted from the
   spectrum until the residual shows no hint of the satellite. While
   this would work for some science cases, it will be more difficult
   or even impossible in other situations (e.g.\ a program studying
   stellar abundances or binary stars).

   Finally, the last resort of mitigation consists in repeating the
   observations that have been damaged by a satellite. In some cases,
   this will be immediately obvious and easily detected when
   controlling the quality of the data. In other cases, the effect
   will be more subtle. Some observations will be simple to
   re-acquire. Others will be lost forever (a short transient
   phenomenon, like the optical counterpart of a gravitational wave
   event).

   Overall, no mitigation method will single-handedly work for all
   instruments and all science cases. Moreover, each of these
   mitigations comes with a cost that should be carefully compared to
   the cost of the observing time loss: in many cases, repeating the
   observation may be cheaper (economically and scientifically) than
   protecting it.

   The way forward is to improve the situation at each step, starting
   with a collaboration with the satellite operators to make the
   satellite less bright, continuing with smarter scheduling of the
   observations, where the work presented in this paper will help
   thanks to the fast and accurate information it provides, shutter
   control where possible, and accounting for the inevitable remaining
   trails in the data.
   
   \begin{acknowledgements}
      This work originated as part of the \textsc{SatCon1} and
      \textsc{Dark \& Quiet Skies} workshops. We thank the organizers
      of these workshops.
   \end{acknowledgements}

   \bibliographystyle{aa}

   \begin{appendix}
     
     \section{Derivation of key expressions}
     \label{sec:ap:key_expressions} 
     In the following paragraphs we first show the derivation of
     Eqn.\,\ref{eq:probdens}, the probability density function
     $P(\phi,i,h_\mathrm{sat})$ (\S\ref{sec:ap:probadens}). We later
     get into some geometric details on the relation between the
     geocentric and topocentric positions of the satellites, impact
     angle of the line of sight on a shell, and the apparent,
     observed, angular velocity of satellites (\S\ref{sec:ap:geopos},
     \S\ref{sec:ap:impact} and
     \S\ref{sec:ap:apparentangvel}). Finally, in
     \S\ref{sec:ap:constedge}), we provide some equations on how to
     locate on the sky the intersection of constellation shell
     boundaries with the local meridian (Eqn.~\ref{eq:boundary} and
     some related expressions).
 
     \subsection{Probability density}\label{sec:ap:probadens}
     The analytical probability density function
     $P(\phi,i,h_\mathrm{sat})$ displayed in Eqn.\,\ref{eq:probdens}
     is at the core of many of the simulations included in this
     paper. Let us get into its derivation, down to some degree of
     detail. We will also prove that its integral is equal to unity.

\begin{figure}[t]
       \includegraphics[width=0.48\textwidth]{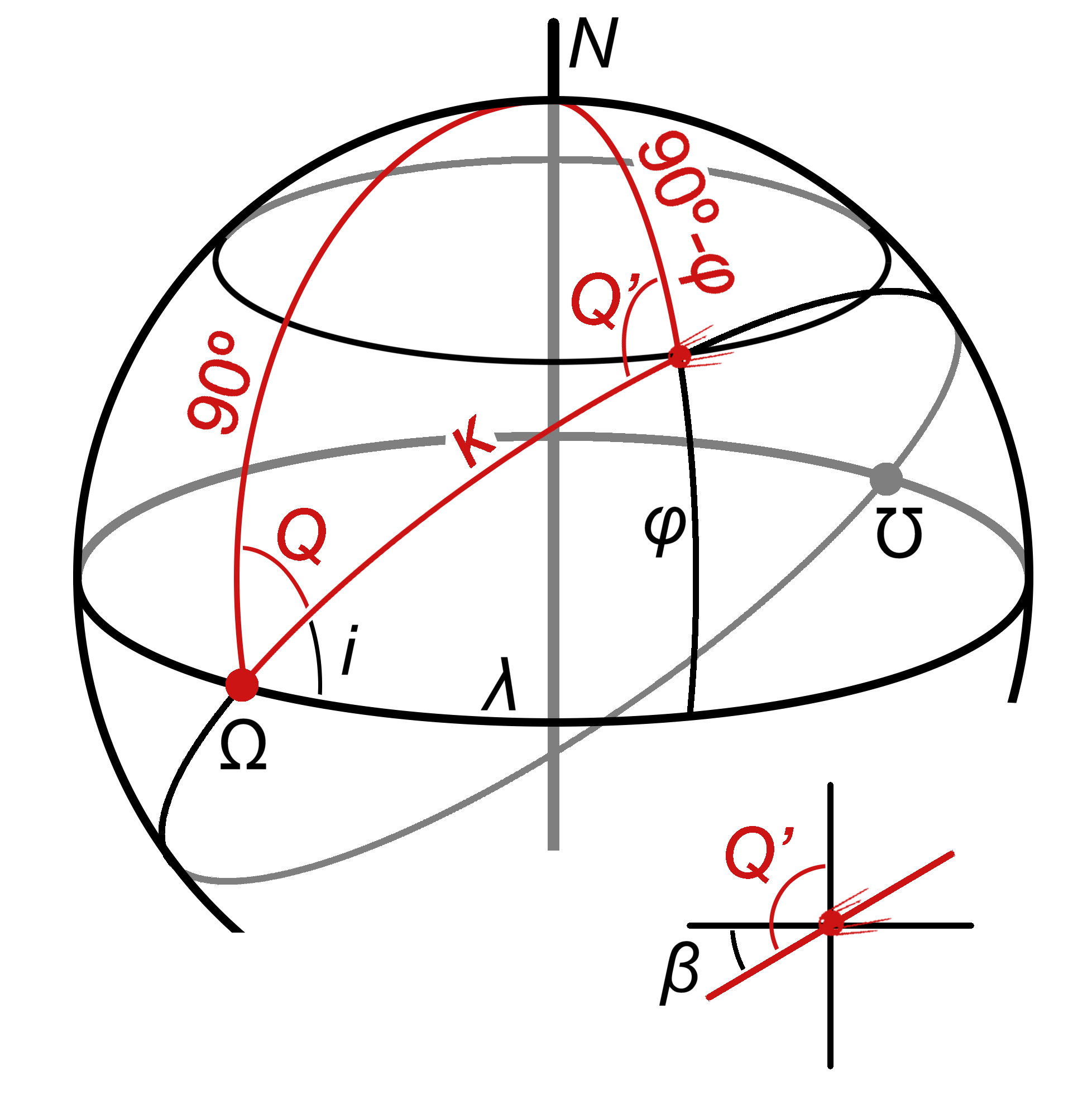}
       \caption{The spherical triangle to obtain the angle $\beta$
         between the orbit and the parallel of latitude $\phi$. Angle
         $\kappa$ indicates the longitude of the satellite measured
         from the ascending node along its orbit. With $\lambda$ we
         indicate the difference between the longitude of the
         satellite and that of the ascending node.}
       \label{fig:triangle}
     \end{figure}

     For a one single satellite in a circular orbit with inclination
     $i$ and orbital altitude $h_\mathrm{sat}$ we first obtain the
     angle $\beta$ between the orbit and any parallel, as a function
     of latitude $\phi$. At the nodes the parallel is the equator and
     we have $\beta = i$. The general situation is depicted in
     Fig.\,\ref{fig:triangle}. In that diagram $Q + i = 90^{\circ}$
     and $Q' = 90^\circ + \beta$. Our problem is finding $Q'$ as a
     function of known quantities. This is solved via the law of
     sines:

     \begin{eqnarray}
       \sin 90^\circ \sin Q = \sin(90^\circ - \phi) \sin Q';
       \\
       \sin(90^\circ - i) = \sin (90^\circ - \phi) \sin (90^\circ + \beta);
       \\
       \cos \beta = \frac{\cos i}{\cos \phi}. \label{eq:cosbeta}
     \end{eqnarray}
     
     The only satellite inside our orbit induces a linear density of
     satellites per angular unit along the orbit itself that is
     $\Lambda = 1/(2\pi)$. We have assumed that the satellite is
     uniformly distributed in time along its trajectory, what is true
     if the orbit is circular. Now we have to transform this angular
     linear density (satellites per unit angle along the orbit) into
     an angular surface density (satellites per unit solid angle on
     the orbital shell).

     It is worth noting that one single orbit, even considering that
     its only satellite is uniformly distributed along it with linear
     density $\Lambda$, in rigour will not induce an uniform surface
     density over the shell nor in longitude nor in latitude. In
     longitude, the density will be higher around the nodes and lower
     at the longitudes corresponding to extreme latitudinal excursions
     (at $90\degr$ from any node). In latitude we have more or less
     the opposite: the latitudinal density will be higher where the
     orbit is tangent to parallels (extreme latitude excursions) and
     lower at the equator. In any case, it is clear that the density
     will be zero outside the range $-i < \phi < i$.

     \begin{figure}[t]
       \includegraphics[width=0.48\textwidth]{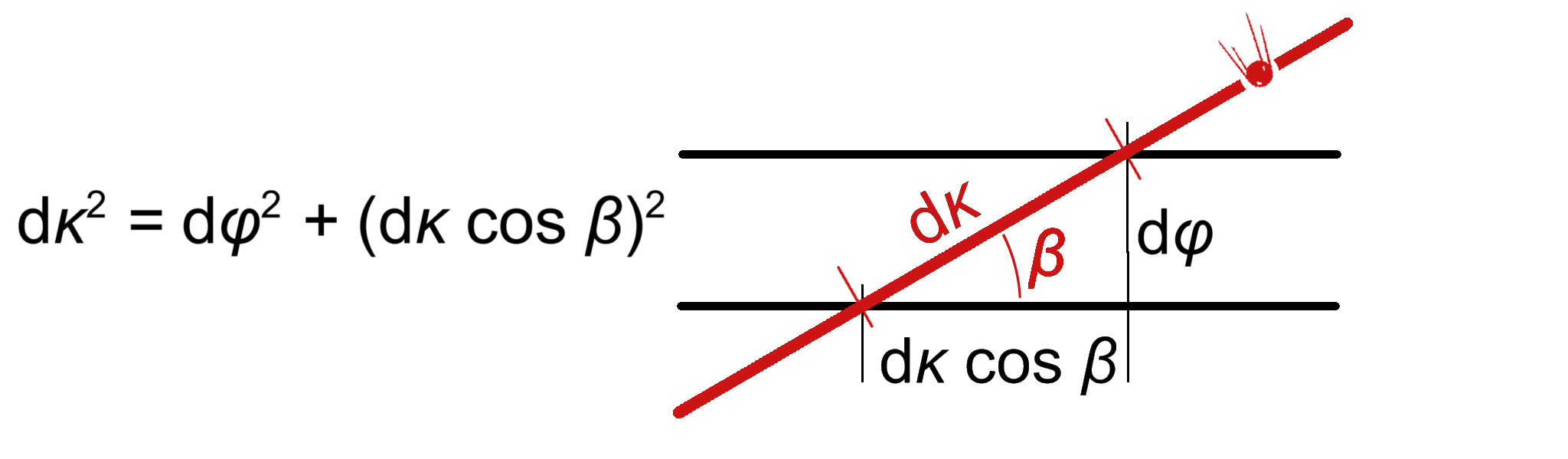}
       \caption{From angular linear density to angular latitudinal
         density.}
       \label{fig:differentials}
     \end{figure}

     However, heterogeneity of density in longitude is not important
     to our purposes, because it will smooth out after many
     iterations, due to the Earth rotation and to random
     initialization conditions. Also, in normal cases we will have
     many satellites inside the same orbit and many similar orbits
     inside one shell (with different longitudes of the ascending
     node), in such a way that the smoothing in longitude will be very
     efficient and fast. So, we assume, and impose, that the final
     density distribution has to have cylindrical symmetry and we can
     concentrate on the latitudinal behaviour.

     Let us depict a latitude band of infinitesimal angular width
     d$\phi$ at latitude $\phi$, as shown in
     Fig.\,\ref{fig:differentials}. The arc of the orbit span crossing
     the band is d$\kappa$ and this relation holds:

     \begin{equation}
       \mathrm{d}\kappa^2 = \mathrm{d}\phi^2 + (\mathrm{d}\kappa \cos \beta)^2.
     \end{equation}
     Then:
     \begin{equation}
       \mathrm{d}\kappa^2 = \frac{\mathrm{d}\phi^2}{1 - \cos^2 \beta} 
       ~ \Rightarrow ~ \mathrm{d}\kappa = \frac{\mathrm{d}\phi}{\sin \beta}.
     \end{equation}
     Introducing density $\Lambda$ to pass from arc $\kappa$ to
     satellites per unit arc in latitude $p=\Lambda \kappa$:
     \begin{equation}
       \mathrm{d}p = \Lambda \mathrm{d}l = \Lambda \frac{\mathrm{d}\phi}{\sin \beta} = 
       \frac{\mathrm{d}\phi}{2 \pi \sin \beta}.
     \end{equation}
     Next, we obtain the expression for the latitudinal angular
     probability density due to one single orbit with one single
     satellite:
     \begin{equation}
       \frac{\mathrm{d}p}{\mathrm{d}\phi} = \frac{1}{2 \pi \sin \beta} = \frac{1}{2 \pi  \sqrt{1-\cos^2 \beta}}.
     \end{equation}
     But, as shown in Eqn.\,\ref{eq:cosbeta}, $\cos \beta = \cos i / \cos \phi$ and, thus:
     \begin{equation}
       \frac{\mathrm{d}p}{\mathrm{d}\phi} =  
       \frac{\cos \phi}{2 \pi \sqrt{\cos^2 \phi - \cos^2 i}}.
     \end{equation}
     Note that this density takes into account only one side of the
     orbit, but each parallel is crossed twice by the same orbit, so
     that the final true latitudinal angular density is twice this
     value, $p_2 = 2p$:
     \begin{equation}
       \frac{\mathrm{d}p_2}{\mathrm{d}\phi} = \frac{\cos \phi}{\pi
         \sqrt{\cos^2 \phi - \cos^2 i}}.
     \end{equation}
     This latitudinal angular density would be measured in fractions
     of the sample per unit angle in latitude. We now divide by the
     length of a parallel at latitude $\phi$, $2\pi \cos\phi$, to
     transform this angular latitudinal density into true density per
     unit solid angle, and label the resulting probability density
     function as $c(\phi,i)$:
     \begin{equation}
       c(\phi,i) = \frac{\mathrm{d}p_2}{\mathrm{d}\Omega} = 
       \frac{1}{2 \pi^2  \sqrt{\cos^2 \phi - \cos^2 i}}.
     \end{equation}
     This density per unit solid angle can be converted into the true,
     physical surface density, in units of fractions of the sample per
     surface unit on the shell, by dividing by the radius of the shell
     squared, $(R_\oplus + h_{\mathrm{sat}})^2$. Also, by using that
     $\cos^2 \phi - \cos^2 i = \sin^2 i - \sin^2 \phi$, and that both
     expressions are differences of squares, we end up with four
     equivalent formulations of our probability surface density
     function:
     
     \begin{align}
       P(\phi,i,h_{\mathrm{sat}}) = \frac{\mathrm{d}p_2}{\mathrm{d}S} =
       \frac{1}{(R_\oplus +
         h_{\mathrm{sat}})^2}\frac{\mathrm{d}p_2}{\mathrm{d}\Omega}
       \\ = \frac{1}{2 \pi^2 (R_\oplus + h_{\mathrm{sat}})^2
         \sqrt{\cos^2 \phi - \cos^2 i}} \label{eq:firstform}\\ =
       \frac{1}{2 \pi^2 (R_\oplus + h_{\mathrm{sat}})^2 \sqrt{(\cos\phi
           + \cos i) (\cos \phi - \cos i)}} \\ = \frac{1}{2 \pi^2
         (R_\oplus + h_{\mathrm{sat}})^2 \sqrt{\sin^2 i - \sin^2 \phi}}
       \\ = \frac{1}{2 \pi^2 (R_\oplus + h_{\mathrm{sat}})^2
         \sqrt{(\sin i + \sin \phi)(\sin i - \sin \phi)}}.
     \end{align}
     Of the four options, we have elected using the last one.

     $P(\phi,i,h_{\mathrm{sat}})$ is a probability density
     distribution and its integral over the space occupied by the
     sample is unity, as stated in the main text of this article. This
     may be shown from any of the four forms of the function, but is
     easier from the first one, Eqn.\,\ref{eq:firstform}. Noting
     $R_\mathrm{sat} = R_\oplus + h_{\mathrm{sat}}$, we have:

     \begin{align}
       \int_{\mathrm{shell}}\frac{{\mathrm{d}}S}{2\pi^2R_\mathrm{sat}^2\sqrt{\cos^2\phi - \cos^2 i}} \\
       = \int_{\theta=0}^{\theta=2\pi}\int_{\phi=-i}^{\phi=+i}
       \frac{R_\mathrm{sat}^2 \cos\phi \,{\mathrm{d}}\phi \,\mathrm{d}\theta}
            {2\pi^2R_\mathrm{sat}^2\sqrt{\cos^2\phi - \cos^2 i}} \\
            = \frac{1}{\pi}\int_{-i}^{+i}\frac{\cos \phi \,\,\mathrm{d}\phi}{\sqrt{\cos^2\phi - \cos^2 i}} ~.
     \end{align}
     This last integral is reduced to a straightforward arcsine with
     the variable change $x=\sin\phi$:
     
     \begin{align}
       \frac{1}{\pi}\int_{x=-\sin i}^{x=\sin i}\frac{\mathrm{d}x}{\sqrt{\sin^2 i - x^2}} =
       \frac{1}{\pi}\left[ \arcsin{\frac{x}{\sin i}} \right]_{-\sin i}^{\sin i} = 1.
     \end{align}
     
     \subsection{Geocentric and topocentric position of the satellites}
     \label{sec:ap:geopos}
     The number of trails affecting an exposure, given by
     Eqn.~\ref{eq:parabolic} in \S\ref{ssec:traildensity}, relies on
     the apparent angular velocity of the satellites at that position
     in the sky.

     To compute that velocity, we first have to relate the geocentric
     longitude and latitude of the satellite ($\theta$, and $\phi$) to
     the topocentric position vector of a satellite (pointing
     direction of the telescope, given by $\vec{OS}$), which is
     characterized by its right ascension (more precisely, the hour
     angle) and declination, or equivalently by its azimuth and
     elevation.

     Writing the vectorial relation between the centre of Earth C, the
     observer O and the satellite S,
     \begin{equation}
       \vec{CS} = \vec{CO} + \vec{OS} ~, \label{eq:apvector}
     \end{equation}
     in geocentric rectangular coordinates ($x$ equatorial at the
     meridian of the observer, $z$ to the pole, and $y$ completing the
     referential) using the longitude and latitude of the observer
     ($\theta_{\mathrm o}$, $\phi_{\mathrm o}$) and of the satellite
     ($\theta$, $\phi$), Eqn.~\ref{eq:apvector} becomes
     \begin{align}
       R_\mathrm{sat} \cos \theta \cos \phi = d x_{OS} + R_\oplus  \cos \phi_{\mathrm o} \label{eq:apCS1}\\ 
       R_\mathrm{sat}  \sin \theta \cos \phi = d y_{OS}  \label{eq:apCS2}\\
       R_\mathrm{sat}  \sin \phi = d z_{OS} + R_\oplus   \sin \phi_{\mathrm o}  \label{eq:apCS3},
     \end{align}
     where $R_\oplus$ is the radius of the Earth, $R_\mathrm{sat} =
     R_\oplus + h_\mathrm{sat}$ is the radius of the satellite orbit,
     $d = |OS|$, and $(x_{OS}, y_{OS}, z_{OS})$ the unit vector so
     that $\vec{OS} = d (x_{OS}, y_{OS}, z_{OS})$. The coordinates are
     obtained from the hour angle and declination of the satellite. We
     eliminate $\theta$ and $\phi$ by summing quadratically these
     three equations:
     \begin{equation}
       d^2 + 2 R_\oplus ( x \cos \phi_\mathrm{o} + z \sin \phi_\mathrm{o} ) d - (h_\mathrm{sat}^2 + 2 R_\oplus h_\mathrm{sat}) = 0~.
       \label{eq:dsquared}
     \end{equation}
     Solving this quadratic equation leads to two solutions for
     $d$. The positive one is the distance to the satellite (the
     negative one is the distance to the satellite shell in the
     opposite direction, below ground).

     Once $d$ is computed, Eqn.~\ref{eq:apCS3} can be solved for
     $\phi$, and then Eqn.~\ref{eq:apCS1} and \ref{eq:apCS2} for
     $\theta$. We have now the geocentric longitude and latitude
     ($\theta$, and $\phi$) of the satellite that corresponds to a
     specific, topocentric pointing direction on the local sky.

     \subsection{Impact angle of the line of sight with the shell}
     \label{sec:ap:impact}
     This is the angle $\alpha$ between the line of sight and the
     normal to the shell, i.e. the angle $\widehat{CSO}$. From the
     cosine law in triangle CSO, we have:
    \begin{equation}
      \cos \alpha = \frac{R_\mathrm{sat}^2 + d^2 - R_\oplus^2}{2 d
        R_\mathrm{sat}}~.\label{eg:apalpha}
    \end{equation}

    \subsection{Apparent angular velocity}\label{sec:ap:apparentangvel}
    A satellite observed in a given detection may either be on the
    north-bound half of its orbit, or on the south-bound
    half. Considering the right spherical triangle including the
    satellite, the ascending node of its orbit and the equator (see
    Fig. \ref{fig:triangle}), we obtain the longitude difference
    $\lambda$ between the longitude of the satellite and that of the
    ascending nodes from:
    \begin{equation}
      \sin \lambda = \frac{\tan \phi }{\tan i} ~,
    \end{equation}
    from which we get the longitudes of the ascending nodes for both
    possible orbits:
    \begin{align}
      \Omega_\mathrm{N} = \theta - \lambda ~,\\
      \Omega_\mathrm{S} = \theta + \lambda + \pi .
    \end{align}

    The geocentric velocity vector of a satellite is obtained from:
    \begin{equation}
      \vec{v}_\mathrm{sat} = v ~ \vec{N} \times \vec{OS},
    \end{equation}
    where $v$ is obtained from elementary celestial mechanics as $ v =
    \sqrt{GM_\oplus/ R_\mathrm{sat}}$, $\vec{N}$ is the unit vector
    normal to the orbit (built from the longitude of the ascending
    node and $i$), $\vec{N} = \vec{CS}/|CS|$ is the unit vector
    pointing from the centre of the Earth to the satellite and
    $\times$ marks the cross product. This is repeated for both
    $\Omega_\mathrm{N}$ and $\Omega_\mathrm{S}$.

    The topocentric velocity vectors are obtained by subtracting the
    geocentric velocity vector of the observer from that of the
    satellite. The components of these vectors perpendicular to the
    line of sight OS are the two apparent angular speeds
    $\omega_\mathrm{N}$ and $\omega_\mathrm{S}$.

    \subsection{Declination of the constellation edges}\label{sec:ap:constedge}
    \begin{figure}[t]
      \includegraphics[width=0.48\textwidth]{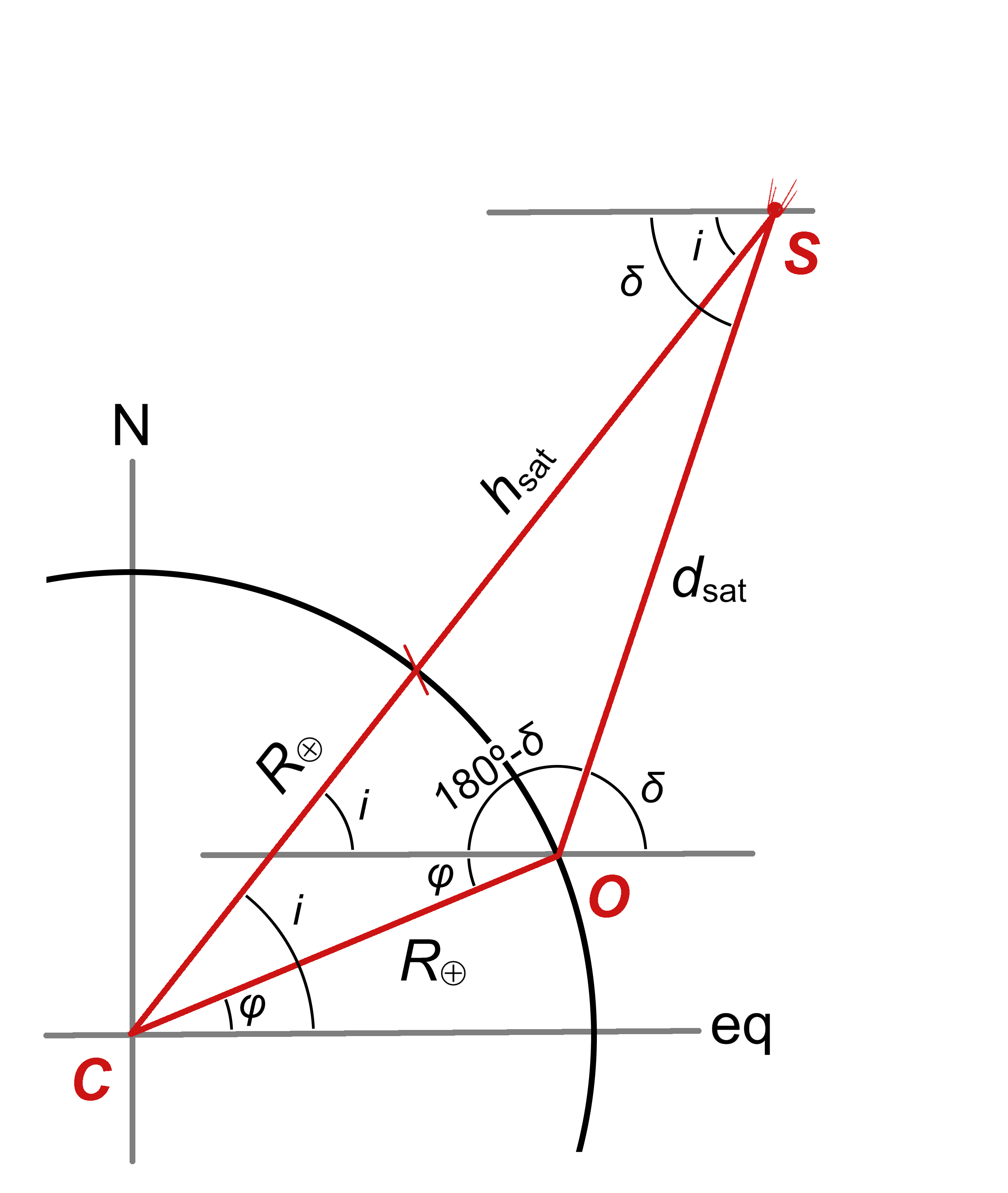}
      \caption{Geometry of shell boundaries at the local meridian. C:
        Earth centre. O: Observatory. S: Intersection of the northern
        shell boundary with the local meridian.}
      \label{fig:boundary}
    \end{figure}

    Equation~\ref{eq:boundary} relates observatory latitude $\phi$ and
    shell inclination $i$ to the declination $\delta$ at which the
    shell boundary cuts the local meridian. It is deduced from
    elementary geometry (plane sinus theorem) applied to triangle COS
    in Fig.\,\ref{fig:boundary}:
    \begin{align}
      \frac{\sin[180^{\circ}+(\phi-\delta)]}{R_{\oplus}+h_{\mathrm{sat}}} = \frac{\sin(\delta-i)}{R_\oplus} \Rightarrow \\
      \sin{(\delta - \phi)} = \frac{R_{\oplus}+h_\mathrm{sat}}{R_{\oplus}} \sin{(\delta - i)}.
    \end{align}
    The same elementary procedure applied to other combinations of
    sides and angles of triangle COS leads to two additional
    relations:
    \begin{align}
      \sin(\delta-\phi)=\frac{R_\oplus+h_\mathrm{sat}}{d_\mathrm{sat}}\sin(i-\phi) \\
      \sin(\delta-i) = \frac{R_\oplus}{d_\mathrm{sat}} \sin(i-\phi).
    \end{align}
    To use these two relations it is necessary to introduce the
    distance $d_\mathrm{sat}$ from the observatory to the shell
    boundary at the meridian, that is also easily deduced from
    Fig\,\ref{fig:boundary} through the plane cosinus theorem:
    \begin{equation}
      d_{\mathrm{sat}}^2 = 2 R_\oplus\left( R_\oplus + h_\mathrm{sat}\right) \left[ 1 - \cos \left( i-\phi \right) \right] + h_{\mathrm{sat}}^2.
    \end{equation}

    \section{Index of symbols}
    \label{sec:ap:index}
    This is a list of the symbols used in this paper, together with a
    short definition.
 
    \begin{itemize}
    \item $\alpha$: impact angle between the line-of-sight and the
      spherical shell at the satellite
    \item $\alpha_\odot$: solar phase angle, angle
      Sun-Satellite-observer
    \item $\chi$: airmass
    \item $\delta$: declination angle
    \item $\delta_\mathrm{sat}$: density of satellites in a field of
      view, in satellites per unit solid angle (n/sq.deg).
    \item $\delta_\mathrm{trail}$: density of satellite trails
      crossing a the field of view, in satellite per linear degree per
      unit of time (deg$^{-1}$\,s$^{-1}$)
    \item $\kappa$: angular coordinate of the satellite measured along
      its orbit, from the ascending node (sometimes called {\em phase}
      in engineering, but we have avoided that denomination to prevent
      confusions with what we normally understand by phase angle
    \item $\Lambda$: linear density of satellites per unit angle along the
      orbit
    \item $\lambda$: difference of longitudes of the satellite and the
      ascending node of its orbit (longitude of the satellite measured
      from the ascending node)
    \item $\omega_\mathrm{sat}$: apparent angular velocity of the
      satellite as seen by the observer
    \item $\rho(\phi,i,h_{\mathrm {sat}}) = N_\mathrm{sat} b$: density
      of satellites at latitude $\phi $
    \item $\theta, \phi$: geocentric longitude and latitude of the
      satellite
    \item $P(\phi,i,h_{\mathrm {sat}})$: probability density function
      of finding a satellite (with $i, h_{\mathrm {sat}}$) at latitude
      $\phi$
    \item $d$, $d_\mathrm{sat}$: distance between the observer and the
      satellite
    \item $d_{\mathrm{sat}\odot}$: distance from the Sun to the
      satellite, $\sim 1$~AU
    \item $h_\mathrm{sat}$: Altitude of the satellites' orbit, km
    \item $i$: orbital inclination of the satellites' orbit, degrees
    \item $k$: atmospheric extinction coefficient, mag/airmass
    \item $L_\mathrm{fov}$: diameter of the (circular) field of view,
      degrees
    \item $l$: latitude of the observer
    \item $m_\mathrm{1000km}$, $m_\mathrm{500km}$: zenithal magnitude
      of the satellite normalized to a distance $d_\mathrm{sat} =
      1000$~km, 500~km
    \item $m_\mathrm{eff}$: effective magnitude of the satellite: the
      magnitude of a static, point-like object that, during the
      exposure time considered, would produce the same accumulated
      intensity in a resolution element than the satellite crossing
      over a resolution element
    \item $m_\mathrm{sat}$: (visual) magnitude of the satellite
    \item $m_{\odot}$: magnitude of the Sun
    \item $n_\mathrm{1}$: number of satellites in a single orbital
      plane
    \item $n_\mathrm{plane}$: number of orbital planes in the
      constellation shell
    \item $n_\mathrm{sat}$: number of satellites present in the field
      of view
    \item $n_\mathrm{trail}$: number of satellite trails crossing the
      field of view during the exposure
    \item $N_\mathrm{sat}$: total number of satellites in the
      constellation shell
    \item $p$: geometric albedo of the satellite
    \item $p(S)$: probability of finding a satellite in region $S$
    \item $r$: angular size in the plane of sky of a detector's
      resolution element or pixel
    \item $R_\oplus$: radius of the spherical Earth,  % 6378.1371km
    \item $R_\mathrm{sat}$: radius of the (spherical) satellite or, in
      other contexts, radius of the (circular) orbit of a satellite
    \item $t_\mathrm{exp}$: exposure time, in seconds
    \item $t_\mathrm{eff}$: effective exposure time for a satellite,
      the duration it takes the satellite to cross a resolution
      element of the detector
    \end{itemize}
  \end{appendix}
   
\end{document}